\documentclass[nofootinbib,twocolumn,preprintnumbers,superscriptaddress,aps]{revtex4-2}
\usepackage{amsmath}
\usepackage{amssymb}
\usepackage{graphicx}
\usepackage{booktabs}
\usepackage{color}
\usepackage{hyperref}
\usepackage{xspace}
\usepackage{multirow}
\usepackage{array}
\usepackage[normalem]{ulem}
\usepackage{enumitem}
\usepackage[table]{xcolor}
\makeatletter
\g@addto@macro\bfseries{\boldmath}
\makeatother

\newcommand{\as}{\alpha_s}

\newcommand{\cS}{\mathcal{S}}
\newcommand{\order}[1]{{\cal O}\!\left(#1\right)}

\newcommand{\GeV}{\,\text{GeV}}
\newcommand{\TeV}{\,\text{TeV}}
\newcommand{\jet}{\text{jet}}

\newcommand{\DR}{\Delta R}

\newcommand{\Jhard}{\mathcal{J}_{\rm hard}}
\newcommand{\JhardIRC}{\mathcal{J}_{\rm hard+IRC}}

\newcommand{\lnptmin}{\ln p_{t,\min}}
\newcommand{\lnptmax}{\ln p_{t,\max}}
\newcommand{\ptmin}{p_{t,\min}}
\newcommand{\ptmax}{p_{t,\max}}

\newcommand{\NC}{N_\text{\textsc{c}}}

\newcommand{\nohat}{}

\colorlet{shadecolor}{gray!30}
\newcommand{\ok}{$\checkmark$}
\newcommand{\no}{\color{red}$\times$}
\newcommand{\hc}{\cellcolor{shadecolor}}
\newcolumntype{C}[1]{>{\centering\let\newline\\\arraybackslash\hspace{0pt}}m{#1}}

\definecolor{darkgreen}{rgb}{0,0.4,0}
\definecolor{grey}{rgb}{0.5,0.5,0.5}
\definecolor{orange}{rgb}{0.9,0.5,0.0}

\newcommand{\cut}{\text{cut}}

\usepackage[caption=false]{subfig}

\newcommand{\logbook}[2]{}

\newcommand{\OXaff}{Rudolf Peierls Centre for Theoretical Physics, Parks Road, Oxford OX1 3PU, UK}
\newcommand{\ASCaff}{All Souls College, Oxford OX1 4AL, UK}
\newcommand{\MITaff}{Center for Theoretical Physics, Massachusetts Institute of Technology, Cambridge, MA 02139, USA}

\begin{document}

\title{Flavoured jets with exact anti-$k_t$ kinematics and \\
  tests of infrared and collinear safety}

\preprint{MIT-CTP 5564, OUTP-23-06P}

\author{Fabrizio Caola}           \affiliation{\OXaff}
\author{Rados\l{}aw Grabarczyk}    \affiliation{\OXaff}
\author{Maxwell L. Hutt}
\affiliation{\OXaff}\affiliation{The Blackett Laboratory, Imperial
  College London, Prince Consort Road, London, SW7 2AZ, UK}
\author{Gavin P. Salam}            \affiliation{\OXaff} \affiliation{\ASCaff}
\author{Ludovic Scyboz}           \affiliation{\OXaff}
\author{Jesse Thaler}            \affiliation{\MITaff}

\begin{abstract}
  We propose extensions of the anti-$k_t$ and Cambridge/Aachen
  hierarchical jet clustering algorithms that are designed to retain
  the exact jet kinematics of these algorithms, while providing an
  infrared-and-collinear-safe definition of jet flavour at any fixed
  order in perturbation theory.
  Central to our approach is a new technique called Interleaved Flavour 
  Neutralisation (IFN), whereby the treatment of flavour is integrated
  with, but distinct from, the kinematic clustering.
  IFN allows flavour information to be meaningfully accessed
  at each stage of the clustering sequence, which enables a consistent
  assignment of flavour both to individual jets and to their substructure.
  We validate the IFN approach using a dedicated framework for
  fixed-order tests of infrared and collinear safety, which also reveals
  unanticipated issues in earlier approaches to
  flavoured jet clustering.
  We briefly explore the phenomenological impact of IFN with anti-$k_t$ jets
  for benchmark tasks at the Large Hadron Collider.
\end{abstract}

\maketitle

% 
% For the purpose of Open Access, the authors have applied a CC BY
% public copyright licence to any Author Accepted Manuscript (AAM)
% version arising from this submission.

{\small
\tableofcontents
}

%======================================================================
\section{Introduction}

The use of jet clustering algorithms is essential and ubiquitous at colliders.
Jet algorithms relate collimated sprays of energetic hadrons to the
underlying concept of hard, perturbative quarks and gluons (or, more generally, partons).
In the vast majority of cases, only the kinematics of the
resulting jets are used for analysis.
Insofar as jets are meant to represent the underlying
partonic structure of an event, though, it is natural to ask whether jets can
also reflect the flavour of the underlying partons, for example their
quark or gluon nature.
The question of how to formulate a jet algorithm where the flavours assigned to jets
are infrared and collinear (IRC) safe was first posed in
2006~\cite{Banfi:2006hf,Banfi:2007gu}.
The algorithm developed there, flavour-$k_t$, based on a modification
of the $k_t$ algorithm~\cite{Catani:1991hj,Catani:1993hr,Ellis:1993tq}, appeared to
be successful in this task.
However, one of the characteristics of flavour-$k_t$ was that the kinematics of the
resulting jets depended on the flavour of the underlying constituents
being clustered.

In modern jet usage, where the subsequently developed anti-$k_t$
algorithm~\cite{Cacciari:2008gp} has found widespread applications, a
flavour-induced modification of the jets' kinematics is undesirable.
Notably, it has been found to complicate unfolding
corrections~\cite{Gauld:2020deh}.
Nevertheless, there are situations where IRC-safe flavoured jet algorithms would be highly beneficial.
For example, the question of IRC-safe jet flavour has recently come to the fore in the context of heavy-flavour
jets~\cite{Caletti:2022hnc,Caletti:2022glq,Czakon:2022wam,Gauld:2022lem}.
IRC safety in this instance ensures that flavoured jet cross
sections do not contain any logarithms of the ratio of the jet transverse
momentum $p_t$ to the quark mass $m_q$.
It also makes it possible to use an $m_q = 0$ approximation in fixed-order perturbative
calculations~\cite{Weinzierl:2006yt,Trocsanyi:2015zma,Ferrera:2017zex,Caola:2017xuq,Gauld:2019yng,Gauld:2020deh,Czakon:2020coa,Hartanto:2022qhh,Hartanto:2022ypo,Czakon:2022khx,Gauld:2023zlv},
with an expectation that any missing contributions are suppressed by
powers of $m_q/p_t$.

In this article, we present a new strategy for flavoured jet finding called Interleaved Flavour Neutralisation (IFN), which is designed to combine an
IRC-safe definition of jet flavour with the IRC-safe kinematics of sequential clustering.
We will study IFN with two generalised-$k_t$-style jet algorithms, the
anti-$k_t$ algorithm, used extensively at the Large Hadron Collider
(LHC), and the Cambridge/Aachen
(C/A) algorithm~\cite{Dokshitzer:1997in,Wobisch:1998wt}, widely favoured for jet substructure studies.
In the case of the anti-$k_t$ algorithm, our objectives are similar to
those of the recent ``flavoured anti-$k_t$''~\cite{Czakon:2022wam} and
``flavour dressing''~\cite{Gauld:2022lem} algorithms, which respectively achieve
approximate and exact anti-$k_t$ kinematics.
Like flavour dressing, IFN yields exact anti-$k_t$ (or
C/A) kinematics, but because it integrates flavour information at each 
stage of the clustering sequence, it is a viable candidate for jet substructure studies.
We also carry out a much more extensive set of IRC safety tests than
in any prior work, which support the conclusion that IFN is IRC safe,
at least through order $\alpha_s^6$.
These tests also reveal unexpected and subtle issues in the default formulations of all prior flavoured jet
algorithms.

We focus on the theoretical definition of jet flavour, leaving a study of experimental issues to future work.
The extent to which any IRC-safe flavour algorithm can be adopted
experimentally is an open question.
Even when identifying heavy-flavour jets, where collinear singularities are regulated by a non-zero $m_q$, such algorithms
would typically require the identification of all heavy-flavoured hadrons in
an event.
That is challenging when there are multiple heavy-flavoured hadrons in
a single jet, or when some of the heavy-flavoured hadrons have low
momenta.%
\footnote{Though as we will see in Sec.~\ref{sec:pheno-ttbar} for the $t\bar t$ process, this may be less of an issue than one might fear.}
Despite these experimental subtleties, the underlying question of IRC-safe flavour identification
remains conceptually important.
Jet flavour can provide a valuable tool in a range of theoretical
work,
for example in matching parton showers and fixed-order
calculations~\cite{Hoeche:2009rj}.
One can further anticipate that it will be useful in testing
logarithmic accuracy for flavour-related aspects of parton
showers~\cite{Karlberg:2021kwr,vanBeekveld:2022ukn}.

The remainder of this article is organised as follows.
In Sec.~\ref{sec:reminders}, we review the key features of
widely-adopted jet algorithms, and some of the issues that arise when
flavour tagging is sought.
We also briefly describe existing proposals for those flavoured jet algorithms that aim to achieve all-order IRC
safety~\cite{Banfi:2006hf,Banfi:2007gu,Czakon:2022wam,Gauld:2022lem}.
In Sec.~\ref{sec:genkt-FN}, we outline our general design aims for a
modern flavoured jet algorithm and present a
concrete realisation via IFN.
That section also includes a discussion of some of the subtle considerations
brought about by IRC-safety requirements.
In Sec.~\ref{sec:IRC-tests}, we present the framework that we
developed to explore IRC-safety issues in some depth (a 
substantial extension of the approach developed some time ago for
testing the SISCone jet algorithm~\cite{Salam:2007xv}), which we apply both to our IFN proposal
and to earlier flavoured jet algorithms.
These tests expose unanticipated issues in earlier proposals,
many of them connected with the treatment of initial-state radiation
in a hadron collider context.
In some cases, we identify simple adaptations of the original
algorithms that should make them IRC safe.
In Sec.~\ref{sec:pheno}, we perform three benchmark phenomenological studies
to illustrate the behaviour of various flavoured jet algorithms,
restricting ourselves to the ones that pass our IRC-safety tests. 
In Sec.~\ref{sec:ee-generalisations}, we briefly present the
adaptation of our approach to $e^+e^-$ colliders.
We conclude in Sec.~\ref{sec:conclusions}.

Additional material is presented in the appendices.
In App.~\ref{sec:double-soft-structure},
we review some features of the double-soft quark emission current that we
used for our analyses.
In App.~\ref{sec:IRCapp-IFN} we perform numerical tests to justify some of the design decisions we made for the IFN algorithm.
In App.~\ref{sec:list-ir-risky}, we provide detailed analyses of the
main IRC-safety issues that we encountered in this work.
In App.~\ref{sec:final-summary-plots}, we present summary plots of
IRC-safety tests for those algorithms that we expect to be IRC safe.

%======================================================================
\section{Reminders about existing jet algorithms}
\label{sec:reminders}

In this section, we briefly review standard jet algorithms and their interplay with jet flavour, including the original flavour-$k_t$ approach~\cite{Banfi:2006hf}.
To avoid confusion, we refer to the flavoured anti-$k_t$ algorithm of Ref.~\cite{Czakon:2022wam} as ``CMP'', and the flavour dressing algorithm of Ref.~\cite{Gauld:2022lem} as ``GHS''.
Throughout this section and most of this article, we concentrate on longitudinally invariant hadron-collider algorithms, with a brief mention of an $e^+ e^-$ adaptation in Sec.~\ref{sec:ee-generalisations}.

\subsection{Flavourless kinematic clustering}

Let us start with a reminder of how the generalised-$k_t$ algorithm
works.
It employs distances $d_{ij}$ between each pair of
pseudojets%
\footnote{Recall that a pseudojet may be either a single
  particle or the combination of more than one particle arising from
  an earlier stage of the clustering.}
  $i$ and $j$ and $d_{iB}$
between each pseudojet $i$ and the beam:
\begin{subequations}
  \label{eq:genkt-distances}
  \begin{align}
    \label{eq:genkt-dij}
    d_{ij} &= \min(p_{ti}^{2p}, p_{tj}^{2p})
             \,\frac{\Delta R_{ij}^2}{R^2},
             %\qquad
    \\
    \DR_{ij}^2 &= (y_i-y_j)^2 + (\phi_i - \phi_j)^2,
    \\
    \label{eq:genkt-diB}
    d_{iB} &= p_{ti}^{2p},
  \end{align}
\end{subequations}
where $p_{ti}$, $y_i$ and $\phi_i$ are respectively the transverse
momentum, rapidity and azimuth of $i$
($y_i = \frac12\ln \frac{E_i + p_{zi}}{E_i - p_{zi}}$).
The algorithm has two parameters, the jet radius $R$, which sets the
angular reach of the jets, and the power $p$, which sets the nature of
the algorithm: $-1$, $0$, $1$ respectively for the anti-$k_t$~\cite{Cacciari:2008gp},
Cambridge/Aachen~\cite{Dokshitzer:1997in,Wobisch:1998wt} and $k_t$
algorithms~\cite{Catani:1993hr,Ellis:1993tq}. 
The algorithm starts with all event particles and proceeds as follows:
\begin{enumerate}
\item identify the smallest of the $d_{ij}$ and $d_{iB}$ among all $i$
  and $j$ at this stage of the clustering;
\item if it is a $d_{ij}$, recombine $i$ and $j$ into a single
  new pseudojet and return to step $1$;
\item if it is a $d_{iB}$, declare $i$ to be a jet and remove it from
  the list of pseudojets to be considered at subsequent clustering
  steps; return to step 1.
\end{enumerate}
The clustering stops once no pseudojets are left to be clustered.
Given the resulting jets, it is common to consider only the subset
that pass minimum $p_t$ (and maximum rapidity or
pseudo-rapidity\footnote{The jets may be massive, and as a result
  pseudo-rapidity is not advised~\cite{Gallicchio:2018elx}.})  constraints.

%----------------------------------------------------------------------
\subsection{Flavour via recombination scheme}
\label{sec:flavour-recomb}

A crucial element of the jet definition is the choice of recombination
scheme.
The most common is the (somewhat inappropriately named) $E$ scheme, in
which 4-momenta are simply added.
Flavour is usually not considered within standard jet algorithms, but
it is useful to introduce three potential flavour recombination
schemes:
\begin{itemize}
\item \textbf{Any-flavour scheme}:
 This scheme is relatively close to typical experimental
 practice for $b$- and $c$-tagging.
 Here, any recombination that involves non-zero flavour, e.g.\ $q+g$,
 $\bar q+g$, or $q+\bar q$, yields a flavoured result. 
  From a theoretical point of view, this scheme is collinear unsafe for massless quarks
  due to the collinear divergence of $g \to q\bar q$ splitting.
  For massive quarks, as in the case of $b$ and $c$ production, this scheme is logarithmically sensitive to the quark mass.
  We will further consider this ``any-flavour'' scheme only in a phenomenological context in Sec.~\ref{sec:pheno-ttbar}.

\item \textbf{Net-flavour scheme}:
This is a theoretically better-motivated scheme that considers the net
  flavour in the recombination.
  In this scheme, a $q$ carries flavour, a $\bar q$ carries
  anti-flavour, and a $q\bar q$ carries no flavour.
  This ``net-flavour'' scheme resolves the collinear unsafety for $g\to q\bar q$ splitting.

\item \textbf{Flavour modulo-$2$ scheme}:
Typically for heavy flavour at hadron level, it is not
  conceptually possible to distinguish flavour from anti-flavour,
  e.g.\ because of $B_0{-}\bar B_0$ oscillations.
  In such a situation, one may consider a ``flavour modulo-$2$'' scheme
  (see e.g.\ Ref.~\cite{Banfi:2007gu}).
  Specifically, $b$ and $\bar b$ are
  treated as equivalent while $b\bar b$, $bb$ and $\bar b \bar b$ are
  all considered to be flavourless.
  This scheme also resolves the issue of collinear unsafety for
  $g\to q\bar q$ splitting.
\end{itemize}

\begin{figure}
  \centering
  \includegraphics[width=0.6\columnwidth]{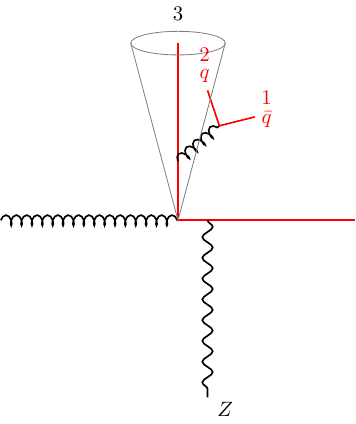}
  \caption{Classic problematic flavour configuration at NNLO.
  A soft gluon at large angle splits to a $\bar q q$ pair (labelled 1 and 2),
  and the flavour of the hard jet (numbered 3) is polluted by the flavour
  of 2, while 1 ends up outside the jet.
  }
  \label{fig:classic-config}
\end{figure}

While the net flavour and modulo-$2$ options ensure that the jet
flavour is unaffected by collinear divergences for $g \to q\bar q$
splittings, they still exhibit IRC safety issues for jet flavour at
higher orders, at least when used with standard jet algorithms.
This occurs at next-to-next-to-leading order (NNLO), as discussed in Ref.~\cite{Banfi:2006hf} and illustrated
in Fig.~\ref{fig:classic-config}
(see App.~\ref{sec:double-soft-structure} for further discussion about the matrix element for this process).
Specifically, when a soft gluon splits to a large-angle $q\bar q$
pair, one or other of the resulting soft quarks can be 
clustered with a hard jet and the net-flavour and modulo-$2$
recombination schemes result in an IRC-unsafe flavour for hard
jets, with the divergence appearing as $\as^2 \ln p_{t,\jet}/m_q$ for a
finite quark mass $m_q$.
This is the classic problem when attempting to obtain IRC-safe jet
flavour.

When considering more than one flavour (e.g.\ all of $udscb$),
flavour recombination is typically applied separately for each
flavour. This may be done either within a single run of the algorithm or (for
algorithms where the flavour does not affect the jet kinematics)
applying the flavour part of the jet clustering in one separate run
of the algorithm for each kind of flavour.

%----------------------------------------------------------------------
\subsection{Existing flavoured jet algorithms}

We now review three jet-flavour definitions that aim to achieve all-order IRC safety
(see Refs.~\cite{Buckley:2015gua,Caletti:2022hnc,Caletti:2022glq} for alternative definitions of jet flavour).

\subsubsection{Flavour-$k_t$}
The flavour-$k_t$~\cite{Banfi:2006hf} algorithm took the approach
of using a net or modulo-$2$ flavour scheme,
while modifying the clustering
distances relative to Eq.~(\ref{eq:genkt-distances}).
Specifically, it modifies the standard $k_t$ ($p=1$) distance when the
softer of $i$ and $j$ is flavoured
\begin{multline}
  \label{eq:flavkt-dij}
  d_{ij}^\text{flav-$k_t$} =  
  [\max(p_{ti},p_{tj})]^{\alpha} [\min(p_{ti},p_{tj})]^{2-\alpha}
  \,\frac{\Delta R_{ij}^2}{R^2}\,,
  \\
  \text{if softer of $i$ and $j$ is flavoured}\,,
\end{multline}
with the parameter $\alpha$ usually taken to be $1$ or
$2$.\footnote{The $\alpha=1$ variant evokes a
  longitudinally-invariant extension of the classic JADE (squared invariant
  mass) clustering distance~\cite{JADE:1986kta,JADE:1988xlj}.
  The well-known drawback of the JADE distance, namely that early in
  the sequence it can cluster soft pairs going in opposite directions,
  is precisely the behaviour needed to resolve the classic jet-flavour
  IRC safety issue of Fig.~\ref{fig:classic-config}. }
This has the consequence that the $d_{ij}$ for the clustering of a soft
flavoured particle with a significantly harder particle is much larger
than the $d_{ij}$ for two similarly soft particles.
As a result, the soft particles cluster first, resolving the original
IRC safety issue of Fig.~\ref{fig:classic-config}.
Note that flavour-$k_t$ also uses a modified $d_{iB}$ distance for
flavoured particles.
The details are best obtained from the original article~\cite{Banfi:2006hf}, however the
essence of the modified beam distance is that one uses the same kind
of construction as in Eq.~(\ref{eq:flavkt-dij}),
\begin{equation}
  \label{eq:flavkt-diB-main-text}
  d_{iB}^\text{flav-$k_t$}  = [\max(p_{ti},p_{tB}(y_i))]^{\alpha} [\min(p_{ti},p_{tB}(y_i))]^{2-\alpha},
\end{equation}
with $p_{tB}(y)$ a rapidity-dependent hardness scale.
In the central region, $p_{tB}(y)$ is of the same order as the overall
event hardness.

Relative to the standard $k_t$ algorithm, the flavour-$k_t$ algorithm
can significantly alter the kinematics of the clustering of hard
flavoured jets.
For example in the presence of a hard $b\bar b$ pair, the
flavour-$k_t$ algorithm can cluster them even when
$\Delta R_{b\bar b} > R$, as observed e.g.\ in
Ref.~\cite{Behring:2020uzq} (see also the discussion in
Sec.~\ref{sec:pheno-WH}).

\subsubsection{Flavour anti-$k_t$ (``CMP'')}

The algorithm of Ref.~\cite{Czakon:2022wam}, there called ``flavour
anti-$k_t$'', will be referred to here as CMP, to avoid ambiguity with
other flavour anti-$k_t$ algorithms.
As in the flavour-$k_t$ algorithm, it is to be used with net-flavour
or modulo-$2$ flavour recombination.
It modifies the anti-$k_t$ ($p=-1$) $d_{ij}$ distance
when $i$ and $j$ are oppositely flavoured
\begin{multline}
  \label{eq:flav-antikt-dij}
  d_{ij}^\text{flav-anti-$k_t$} =
  d_{ij}^\text{anti-$k_t$}
  \times
  \mathcal{S}_{ij}\,,
  \\
  \text{if $i$ and $j$ are oppositely flavoured}\,,
\end{multline}
where
\begin{equation}
  \label{eq:flav-antikt-Sij}
  \mathcal{S}_{ij}
  %&
  = 1 - \Theta(1-\kappa)\cos\left(\frac{\pi}{2}\kappa\right),
  \quad
  % \\
  \kappa
    \equiv \frac{1}{a} \frac{p_{ti}^2 + p_{tj}^2}{2p_{t,\max}^2}\,,
\end{equation}
and $p_{t,\text{max}}$ would typically be a hard scale (see
Ref.~\cite{Czakon:2022wam} for further details).
Throughout this paper, we  use
$p_{t,\text{max}} \equiv p_{t,\text{global-max}}$, where
$p_{t,\text{global-max}}$ is the transverse momentum of the hardest
pseudojet across the event at the given stage of the clustering.\footnote{We are grateful to the
  authors of Ref.~\cite{Czakon:2022wam} for discussions on this
  point.}
In addition to the jet radius, the algorithm has one parameter, $a$,
taken in the range $0.01{-}0.5$ in the original
publication~\cite{Czakon:2022wam}.
Unlike the flavour-$k_t$ algorithm, the CMP algorithm uses a beam
distance that is identical to that of the plain anti-$k_t$ algorithm.

The CMP algorithm resolves the problem in Fig.~\ref{fig:classic-config}
because when particles $1$ and $2$ are both soft, $\kappa$ is very
small.
Specifically, taking dimensions such that $p_{t,\text{global-max}}=1$,
a soft $ij$ quark pair has
$\mathcal{S}_{ij} \sim \kappa^2 \sim \max(p_{ti}^4,p_{tj}^4)$, leading to an
overall $d_{ij} \sim \max(p_{ti}^2,p_{tj}^2) \Delta R^2_{ij}$.
This is much smaller than the anti-$k_t$ clustering distance of a soft
quark with a hard parton, which is of order $\Delta R^2_{ij}$. 
As a result the soft $q\bar q$ pair clusters first and there is no
IRC-safety issue in Fig.~\ref{fig:classic-config}. 
Note that when one or other of $i$ and $j$ is hard, the use of a small
value for the parameter $a$ results in $\kappa$ being large and thus
$\mathcal{S}_{ij}=1$.
As a result, the CMP algorithm behaves like the anti-$k_t$ algorithm
for hard particles.
For $a \to 0$, the algorithm reduces to anti-$k_t$.
However, for finite $a$, the algorithm does sometimes yield jets whose
kinematics differ from those of the anti-$k_t$
algorithm. 

\subsubsection{Flavour dressing (``GHS'')}
\label{sec:ghs}

The algorithm of Ref.~\cite{Gauld:2022lem}, there called ``flavour
dressing'', will be referred to here as GHS.
This algorithm involves three stages: a standard
clustering stage in which flavour is not considered, an
``accumulation'' stage in which flavoured particles accumulate
momentum from non-flavoured ones, and a ``dressing'' stage, which
assigns the flavour to the original anti-$k_t$ jets.
Here, we limit ourselves to sketching the main features of
each of the steps, and refer the reader to the original reference for
the full details.

In the first step, the event is clustered with the standard anti-$k_t$
algorithm.
In this step, one also applies standard jet cuts, e.g.\ on transverse
momentum and rapidity, to the resulting jets.

In the second step, the algorithm runs an ``accumulation'' stage, which follows
a version of C/A clustering (i.e.\ $p=0$ in
Eq.~(\ref{eq:genkt-dij})) with a radius of $R_\text{cut}$, with two
modifications: (i) clustering of flavoured objects with non-flavoured
ones discards the non-flavoured one if the clustering fails to pass a
SoftDrop kinematic cut~\cite{Larkoski:2014wba},
\begin{equation}
  \label{eq:SD-cut}
   \frac{\min(p_{ti},p_{tj})}{(p_{ti}+p_{tj})}
  > z_\text{cut}
    \left(\frac{\Delta R_{ij}}{R_\text{cut}}\right)^\beta,
\end{equation}
where $z_\cut$ and $\beta$ are the usual SoftDrop parameters;
(ii) when two flavoured objects would normally cluster, they are
instead both removed from the accumulation clustering process and each is treated
as a ``flavour cluster'', to be used as an input to the third step of
the algorithm.
Any flavoured clusters that remain at the end of the modified C/A
clustering also serve as inputs to the third step.

The third step is the flavour ``dressing'' itself.
It evaluates flavour-$k_t$ distances (a) between pairs of flavour
clusters ($d_{\hat f_i \hat f_j}$), (b) between each flavour cluster
and the anti-$k_t$ jet, $j_k$, to which the flavoured particle in the
cluster belonged ($d_{\hat f_i j_k}$) and (c) with the beam
($d_{\hat f_i B_\pm}$).
When the smallest distance is a $d_{\hat f_i \hat f_j}$, the flavours
annihilate and $\hat f_i$ and $\hat f_j$ are removed from further
consideration;
when it is a $d_{\hat f_i j_k}$, the flavour of $i$ is assigned to jet
$j_k$ and $\hat f_i$ is removed from further consideration; and when
it is a $d_{\hat f_i B_\pm}$, $\hat f_i$ is simply discarded.
Distance measures involving any flavour clusters $\hat f_i$ or
$\hat f_j$ that were annihilated, assigned or discarded are then
removed from the list, and the procedure repeats until no flavoured
clusters remain.
Besides the standard jet radius, the algorithm has four parameters:
$R_\cut$, associated with the C/A clustering, $\beta$ and
$z_\text{cut}$ for the SoftDrop condition, and the $\alpha$ of the
flavour-$k_t$ distances.

In the configuration of Fig.~\ref{fig:classic-config}, we would have
three flavour clusters ($1$, $2$, $3$), with $2$ and $3$ associated
with a hard jet.
The third step of the algorithm would annihilate the $\bar q$ and $q$
flavours of $1$ and $2$, because they have the smallest flavour-$k_t$
distance, and attribute the flavour of $3$ to the hard jet.
The flavour dressing algorithm never modifies the kinematics of the
original anti-$k_t$ jets, only their flavour.
Note that for events where every anti-$k_t$ jet consists of a single
particle, i.e.\ events where there has been no kinematic
recombination, the flavour of each jet is the same as for the
anti-$k_t$ algorithm.
This is a property that we will seek also in our IFN algorithm.

%......................................................................
\subsubsection{Multi-flavoured events}

A final comment concerns clustering of events with more than one
flavour (e.g.\ tracking both $b$ and $c$ flavour). 
The flavour-$k_t$ algorithm is to be run with all flavours
for which one wants information in the final jets.
The CMP and GHS algorithms are designed for a single flavour at a time
(e.g.\ just the $b$ and $\bar b$ flavour in the event).
However, we note that for the CMP algorithm it is straightforward to
identify potential ways of extending it, for example by using the
distance in Eq.~(\ref{eq:flav-antikt-dij}) whenever a pair has the
potential for at least some cancellation of flavour.
As concerns GHS, since it does not modify the anti-$k_t$ jets'
kinematics, one can simply re-run it again for a second flavour, and
so forth.

%======================================================================
\section{Anti-$k_t$ and C/A jets with interleaved flavour
  neutralisation}
\label{sec:genkt-FN}

In this section, we present the motivation for, and description of, our new flavour neutralisation algorithm.

\subsection{Design aims and core concept}

\begin{figure*}
  \centering
  \subfloat[]{
    \includegraphics[width=.24\textwidth]{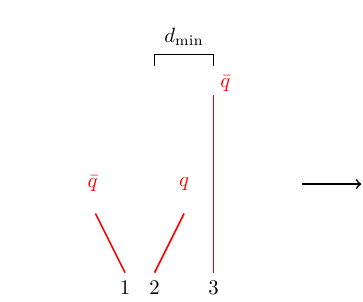}
    \label{fig:FN-illustration_a}
  }
  \subfloat[]{
    \includegraphics[width=.24\textwidth]{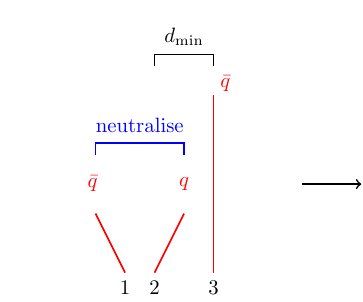}
    \label{fig:FN-illustration_b}
  }
  \subfloat[]{
    \includegraphics[width=.24\textwidth]{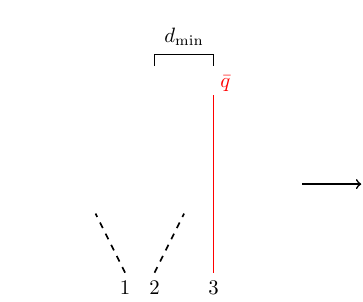}
    \label{fig:FN-illustration_c}
  }
  \subfloat[]{
    \includegraphics[width=.24\textwidth]{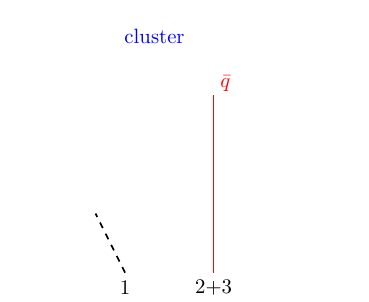}
    \label{fig:FN-illustration_d}
  }
  \hfill
  \caption{
     Illustration of  the flavour-neutralisation approach.
    The event displayed here (a) has the property that there is a soft
    $\bar q q$ pair (particles $1$ and $2$), and a hard $\bar q$
    (particle $3$) with
    $p_{t1} \sim p_{t2} \ll p_{t3}$.
    Additionally, we have all $\DR$ distances of order one, but with
    the constraint that
    $\DR_{23} < R$, while $\DR_{12} > R$, so that within the anti-$k_t$
    algorithm, $2$ and $3$
    cluster into one jet, while $1$ would form a separate soft jet.
    In (b), just before the $2+3$ clustering, the flavour of 1 is used to neutralise the flavour of 2,
    which results in the intermediate stage shown in (c), where
    particles $1$ and $2$ have lost their flavour (as represented by
    the black dashed lines).
    Finally, in (d) the (now) flavourless pseudojet 2 is clustered
    with 3 into a pseudojet 2+3 with the $\bar q$ flavour of just
    particle $3$.}
  \label{fig:FN-illustration}
\end{figure*}

If we consider what is needed for broad usage of a jet flavour algorithm,
we can identify at least four criteria that are necessary, or at least
highly desirable:
\begin{enumerate}
\item \textbf{IRC safety}: Both the kinematics and the flavours of any hard jets should be IRC safe.
\item \textbf{Preserved kinematics}: For a given member of the generalised-$k_t$ algorithm family,
  the flavour algorithm should not modify the jets' kinematics.
\item \textbf{Multi-scale flavour resolution}:
The flavours of the pseudojets should be well defined at any step
  of the clustering, so as to leave open the possibility of using
  flavour information with the full cluster sequence, e.g.\ for jet
  substructure studies.
\end{enumerate}
Additionally, as mentioned at the end of Sec.~\ref{sec:ghs}, it can be beneficial to have the following property:
\begin{enumerate}
\item[4.] \textbf{Single parton consistency}:   For events in which each jet contains exactly one parton,
  the flavour algorithm should not modify the jets' flavours relative
  to the simple generalised-$k_t$ algorithm.
  This ensures that typical leading-order calculations will give the
  same results for the generalised-$k_t$ algorithm and its flavoured
  extension.
  It notably means that one cannot a priori decide to neglect some
  subset of flavour in an event without specifying the jet kinematics.
\end{enumerate}

To achieve these aims,
the core novel idea that we introduce here is that of maintaining the
standard clustering procedure, but modifying the flavour-related
aspects of the recombination scheme at each step of kinematic
recombination.
In particular, our approach uses a global, event-wide treatment of
flavour at each pair-wise clustering step.
By construction, the resulting jets will have identical kinematics as
compared to the original jet algorithm, and we aim to arrange for
the flavour labels associated with the jets to be IRC safe at any stage in the clustering sequence.

There is quite some freedom in such an approach.
The fundamental principle of flavour neutralisation, which we believe can be applied in a
variety of ways, is illustrated in Fig.~\ref{fig:FN-illustration}.
When a pseudojet with non-zero flavour is about to undergo a kinematic
clustering (soft $q$ ($2$), clustering with hard $\bar q$ ($3$) in
Fig.~\ref{fig:FN-illustration_a}), the algorithm needs to establish
whether the flavours of $2$ and $3$ should be combined as per usual
net-flavour summation, or instead whether the flavour of either of the
particles should be ``neutralised'' by some other particle(s) in the
event before allowing the kinematic ($2+3$) clustering to proceed.
For example, in Fig.~\ref{fig:FN-illustration_b}, with a soft $\bar q$
(particle $1$) in the vicinity of the soft $q$ ($2$), the algorithm
may decide to first neutralise the flavours of particles $1$ and $2$,
before moving ahead with the $2+3$ clustering.
If that neutralisation happens, then particles $1$ and $2$ become flavourless,
as illustrated by the black dashed lines in Fig.~\ref{fig:FN-illustration_c}.
This is then followed by the kinematic clustering in
Fig.~\ref{fig:FN-illustration_d}, resulting in a $2+3$ jet that retains
the $\bar q$ flavour of hard particle $3$, as needed for IRC safety.

In general, the IRC safety (or otherwise) of the algorithm
resides in the criteria used to decide whether to neutralise a given
pseudojet's flavour, and if so then with which other pseudojet(s).
As with earlier flavoured clustering algorithms, such a procedure
will need to rely on some measure of the likelihood that a given
flavoured pair came from an effective parent gluon's splitting, versus
the flavour originating from a genuine hard parton.

\subsection{Introducing the IFN algorithm}
\label{sec:IFN-overview}

We now construct a concrete algorithm based on Fig.~\ref{fig:FN-illustration} that integrates jet clustering with flavour neutralisation:  Interleaved Flavour Neutralisation (IFN).
The core of our algorithm is the search for neutralisation candidates
at any given stage of the clustering.
Among the ingredients of that search is a measure of flavour
neutralisation distance $u_{ij}$ between any pair of particles
$i$ and $j$, the softer of which will always be flavoured.
For now, the reader may wish to think of $u_{ij}$ as being a flavour-$k_t$ type
distance, cf.\ Eq.~(\ref{eq:flavkt-dij}), though there are important
further subtleties, discussed below in
Sec.~\ref{sec:neutralisation-distance-choice}.

In defining the algorithm in the next few paragraphs, we shall
frequently make reference to Fig.~\ref{fig:FN-illustration} to
illustrate the function of the different steps, keeping in mind that
the flavour of the final hard jet (made of particles $2$ and $3$)
should ultimately just be that of the hard particle $3_{\bar q}$
without contamination from the flavours of the soft $1_{\bar q}2_q$
pair.

We write the core neutralisation search part of the algorithm in
the style of a computer subroutine $N(i,u_{\max}, C, E)$, taking a
number of arguments as inputs, specifically:
\begin{itemize}
\item the index $i$ of the pseudojet for which to identify potential
  neutralisation partner(s) (e.g.\ $i=2$ in
  Fig.~\ref{fig:FN-illustration_a});
\item a threshold $u_{\max}$ above which to ignore neutralisation
  candidates (e.g.\ in the context of the $2+3$ kinematic clustering
  in Fig.~\ref{fig:FN-illustration_a} this would be $u_{\max}=u_{23}$);
\item a list $C$ of all potential neutralisation candidates, i.e.\ all
  currently flavoured pseudojets in the event ($C=\{1,2,3\}$ in
  Fig.~\ref{fig:FN-illustration_a});
\item a subset $E$ among those flavoured pseudojets to be excluded in
  the neutralisation search because they have already been considered
  in some prior step of the algorithm ($E = \{2,3\}$ in
  Fig.~\ref{fig:FN-illustration_a}, because particles $2$ and $3$ have
  already been considered in that they set $u_{\max}=u_{23}$).
\end{itemize}
The $N(i,u_{\max}, C, E)$ algorithm
is formulated as follows:
%......................................................................
\begin{enumerate}[label=N\arabic*.,ref=N\arabic*]
  
\item \label{alg:N:N0} Create a list $L$ of $u_{ik}$ distances for all
  $k$ among the candidates $C$ that satisfy $u_{ik} < u_{\max}$,
  excluding those in the exclusion set $E$.
  
\item \label{alg:N:find-smallest-uik}
  Identify the $k$ that corresponds to the smallest $u_{ik}$ in the list.
\item If $k$ contains no flavour that can neutralise flavour in $i$
  (e.g. $k$ is a $b$-quark and $i$ is a $c$-quark), remove the
  corresponding $u_{ik}$ from list $L$, and loop back to
  step~\ref{alg:N:find-smallest-uik}.
\item \label{alg:N:recursion}
  Before using $k$ to neutralise flavour in $i$, check to see
  whether there are other pseudojets that could more naturally be
  paired with $k$ in order to neutralise $k$'s flavour.
  Do so through a recursive use of flavour neutralisation, searching
  for neutralisation partners of $k$ by running
  $N(k,u_{ik},C,E\cup{\{k\}})$.
  Sec.~\ref{sec:need-recursion} explains the importance of
  recursion for IRC safety.
  
\item For each flavour currently in $i$, neutralise as much of that
  flavour as one can with any flavour that is still present in
  $k$.\footnote{If working with flavour modulo-$2$, then initial
    flavours are always to be understood as being modulo-$2$, and each
    comparison and/or combination is also to be performed in a
    modulo-$2$ sense.}
  For example, if $i$ has flavour $c\bar b$ and $k$ has flavour
  $b b$, use $k$ to cancel the $\bar b$ flavour, so that the updated
  $i$ has flavour $c$ and the updated $k$ has flavour $b$.
\item If $i$ is now flavourless, exit.
\item Otherwise remove the current $u_{ik}$ from list $L$.
  If any entries are still left in list $L$, loop back to
  step~\ref{alg:N:find-smallest-uik}.
  Otherwise exit.
\end{enumerate}

In our IFN formulation, the flavour
neutralisation search is triggered whenever a clustering is about to
occur for which the softer pseudojet is flavoured, specifically:
\begin{enumerate}[label=I\arabic*.,ref=I\arabic*]
\item \label{alg:main:start}
  When pseudojets $i$ and $j$ recombine in the standard kinematic
  clustering sequence, let $i$ be the pseudojet with lower $p_t$.
  If $i$ is flavourless, then $i+j$ simply takes the flavour of $j$
  and one moves on to the next kinematic jet clustering step.

\item \label{alg:main:get-K}
  Otherwise, identify all pseudojets that currently carry flavour,
  including any flavoured jets declared earlier according to a
  $d_{iB}$ step, and put them into a list $C$ of potential
  neutralisation candidates.
  Initialise the set $E = \{i,j\}$ of particles to be excluded from
  the search for neutralisation candidates.

\item Call the flavour-neutralisation search, $N(i,u_{ij},C,E)$, which
  may use one or more flavoured particles in set $C$ to neutralise
  some or all of the flavour contained in $i$.
  
\item \label{alg:main:final} For any remaining flavour in $i$, apply
  the standard net-flavour (or flavour modulo-$2$) summation of $i$
  with $j$ and move on to the next kinematic jet clustering step.
\end{enumerate}

Interleaving flavour neutralisation at each step of the
clustering is important from the point of view of collinear safety.
To illustrate this, it is helpful to suppose
that particles $i$, $j$ and $k$ all have comparable transverse momenta
and inter-particle distances $\Delta R \sim R$.
In this situation $u_{ij} \sim u_{ik}$.
Consider the case where $j$ undergoes a collinear splitting,
$j \to j_a,j_b$ with $\Delta R_{j_a, j_b} \ll R$.
If one ran flavour neutralisation without clustering, one could 
find oneself in a situation where $u_{ik}<u_{ij}$, but
$u_{ik}>u_{ij_a}$, thus changing the neutralisation sequence.

Now let us examine how this changes if neutralisation is interleaved
with clustering.
The clustering algorithms that we consider are the
anti-$k_t$ and C/A algorithms.
They both have the property that when all
particles have similar transverse momenta, clustering of the collinear
$j_a,j_b$ pair will precede the $ij$ clustering step.
At the $j_a,j_b$ clustering, if the neutralisation search gets
triggered, then $j_a$ and $j_b$ will cluster with normal net-flavour
recombination, since $u_{j_a j_b}$ is much smaller than all other
$u$'s.
When the clustering reaches the $ij$ step, all distances will see the
kinematics of $j$, rather than that of the underlying $j_a$ and $j_b$,
thus ensuring that the algorithm is collinear safe.\footnote{When
  considering collinear splitting in events with a hierarchy of
  energies, the different members of the generalised-$k_t$ family may
  perform the soft and the collinear clusterings in different orders.
  However, when the neutralisation search is, say, comparing neutralisation
  distances involving two soft particles $i$ and $k$ and a hard
  particle $j$ ($u_{ik} \ll u_{ij},u_{kj}$), a collinear splitting of
  any of the soft or hard particles will only modify the $u$'s by a
  factor of order $1$ and it will leave the hierarchies untouched, and
  correspondingly also the resulting neutralisation pattern.
}

\subsection{Choice of neutralisation distance}
\label{sec:neutralisation-distance-choice}

Let us now turn to the $u_{ik}$ flavour neutralisation distance
between a pair of particles $i$ and $k$.
Recall that the softer of the two will always be flavoured, while the
harder one may or may not be.

We write the $u_{ik}$ distance generically with two parameters, $\alpha$
and $\omega$:
\begin{subequations}
  \label{eq:uik-all}
  \begin{align}
    \label{eq:uik-gen}
    u_{ik} &\equiv
             [\max\left(p_{ti}, p_{tk}\right)]^{\alpha}
             [\min\left(p_{ti}, p_{tk}\right)]^{2-\alpha}
             \times
             \Omega_{ik}^2\,,
    \\
    \label{eq:omegaik-gen}
    \Omega_{ik}^2 &\equiv 
                    2\left[
                    \frac1{\omega^2}\left(
                    \cosh (\omega\Delta y_{ik}) -1
                    \right)
                    - \left( \cos \Delta \phi_{ik} - 1 \right)
                    \right],
  \end{align}
\end{subequations}
where $\Delta y_{ik} = y_{i}-y_{k}$ and analogously for $\Delta
\phi_{ik}$.
Let us start with the part related to the transverse momenta.
This is identical to that used in the flavour-$k_t$ algorithm, cf.\
Eq.~(\ref{eq:flavkt-dij}), with the same parameter $\alpha$.
As in typical flavour-$k_t$ studies, we assume
$0 < \alpha \le 2$, and in particular concentrate on $\alpha=1$ and
$\alpha=2$.

\begin{figure}
  \centering
  \includegraphics[width=.6\columnwidth]{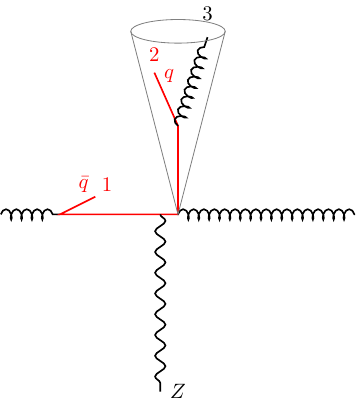}
  \caption{
    NNLO contribution to the $pp\to Z+\text{jet}$ process, that 
    helps illustrate the origin
    of the condition, Eq.~(\ref{eq:omega-constraint}), on the $\omega$
    parameter in the angular part of the $u_{ij}$ distance,
    Eq.~(\ref{eq:uik-gen}). 
    It involves a hard jet with a final-state splitting (where the jet
    constituents, $q$ and a gluon, are labelled 2 and 3 respectively),
    as well as an  initial-state collinear
    splitting ($g \to q \bar q$, with the $\bar q$ labelled 1).
    When $\alpha + \omega < 2$, the initial-state collinear $\bar q$
    ($1$) neutralises the flavour of the $q$ ($2$).
  }
  \label{fig:config-as2-IHC-FmaybeC}
\end{figure}

Next, we examine the angular part of the distance, $\Omega_{ik}^2$,
which involves a parameter $\omega$.
For any $\omega$ of order $1$, in the limit of small $\Delta y_{ik}$ and
small $\Delta \phi_{ik}$, $\Omega_{ik}^2$ reduces to the standard $\Delta R_{ik}^2
= \Delta y_{ik}^2 + \Delta \phi_{ik}^2$.
The reason for using $\Omega_{ik}^2$ rather than the standard
$\Delta R^2$ is to ensure IRC safety as concerns the interplay between
collinear initial-state splittings and splittings elsewhere in the
event.
This is best explained with the help of
Fig.~\ref{fig:config-as2-IHC-FmaybeC}. 
In the anti-$k_t$ and C/A algorithms, particles $2$ and $3$ will
cluster first.\footnote{This would not be the case for the $k_t$
  algorithm, and an investigation of the interplay of $k_t$ clustering
  with IFN is left to future work.}
When $p_{t2} < p_{t3}$, the $2+3$ clustering triggers a flavour
neutralisation search.
The only candidate for flavour neutralisation is particle $1$ and one
should compare the $u_{12}$ and $u_{23}$ distances.
We will suppose that particles $2$ and $3$ have similar $p_{t}$'s and
are at central rapidity.
The initial-state collinear splitting that creates particle $1$
typically results in $y_{1} = \ln p_{t3}/p_{t1} + \order{1}$.
Neglecting $\order{1}$ factors, we then have
\begin{subequations}
  \begin{align}
    u_{12} &\sim
             p_{t2}^\alpha p_{t1}^{2-\alpha} \left(\frac{p_{t3}}{p_{t1}}\right)^\omega
             \sim\;  p_{t3}^{(\alpha+\omega)} p_{t1}^{(2-\alpha - \omega)}\,,
    \\
    u_{23} &\sim
             p_{t3}^\alpha p_{t2}^{2-\alpha} \Delta R_{23}^2
             \,\sim\;
             p_{t3}^2 \Delta R_{23}^2\,.
  \end{align}
\end{subequations}
where in the rightmost part of each equation we have exploited $p_{t2}
\sim p_{t3}$.
One immediately observes that if $\alpha + \omega < 2$, then in the
initial-state collinear limit, where $p_{t1} \ll p_{t3}$, one has
$u_{12} \ll u_{23}$.
This causes particle $1$ to neutralise the flavour of particle
$2$, even when $1$ is arbitrarily collinear, resulting in a
flavourless hard jet.
In contrast, when the initial-state splitting is absent, the hard jet
will be flavoured.
Thus, the algorithm would be unsafe with respect to initial-state collinear
splittings.
On the other hand, if we take
\begin{equation}
  \label{eq:omega-constraint}
  \alpha + \omega > 2 \,,
\end{equation}
then $u_{12}$ will always be parametrically larger than $u_{23}$ in
the limit $p_{t1} \to 0$, thus effectively forbidding neutralisation
of $1$ and $2$; see App.~\ref{sec:IFN-omega} for further discussion.\footnote{We have also explored the border case of
  $\alpha + \omega = 2$ and find that it diverges.
  This is relevant in particular to the case of $\alpha=1$ and $\omega=1$,
  for which $u_{ik}$ coincides with the $ik$ squared invariant mass
  when $i$ and $k$ are massless, i.e.\ a JADE-like
  distance~\cite{JADE:1986kta,JADE:1988xlj}.
  An issue to be aware of with an invariant-mass distance in a
  hadron collider context is that the invariant mass between an energetic
  initial-state collinear emission and a hard final-state particle is
  commensurate with that between two well separated hard final-state
  particles.
  Furthermore, a potential solution to this issue, i.e.\ clustering
  initial-state collinear emissions early, via their small invariant
  mass with the beam, involves ambiguities in the identification of
  the beam energy.
}
In practice, we will nearly always take
\begin{equation}
  \label{eq:default-omega}
  \text{default: }\quad\omega = 3 - \alpha\,,
\end{equation}
and where not explicitly stated in plots, this will be the choice that
we adopt.

IRC-safety subtleties connected with the large $\Delta y_{ij}$
behaviour of normal $\Delta R_{ij}^2$ distances are relevant for all
flavour algorithms, though sometimes the issues appear only at orders
beyond $\as^2$.
Further discussion of this point is provided in
Apps.~\ref{sec:IFN-omega}, \ref{sec:IRCapp-flavkt}
and \ref{sec:CMP-IHC-DS}.
Note also that the original formulation of the $k_t$ algorithm for
hadron colliders~\cite{Catani:1993hr} foresaw the possibility of an
angular distance $\Omega_{ik}^2$ with $\omega=1$, though this does not
have IRC safety implications for the kinematic aspects of normal jet
clustering.
%

%----------------------------------------------------------------------
\subsection{Need for recursion}
\label{sec:need-recursion}

\begin{figure}
  \centering
  \includegraphics[width=.6\columnwidth]{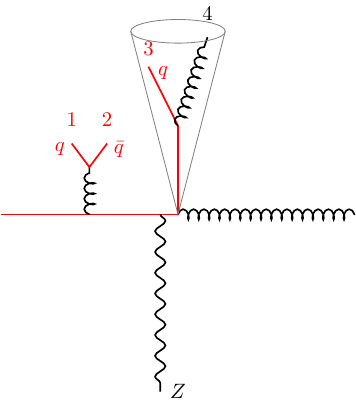}
  \caption{
    N$^3$LO contribution to the $Z+\text{jet}$ process that helps illustrate the need for
    recursion in step \ref{alg:N:recursion} of the flavour neutralisation search. 
    It involves a hard jet with a non-collinear splitting (flavoured
    $3$ and flavourless $4$) and a flavoured initial-state double-soft
    pair (labelled $1$ and $2$).
    Without recursion, particle $2$ can end up neutralising the flavour
    of $3$.
  }
  \label{fig:need-for-recursion}
\end{figure}

A key element of IFN is the recursion in step \ref{alg:N:recursion} above.
The need for recursion can be illustrated with the help of
Fig.~\ref{fig:need-for-recursion}.
Again considering the anti-$k_t$ or the C/A clustering algorithms, the
first clustering step is that of particles $3$ and $4$.
If $p_{t3} < p_{t4}$, their clustering triggers a flavour
neutralisation search.
That search will identify particle $2$ (from a large-angle soft pair)
as a potential neutralisation candidate.
With $\alpha=1$, we will have $u_{23} \ll u_{34}$, while for
$\alpha=2$, $u_{23} \sim u_{34}$.
Either way, without recursion, it would be possible for $2$ to
neutralise the flavour of $3$, which would ultimately result in the
hard jet being flavourless.
In the absence of the $(1,2)$ pair, the hard jet would be flavoured.
This would induce an infrared divergence.

The recursive aspect of the algorithm resolves this problem as
follows: when $2$ is identified as a neutralisation candidate for $3$,
the recursive search that is triggered in step~\ref{alg:N:recursion}
identifies particle $1$ as a 
neutralisation candidate for $2$.
For both $\alpha=1$ and $\alpha=2$, we have $u_{12} \ll u_{23}$,
and so particles $1$ and $2$ will neutralise.
When the algorithm exits the recursion step, there are no longer any
remaining flavoured particles to neutralise the flavour of particle
$3$.
Thus the hard jet will retain its net flavour, resolving this IRC
safety issue (see e.g.\ App.~\ref{sec:IFN-non-recursive}).

%----------------------------------------------------------------------
\subsection{Further comments}

We conclude this section with a few general comments about the IFN algorithm.

A first comment concerns single parton consistency, as discussed in Sec.~\ref{sec:genkt-FN}.
A potentially useful characteristic of IFN, shared with GHS, is that for
configurations where each jet contains no more than one particle, the
flavours of those jets are identical to those in standard anti-$k_t$.
This is trivial, because for such configurations there is never a
situation where two particles would cluster together and so the
flavour neutralisation part of the algorithm is never triggered.
Thus any leading-order jet calculation, for an arbitrary number of
final-state jets, will give identical jets and flavours for those jets
in the anti-$k_t$ and its IFN extension.

A second comment concerns the fact that unlike the flavour-$k_t$
algorithm, the flavour-related part of our IFN algorithm has no
specific treatment of beam distances for flavoured particles
(the CMP algorithm has similarities in that it leaves the anti-$k_t$
beam-distance untouched for flavoured particles).
This means that particular care is needed around the potential for
long-distance clusterings, as discussed in
Sec.~\ref{sec:neutralisation-distance-choice}.
Nevertheless, even algorithms with beam distances can suffer from long
distance clustering when using standard $\Delta R_{ij}^2$ type angular
measures, as discussed in App.~\ref{sec:IRCapp-flavkt}. 

A third comment concerns events with more than one flavour, e.g.\ both
$c$ and $b$ flavour.
One possibility is to consider all flavours within a single IFN run.
Suppose $i$ has flavour $b$ and is about to cluster with $j$.
This triggers a search for candidates to neutralise $i$'s flavour.
The search may find a particle $k$ with flavour $c\bar b$ (which could
have arisen, for example, through earlier clusterings).
The recursion of the IFN algorithm may then identify some other
particle with flavour $\bar c$, which neutralises the $c$ component of
$k$'s flavour.
Thus $c$ flavour elsewhere in the event is affected by the $b$ flavour
in the $i+j$ clustering.
Alternatively, one could choose to run the IFN algorithm first for the
$b$ flavour, then for the $c$ flavour.
In that case, the flavour neutralisation steps for $b$ flavour have no
side effects on those for $c$ flavour.
Consequently, the output of the algorithm can be different according to
whether one runs it for all flavours at once, or separately a flavour
at a time.
In those of our studies below that include multiple flavours (the IRC
safety tests of Sec.~\ref{sec:IRC-tests} for the IFN algorithms,
and the phenomenological study of Sec.~\ref{sec:pheno-Zj}), we
treat all flavours at once.

In the discussion so far, we have always described the IFN algorithm as
happening at the same time as the kinematic clustering.
However, because IFN preserves the kinematic clustering sequence, the
neutralisation steps can also be run as an add-on.
Here, one loops again through each step of the kinematic clustering
and updates the flavour information.
This may be more convenient in cases where one already has a jet
collection (and associated clustering sequence) defined.

A final comment concerns the ``bland'' option of
flavour-$k_t$~\cite{Banfi:2006hf}, which sets to infinity any
clustering distances that would lead to flavours that are inconsistent
with a single partonic flavour (e.g.\ $bb$ or $\bar c b$).\footnote{
  This approach was adopted also in the IRC-unsafe ``QCD-aware''
  clustering algorithm~\cite{Buckley:2015gua}, without any clustering
  distance modification.  }
One could imagine a similar bland extension for our flavour
neutralisation distances, but we leave the study of this question to
future work.

%======================================================================
\section{IRC safety: discussion and tests}
\label{sec:IRC-tests}

Given the considerable subtlety of IRC safety for jet flavour, it is
important to design tests to help build confidence in the IRC safety
of any new algorithm.
Subtle IRC-safety problems have arisen in the past in the context of
cone-type jet algorithms, which ultimately led to
the construction of an automated testing framework, used to verify the
IRC safety of the SISCone algorithm~\cite{Salam:2007xv}.
Here we adapt and substantially extend that framework.
The framework is available on request from the authors.

%----------------------------------------------------------------------
\subsection{Methodology}
\label{sec:IRC-methodology}

\begin{figure}
  \centering
  \includegraphics[width=1.\columnwidth]{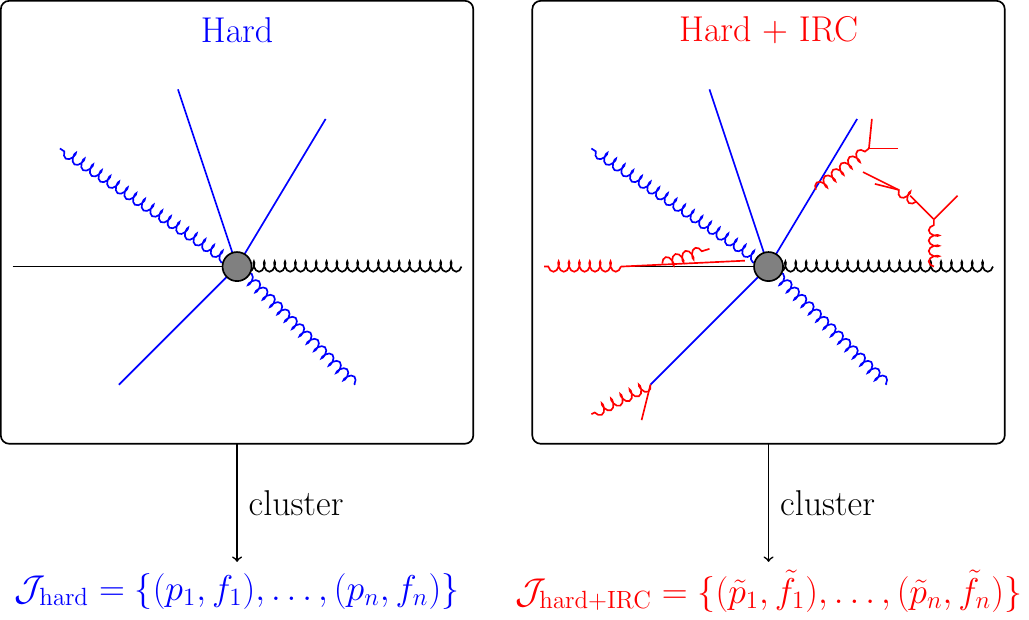}
  \caption{
    On the left, hard particles (in blue) are generated, some with flavour, at central
    rapidities. The event is clustered with a given jet algorithm,
    resulting in a set of ``hard'' jets $\Jhard$, with
    kinematics $\{p_i\}$ and associated flavours $\{f_i\}$.
    On the right, additional IRC radiation is added to the event as explained in the
    main text. This modified event is then clustered with the same
    jet algorithm, and the resulting set of ``hard+IRC'' jets $\JhardIRC$
    is compared against the original set of hard jets (and similarly for each hard
    step of the underlying clustering sequence).
    The sets agree
    if both the kinematics and the flavours of the various jets (and
    hard clustering steps) are
    identical. In the limit where the extra radiation becomes soft
    and collinear, the rate of failed events (where $\Jhard \neq \JhardIRC$)
    should go to zero for an algorithm that is IRC safe.
    The right-hand figure also serves to illustrate some of the
    classes of IRC additions that we make, though in practice we do
    not go beyond sixth order in $\as$, i.e.\ we do not simultaneously
    add as many emissions as are shown.  }
  \label{fig:IRtests-framework}
\end{figure}

Our approach is illustrated in Fig.~\ref{fig:IRtests-framework}, which goes beyond
the tests performed in the more recent literature.
We begin by generating a random ``hard'' event, with some number of
particles (flavoured or not), and run the clustering with the jet
definition that we wish to test.
This results in a set of hard jets,
$\Jhard = \{ (p_1, f_1), \dots, (p_n, f_n)\}$ with kinematics
$\{p_1, \dots, p_n\}$ and associated flavours $\{f_1, \dots,
f_n\}$.
Note that here, we do not force the total 4-momentum (or even
transverse momentum) of the hard
event to be balanced, i.e.\ it is as if the events have neutrinos,
leptons or isolated photons that would balance the momentum but do not take part in the
clustering. 
We then construct a modified ``hard+IRC'' event, where we add soft
emissions and collinear splittings up to some given order in $\as$.
We cluster that modified event and verify whether the hard jets in the
modified event, $\JhardIRC$ coincide with the hard jets in the
original event, both in terms of kinematics and flavour.\footnote{In
  the modified event, we also identify each step in the clustering
  sequence that involves clustering of two hard particles, and compare
  its kinematics and flavour to that of the corresponding step for the
  unmodified event.}
We then examine the rate of failure as a function of the logarithmic
momentum range ($L$) of IRC additions.
For an IRC-safe algorithm, we expect that failure rate to vanish as a
(possibly fractional) power of the momentum scale of the IRC additions.

Ideally, we would consider all possible IRC insertions.
There are two logarithms per order in $\as$, and we have found that it
is important to explore configurations at least up to $\as^4$.
The smallest non-IRC-safe contribution would be a term independent of
$L$, and at $\as^4$ that would imply identifying one event in $L^8$
that fails.
We will return to the question of the meaning and range of $L$ below,
but for now let us consider $L=30$.
That would imply identifying failures at the level of one event in
$30^8 \simeq 6.6\cdot 10^{11}$, which is prohibitive.
Note, however, that the only contributions that give the maximum
number of logarithms are those that exclusively involve the emission
of simultaneously soft and collinear gluons, which are not the most
likely configuration for triggering flavour-related IRC safety issues.

Consequently, we take a more targeted approach, in which we allow up to one
logarithm per order in $\as$, prioritising configurations that are
potentially non-trivial from the point of view of flavour.
We do so by omitting single soft-gluon divergences unless they involve
a subsequent splitting to a pair of commensurate-angle partons.

\subsection{Classes of IRC emissions}

\begin{figure*}
  \centering
  \subfloat[FHC:  final-state hard-collinear]{
    \includegraphics[scale=0.5]{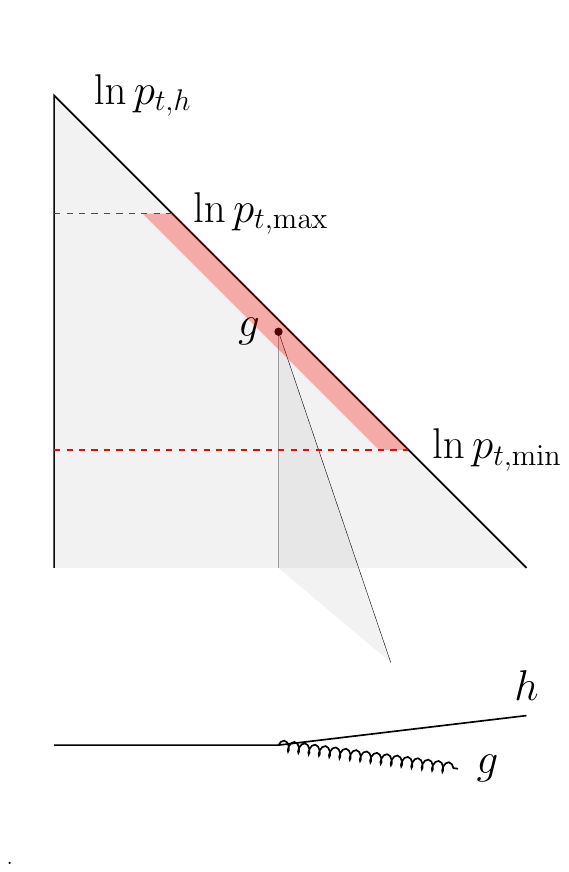}
    \label{fig:IRtests-sampling_a}
  }\qquad \qquad
  \subfloat[IHC:  initial-state hard-collinear]{
    \includegraphics[scale=0.5]{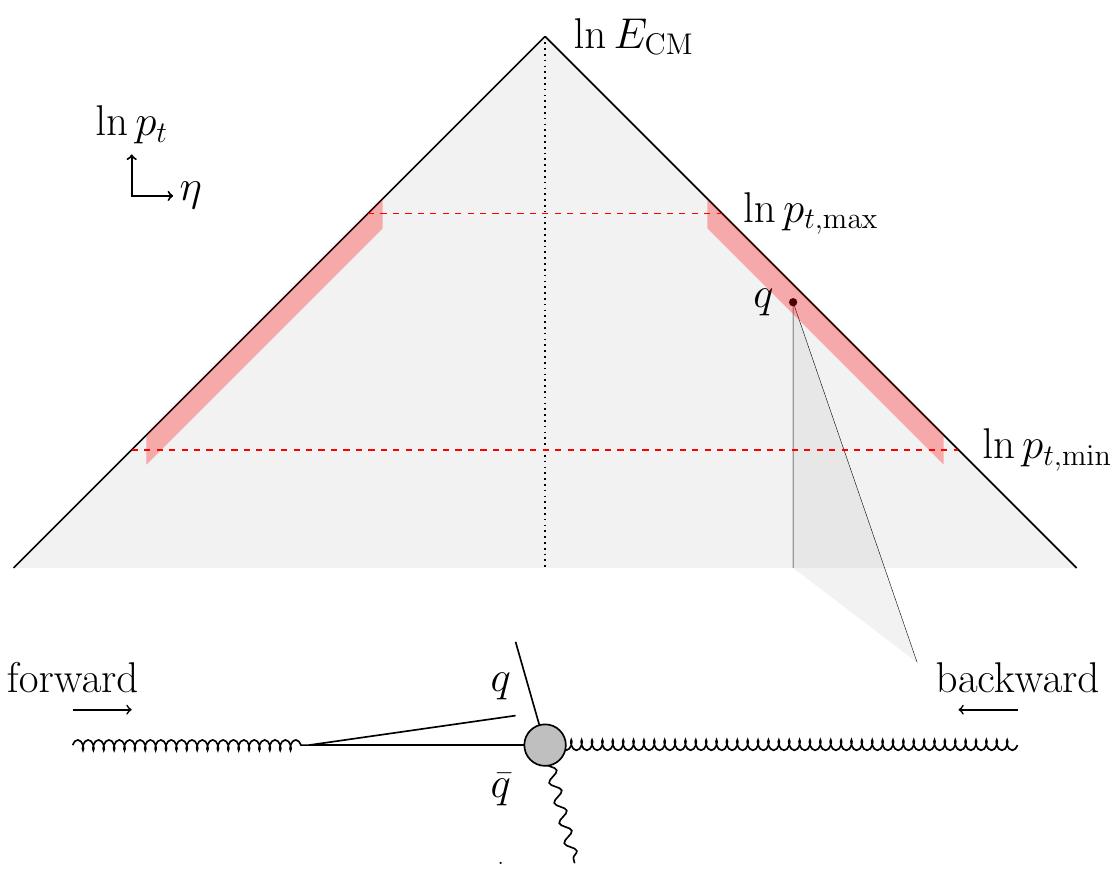}
    \label{fig:IRtests-sampling_b}
  }
  \\
  \subfloat[FDS:  final-state double-soft]{
    \includegraphics[scale=0.5]{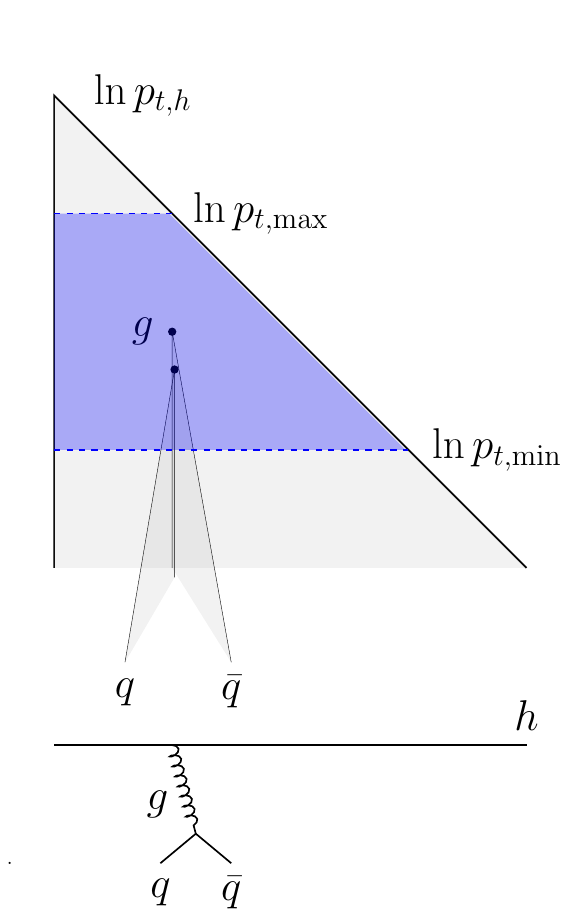}
    \label{fig:IRtests-sampling_c}
  }\qquad \qquad
  \subfloat[IDS: initial-state double-soft]{
    \includegraphics[scale=0.5]{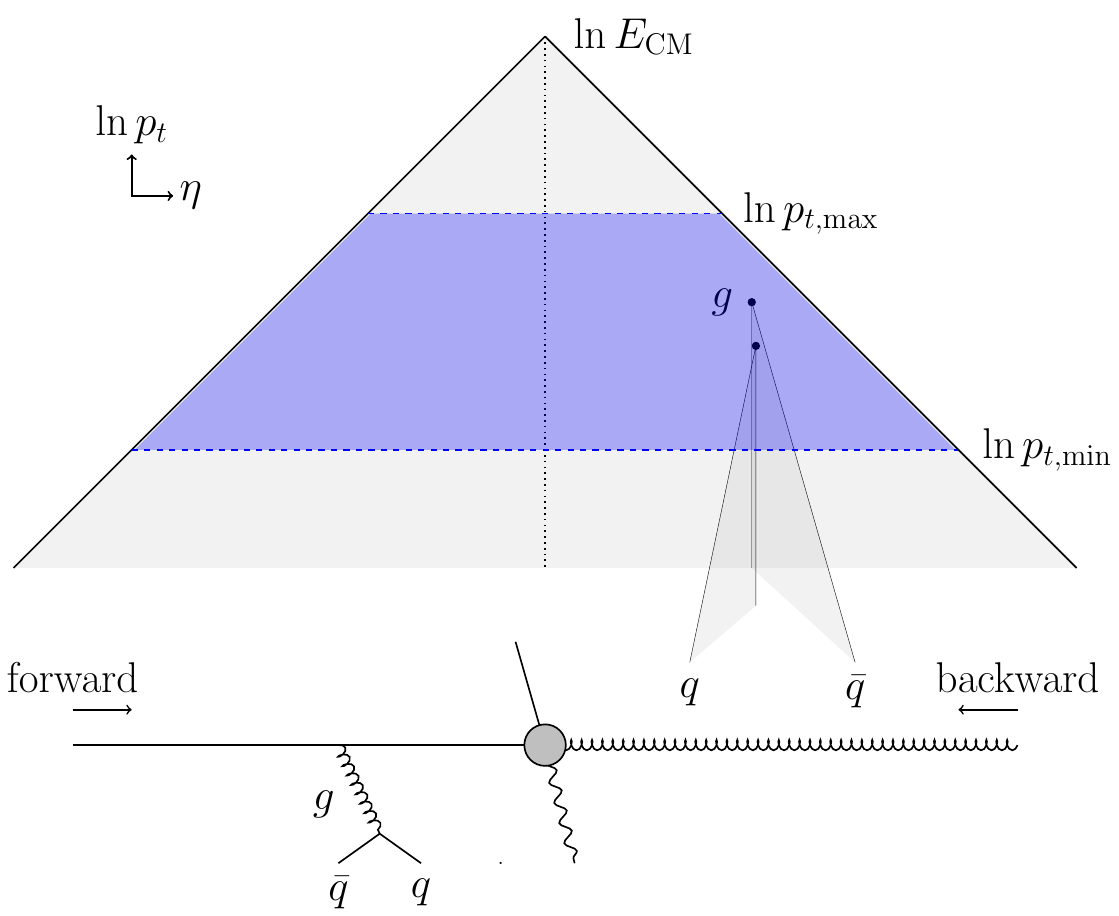}
    \label{fig:IRtests-sampling_d}
  }
  \caption{
    Illustration of the emissions sampled in the ``hard+IRC'' event on
    a Lund diagram~\cite{Andersson:1988gp}.
    The left column shows emissions from a final-state jet.
    The right column shows initial-state radiation from the beams, where an emission collinear to the
    forward beam (coming from the left on the diagram) ends up at positive
    rapidities (right-hand half of the Lund plane), and
    vice-versa for an emission collinear to the backward beam.
    The top row shows hard-collinear splittings from (a, FHC) hard final-state particles or (b, IHC) the beams.
    The bottom row shows double-soft pairs, flavoured or not, being emitted from (c, FDS) hard final-state particles or (d, IDS) the beams.  
    In a bin defined by $\ln p_{t,{\min}} < \ln p_t < \ln p_{t,{\max}}$, we sample any
    additional radiation in slices in the Lund planes (both for
    initial- and final-state radiation).
    We typically choose $\lnptmin = 3 \lnptmax$, with $\lnptmax < 0$.
  }
  \label{fig:IRtests-sampling}
\end{figure*}

The specific IRC emissions included in our testing framework are shown in Fig.~\ref{fig:IRtests-sampling} and described below:
\begin{itemize}
\item \textbf{Final-state hard-collinear (FHC) emission}:
  we perform a (hard) collinear splitting of a randomly chosen final-state
  particle.
  We uniformly sample the logarithm of the transverse momentum of the
  splitting.
  We also uniformly sample the longitudinal momentum fraction of the
  splitting.
  This is consistent with our choice not to include the soft gluon emission
  divergence as part of the FHC class.
  For all flavour
  combinations ($q \to qg$, $g \to q\bar q$, $g \to gg$), an FHC
  branching is associated with one power of $\as$ and one logarithm of
  the IRC scale.
  For readers in the habit of using a
  Lund-diagram~\cite{Andersson:1988gp} representation of
  soft-collinear phase space, this corresponds to a strip close to the
  hard-collinear boundary in the Lund leaf of the emitter
  (Fig.~\ref{fig:IRtests-sampling_a}; that figure shows a shaded
  logarithmic transverse momentum range, which we further discuss below).
  Note that sensitivity to soft gluon emission will still be present
  in the analysis, but will be obtained through the double-soft
  (FDS/IDS) contributions below.

\item \textbf{Initial-state hard-collinear (IHC) emission}:
  we perform a hard-collinear splitting
  of the beam (Fig.~\ref{fig:IRtests-sampling_b}).
  Again, we sample the longitudinal momentum fraction of the splitting
  uniformly, bringing one order of $\as$ and one logarithm.

\item \textbf{Final-state double-soft (FDS) pair}, i.e.\ the addition of a
  $g \to gg$ or $g \to q\bar q$ pair.
  We choose one emitter randomly among the final-state hard particles,
  and uniformly sample the logarithm of the transverse momentum
  $\ln p_{t,g}$ and the rapidity $y_g$ of the intermediate gluon with
  respect to the emitter.
  This corresponds to uniform sampling of the bulk of the Lund leaf
  for that emitter (Fig.~\ref{fig:IRtests-sampling_c}) and brings
  one power of the coupling and two powers of the logarithm.
  We allow the intermediate gluon to split and distribute the
  kinematics of the resulting pair in such a way as to correctly
  reproduce the asymptotic behaviour of the $q\bar q$
  double-soft matrix element in those kinematic regions
  where the splitting is asymmetric (either substantially different
  momentum fractions or rapidities, as elaborated upon in
  App.~\ref{sec:double-soft-structure}).
  Note that even for $g \to gg$ splittings, we use the asymptotic matrix element for $g \to q\bar q$.%
   \footnote{This might seem surprising at
    first sight, since the production of a soft $gg$
    pair has a qualitatively different structure from that of a soft
    $q\bar q$ pair when the pair is well separated.
    For example, for emission of a double soft gluon pair from a quark line,
    the component with the $C_F^2$ colour factor corresponds to
    independent emission, with logarithmic divergences both in the
    ratio of the gluon transverse momenta and the rapidity
    separation.
    However these contributions would be associated with the double
    (soft-collinear) logarithms that we are deliberately leaving out.
    Similarly the soft singularity in the $C_F C_A$ term would also
    bring an extra soft-gluon logarithm that is beyond what we aim to
    sample.
    }
  With the branching to a pair, we gain an extra power of the coupling
  and no logarithms, giving in total two powers of the coupling and
  two powers of the logarithm.
  We do not include the collinear divergence when $\theta_{q\bar q}
  \ll \theta_{hq} \sim \theta_{h\bar q}$, where $h$ is the hard
  particle.
  That would bring three powers of logarithm for two powers of the
  coupling.
  That said, we still have configurations with
  $\theta_{q\bar q} \ll \theta_{hq}$, but those are generated by a different sequence, namely FDS production of a pair
  of soft gluons, followed by nested FHC branching of one of those
  gluons to a $q\bar q$ pair.

\item
  \textbf{Initial-state double-soft (IDS) pair}:
  these are generated similarly to FDS, but with respect to the
  forward or backward beam (Fig.~\ref{fig:IRtests-sampling_d}).
  Note that both the IDS and FDS mechanisms include a subset
  of phase space where the double-soft pair is not collinear but
  instead at large angles, i.e.\ the configuration of
  Fig.~\ref{fig:classic-config}.
  The matrix element for large-angle double-soft production is the
  same as collinear double-soft production (up to complications of
  colour factors), a consequence of longitudinal boost invariance of
  soft production in the (leading-$\NC$) colour dipole rest frame.
  As a result, the IDS component in particular is guaranteed to fully
  cover the soft large-angle double-soft phase space.
\end{itemize}

For each emission, we need to choose the range in $\ln p_t$, which we
define as $\lnptmin < \ln p_t < \lnptmax$ (we choose our dimensions
such that $\lnptmax$ and $\lnptmin$ are always negative).
One potential difficulty is that of proximity (or overlap) between the
momentum scale of the IRC additions and the momentum scales of the
hard event, because when there is proximity or overlap, the IRC
additions can legitimately modify the hard event. 
If one keeps the range of hard scales fixed and takes IRC scales to
zero, one expects the rate of such legitimate modifications to vanish.
There should also be a suitably generous factor between the upper ($\lnptmax$) and
lower ($\lnptmin$) edges of the IRC additions' $\ln p_t$ range, so as
to be sensitive to IRC-unsafety mechanisms that work across a
hierarchy of scales.

\subsection{Implementation details}

In our testing framework, the FHC, IHC, FDS, and IDS emissions are applied in a nested way (with some caveats, see
below) so as to generate configurations up to order $\as^6$.
We could have imposed angular ordering between nested emissions, but
choose not to, so as to ensure that we retain sensitivity to cases
where strict angular ordering might miss a potentially divergent part
of phase space.
We also allow for FHC emissions from the individual $q$, $\bar q$ or
$g$ of the FDS and IDS pairs (again with no angular ordering
condition).

Nesting is essential in order to test collinear splitting chains.
It also partially alleviates worries of missing relevant IRC-unsafe
configurations as a result of only considering configurations with one
logarithm per power of the coupling.
For example, as indicated above, $g \to q\bar q$ collinear splitting
from a soft-collinear gluon ($\as^2 L^3$) is not generated directly,
but the underlying $L^3$ divergence does appear in our framework, in
the context of FDS (double gluon) emission followed by the nested
collinear splitting of one of the soft gluons to $q\bar q$.
This is not exactly the same configuration as the $\as^2 L^3$
divergence, since it involves an extra soft gluon.
In effect, if there is a problem in an $\as^2 L^3$
$h \to hg \to hq\bar q$ configuration, we are making the assumption
that we will still detect it in at least some of the
$h \to hgg \to h gq\bar q$ configurations.
For technical reasons, there are some nestings that we are missing: (a)
nesting of one double-soft emission from a prior double-soft emission;
and (b) insertion of double-soft emissions on more than one of the
descendants of a collinear splitting.
These limitations may be addressed in future work.

One final potential concern is that by allowing strongly
angular-disordered configurations, we might mistakenly declare an
algorithm to be IRC unsafe.
To guard against that risk, when we identify an IRC-unsafe
configuration, we further investigate it to establish whether it is
genuine.
All IRC failure classes that we identified were genuine, as
documented in App.~\ref{sec:list-ir-risky}.

We close this description with some final technical details:
\begin{itemize}
\item We generate a random number of hard particles, and randomly
  sample their flavours.
  The maximum number of hard particles we consider is $8$.
  In the figures targeting specific IRC-unsafe configurations,
  the hard particles' momenta are chosen randomly
  uniformly between $100\GeV$ and $1\TeV$, and their rapidities
  uniformly in the range $|y|<1.5$.
  In our final IRC safety tests for the IFN algorithms, we use a
  broader range, with the hard particles' momenta chosen randomly
  uniformly between $1\GeV$ and $1\TeV$, and their rapidities taken
  uniformly within $|y|<2.5$.
  Plots with results will always indicate the ranges used.

\item The $\lnptmax$ scale is typically scanned in the range $-3$ down
  to $-42.5$ (with $p_t$ expressed in units of GeV).
  When we show a failure rate it will always include the
  $\ln^n \ptmax/\ptmin$ measure that arises from the $n$ logarithmic
  integrations at a given order $n$, but it will not include overall
  constants such as colour factors.
  We will refer to this as a phase-space weighted failure rate.

\item We generally use $\lnptmin = 3\lnptmax$.
  However, in the tests of our IFN algorithms, we have also carried
  out a subset of tests with a larger ratio $\lnptmin = 5\lnptmax$.
  Results are consistent between the two sets of tests.
  \logbook{9e5e8eb}{
  see in IRSafety-prec/results-2023-04-12/neutralisers:
  res-antikt-hp-jadea2-as-4-nhard-6-1GeV-15-lnptratio-5.0.le.lnpt.le.-9-ddreal.dat
  res-antikt-hp-jadea2-as-4-nhard-6-1GeV-26-lnptratio-5.0.le.lnpt.le.-15-qdreal.dat
  res-antikt-hp-jadea2-as-4-nhard-6-1GeV-9-lnptratio-5.0.le.lnpt.le.-3-double.dat
  res-antikt-hp-maxscale-as-4-nhard-6-1GeV-15-lnptratio-5.0.le.lnpt.le.-9-ddreal.dat
  res-antikt-hp-maxscale-as-4-nhard-6-1GeV-26-lnptratio-5.0.le.lnpt.le.-15-qdreal.dat
  res-antikt-hp-maxscale-as-4-nhard-6-1GeV-9-lnptratio-5.0.le.lnpt.le.-3-double.dat
  }
  
\item The jet radius is sampled randomly between $0.3$ and $1.57$.

\item The jet algorithms are coded as plugins to 
  FastJet~\cite{Cacciari:2011ma} version 3.4.1 using techniques that allow for more
  accurate handling of large rapidities and very small rapidities.
  Ultimately, however, we found that in order to fully explore the
  phase space it was also necessary to use higher-precision numerical
  types from the \texttt{qd} package~\cite{hida2000quad}, up to four
  times normal double precision.
  This was achieved with the help of a suitably converted version of
  the \texttt{fjcore} form of the FastJet package.
  Many of the techniques that we explored were inspired by, adapted
  from and sometimes fed back to the PanScales
  project~\cite{Dasgupta:2020fwr,Hamilton:2020rcu,vanBeekveld:2022ukn}.
  
\item %
  In practice, the framework can operate in two modes: it can sample
  randomly across the available configurations at any given order,
  which is useful to systematically check whether there are any IRC-unsafe configurations for a given algorithm; alternatively, it can
  focus on a specific class of configuration, which is useful when
  trying to understand the detailed nature of any IRC safety issues
  that have appeared.
\item For flavour jet algorithms where the flavour of the cluster
  sequence is meaningful (i.e.\ all algorithms except GHS), we test
  not just the flavour and kinematics of the final jets, but
  additionally those of all steps in the hard+IRC clustering sequence
  that correspond to steps in the hard clustering sequence.
\item Some algorithms (such as flavour-$k_t$ and IFN) can naturally
  handle multiple flavours at a time, while others (such as CMP and
  GHS) are designed around a single flavour at a time.
  Most of our tests will be carried out with one flavour.
  For the higher-order IFN tests that go into our summary,
  Table~\ref{tab:IRC-test-results}, we use six light flavours, so
  as to ensure that we do not accidentally miss IRC issues that would
  arise only for multi-flavoured configurations.
  Plots will always be labelled with the number of light flavours
  used.
\end{itemize}

%----------------------------------------------------------------------
\subsection{Results}
\label{sec:IRC-results}

\begin{figure*}
  \centering
  \subfloat[]{
    \includegraphics[width=0.32\textwidth]{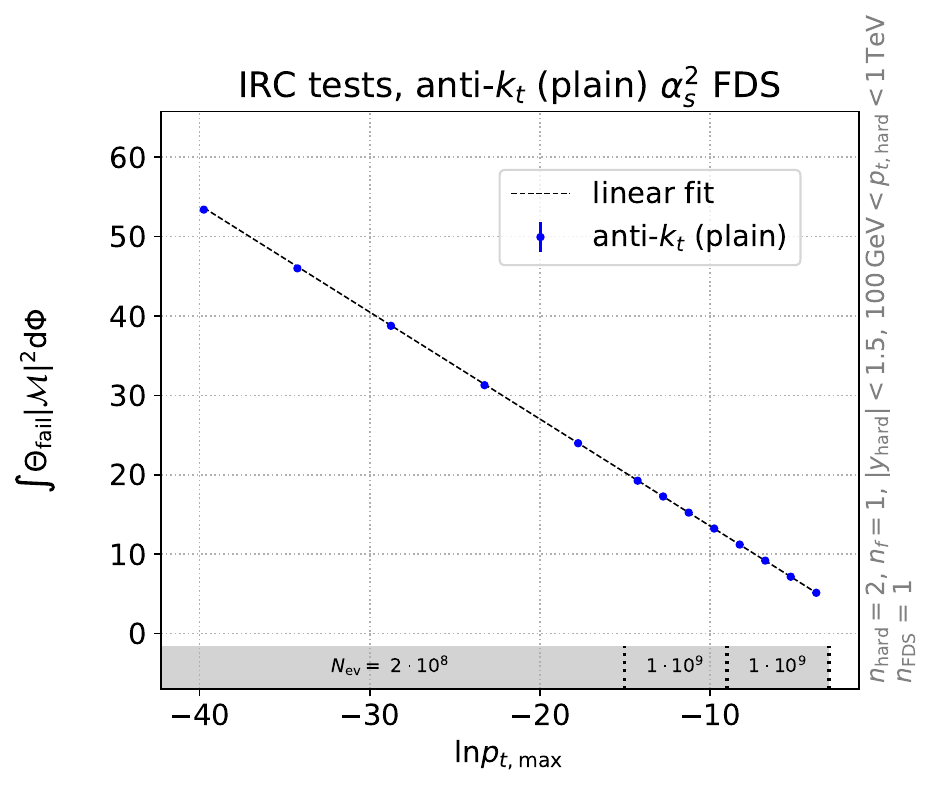}
  }
  \subfloat[]{
    \includegraphics[width=0.32\textwidth,page=2]{figures/plot-akt-IFN-full.pdf}
  }
  \subfloat[]{
    \includegraphics[width=0.32\textwidth,page=4]{figures/plot-akt-IFN-full.pdf}
  }
  \caption{
    Example results from our numerical IRC safety tests showing (a) the IRC unsafety of standard anti-$k_t$ at order $\as^2$, and (b,c) the IRC safety of anti-$k_t$ with Interleaved
    Flavour Neutralisation (IFN) at orders $\as^2$ and $\as^4$, respectively.
    In all plots, the phase-space weighted
    failure rate is shown as a function of the maximum hardness of the extra
    radiation $\ln p_{t,\rm{max}}$, where we sample values $3\ln
    p_{t,\rm{max}} < \ln p_t < \ln p_{t,\rm{max}}$ for any bin of $\ln
    p_{t,\rm{max}}$.
    In (a), the integrated failure rate is plotted on a linear scale for standard anti-$k_t$ in the FDS
    configuration from Fig.~\ref{fig:classic-config} with one double-soft pair and 2 hard particles in the event.
    The classic anti-$k_t$ IRC safety issue is confirmed numerically by the linear divergence $\as^2 \ln p_{t,\rm{max}}$, from one soft gluon splitting to a
    flavoured pair at large angle.
    In (b) and (c), the integrated failure rate is plotted on a logarithmic scale for anti-$k_t$ with IFN for parameters $(\alpha=1,
    \omega=2)$ and $(\alpha=2, \omega=1)$, for all configurations that contribute at (b) order $\as^2$ and (c) order $\as^4$, with up to 8 hard particles in the event and up to 6 flavours.
    The total failure rate goes to zero as $\ln p_{t,\rm{max}} \to -\infty$, implying the IRC safety of anti-$k_t$ with IFN to the tested order and accuracy.
    The absence of points below $\lnptmax \simeq -15$ signals no IRC safety failures out of the
    $5\cdot 10^9$ events studied for lower $\lnptmax$ values
    (the number of events in each of three regions of $\lnptmax$ is
    indicated in the shaded bands at the bottom of the plot).
    The plots also include moderately small values of $\lnptmax$ so
    as to better show the overall scaling behaviour.
  }
\label{fig:IRtests-num}
\end{figure*}

Sample results from our numerical IRC-safety tests are illustrated in
Fig.~\ref{fig:IRtests-num}.
The left-hand plot shows the $\as^2$ FDS contribution to
the phase-space weighted failure rate for the plain anti-$k_t$ algorithm.
The failure rate grows linearly with $\lnptmax$, consistent with the
expectation of an $\as^2 L$ divergence.

The middle plot of Fig.~\ref{fig:IRtests-num} shows the phase-space
weighted failure rate for the anti-$k_t$+IFN algorithm (both with
$\alpha=1$ and $\alpha=2$) at order $\as^2$.
These results are summed over all sampled
configurations (IDS, FDS, IHC$^2$, FHC$^2$, IHC$\times$FHC).
One sees that the failure rate vanishes as $\ptmax \to 0$,
consistently with a power law, implying no IRC divergence.
The right-hand plot shows the results for anti-$k_t$+IFN at order
$\as^4$.
Again the plot indicates that the phase-space weighted failure rate
vanishes as $\ptmax$ is reduced, consistently with a power law.

For more negative values of $\lnptmax$, no points are shown
simply because we observed no failures with anti-$k_t$+IFN.
The grey band at the bottom of the plots shows how the test is broken
up into different regions with the number of events used for each
region ($5\times 10^9$ in the lowest region).
The regions each involve a different underlying numerical precision
type in the code, and one of the limiting factors in our tests is the
speed of the code in the lowest region where we are using four times
the precision of a standard IEEE double type.%
\footnote{\label{footnote:lnptMaxMin}In interpreting these results, one should keep in mind that
  the integration volume at order $\as^n$ is effectively
  $(\lnptmax-\lnptmin)^n$.
  Defining $\lnptmax \equiv L$ and $\lnptmin = (1+c)L$ ($c=2$ in
  Fig.~\ref{fig:IRtests-num}), that corresponds to $(cL)^n$.
  Assuming that the failure rate for a $L^p$ divergence goes as
  $(cL)^p$, then with $N$ events in a given bin, we expect to observe
  a non-zero failure rate down to $L\sim -N^{1/({n-p})}/c$.
  Fig.~\ref{fig:IRtests-num} was generated with $N \sim 10^9$ events
  per point and $c=2$.
  We therefore expect that up to $n=4$ we should observe all main classes
  of failure (i.e.\ any case with $p\ge 0$) for $L\gtrsim -90$, i.e.\
  over the full range of $L = \lnptmax$ in
  Fig.~\ref{fig:IRtests-num} (keeping in mind that order $1$ factors
  that we have neglected can have a significant impact on the range).
  For $n=5$, failures with a $p=0$ structure would be observed only
  for $L \gtrsim -30$, while for $n=6$, this would reduce to $L
  \gtrsim -15$.
  Note that it is still useful to explore the full range of $L$, even
  at high orders $n$, because there can be cases where an IRC
  divergence appears for the first time already with some number of
  logarithmic enhancements (e.g.\ the $L^3$ divergence that appears at
  $\as^4$ in the GHS algorithm, as discussed in
  App.~\ref{sec:GHS-trouble-as4}).
  A final comment concerns the overall normalisation of the
  result of the phase-space weighted failure rate. This can appear
  large at moderate values of $L$ where there are still failures,
  because it includes the $(cL)^n$ integration volume. }
Overall,
the results provide a strong indication of IRC safety for the
anti-$k_t$+IFN algorithm developed in this paper.

\begin{table*}[t]
\begin{tabular}{c|c|C{1.6cm}|C{1.6cm}|C{1.6cm}|C{1.6cm}|C{1.6cm}|C{1.6cm}}
\multicolumn{2}{c|}{order relative to Born} & anti-$k_t$ & flav-$k_t$ ($\alpha=2$) & CMP & GHS$_{\alpha,\beta}$\newline$(2,2)$ & anti-$k_t$+IFN$_{\alpha}$ & C/A+IFN$_{\alpha}$ \\\hline\hline
\multirow{2}{*}{$\quad\alpha_s\quad$}
& \,FHC\, & {\ok} & {\ok} & {\ok} & {\ok} & {\ok} & {\ok}\\\cline{2-8}
& \,IHC\, & {\ok} & {\ok} & {\ok} & {\ok} & {\ok} & {\ok}\\\hline\hline
\multirow{5}{*}{$\quad\alpha_s^2\quad$}
& \,FDS\, & {\no$_\text{\ref{sec:flavour-recomb}}$} & {\ok} & {\ok} & {\ok} & {\ok} & {\ok}\\\cline{2-8}
& \,IDS\, & {\no$_\text{\ref{sec:flavour-recomb}}$} & {\ok} & {\ok} & {\ok} & {\ok} & {\ok}\\\cline{2-8}
& \,FHC$\times$IHC\, & {\ok} & {\ok} & {\ok} & {\ok} & {\ok} & {\ok}\\\cline{2-8}
& \,IHC$^2$\, & {\ok} & {\ok} & {\no$_\text{\ref{sec:CMP-trouble-as2}}$} & {\ok} & {\ok} & {\ok}\\\cline{2-8}
& \,FHC$^2$\, & {\ok} & {\ok} & {\ok} & {\no$_\text{\ref{sec:GHS-trouble-as2}}$} & {\ok} & {\ok}\\\hline\hline
\multirow{2}{*}{$\quad\alpha_s^3\quad$}
& \,IHC$\times$IDS\, & {\hc} & {$\sim_\text{\ref{sec:IRCapp-flavkt}}$} & {\no$_\text{\ref{sec:CMP-IHC-DS}}$} & {$\sim_\text{\ref{sec:IRCapp-flavkt}}$} & {\ok} & {\ok}\\\cline{2-8}
& \,rest\, & {\hc} & {\hc} & {\hc} & {\hc} & {\ok} & {\ok}\\\hline\hline
\multirow{2}{*}{$\quad\alpha_s^4\quad$}
& \,IDS$\times$FDS\, & {\hc} & {\hc} & {\hc} & {\no$_\text{\ref{sec:GHS-trouble-as4}}$} & {\ok} & {\ok}\\\cline{2-8}
& \,rest\, & {\hc} & {\hc} & {\hc} & {\hc} & {\ok} & {\ok}\\\hline\hline
\multirow{1}{*}{$\quad\alpha_s^5\quad$}
& \,\, & {\hc} & {\hc} & {\hc} & {\hc} & {\ok} & {\ok}\\\hline\hline
\multirow{1}{*}{$\quad\alpha_s^6\quad$}
& \,\, & {\hc} & {\hc} & {\hc} & {\hc} & {\ok} & {\ok}\\\hline\hline
\end{tabular}
  \caption{
    Summary of the IRC safety test results.
    Red crosses ({\no}) indicate a clear failure of IRC safety.
    Checkmarks (\ok) signify that the algorithm passes numerical tests at that order or for that configuration.
    The tilde ($\sim$) for flavour-$k_t$ (and by extension GHS, which uses flavour-$k_t$
    distances) indicates marginal convergence, though one expects 
    divergent behaviour at higher orders.
    For algorithms that fail or are marginal at a given order, we display greyed-out
    boxes at higher orders, since those higher orders are also bound to
    fail.
    In a few cases, we have identified a new class of problem that
    only arises at higher order and we explicitly mark these with a red cross.
    The GHS parameters here are set to $\alpha=2,\beta=2$.
    The IFN procedure is tested both for the anti-$k_t$ and C/A
    algorithms, and the IFN parameters are chosen as $\alpha \in
    \{1,2\}$ with $\omega=3-\alpha$ 
    (tests are successful for both sets of parameters).
    Detailed discussions of the issues identified are linked to from
    the relevant table cells.
    Plots in support of the IRC safety conclusion for the IFN
    combinations are to be found in
    App.~\ref{sec:final-summary-plots}, specifically
    Figs.~\ref{fig:IRC-safety-summary-IFN-akt} and
    ~\ref{fig:IRC-safety-summary-IFN-cam}, as are plots
    (Figs.~\ref{fig:IRC-safety-summary-FlavKtOmega} and
    \ref{fig:IRC-safety-summary-CMPOmega}) supporting the IRC safety
    of our modified versions of the flavour-$k_t$ and CMP algorithms,
    respectively flavour-$k_{t,\Omega}$ and CMP$_\Omega$, which are
    discussed in the text. (They are not shown in the table, because
    we have run them with lower statistics.)}
  \label{tab:IRC-test-results}
\end{table*}

Table~\ref{tab:IRC-test-results} summarises the results of our testing framework applied to a range of jet algorithms.
At lowest order, we organise the results according to the class of
divergence being probed, as indicated in the second column of the
table, while at higher orders we limit the breakdown to configurations
that have turned out to be of specific interest.
The corresponding failure rate plots for the IFN algorithms (with the anti-$k_t$ and
C/A algorithms) are given in App.~\ref{sec:final-summary-plots}.

Algorithms whose failure rate goes down as the extra radiation becomes
softer/more collinear are indicated by a checkmark (\ok).
Algorithms that develop a divergence (a non-vanishing integrated
failure rate as $\ln p_{t,\rm{max}} \to -\infty$) for a given target
configuration, are marked by a red cross ({\no}).
For each case that has shown a divergence, we have examined a few
events where there is a clear failure and developed an analytic
understanding of the nature of the problem.
We will briefly discuss each issue here, while the table also
links to the relevant part of App.~\ref{sec:list-ir-risky} with
further analytic and numerical studies.

The first two rows of Table~\ref{tab:IRC-test-results} emphasise that at order $\as$ with just one emission (FHC or IHC), there are no divergences for any jet algorithm with IRC safe kinematics, even without a special treatment of flavour.
The classic IRC safety problem of standard anti-$k_t$ only shows up at order $\as^2$, as highlighted in the next two rows of Table~\ref{tab:IRC-test-results} for a configuration with one double-soft pair (see Fig.~\ref{fig:classic-config}).
That problem arises in both the FDS and the IDS channels, and in each
case it appears for the subset of events where the FDS or IDS pair is
at large angles.
From the table, it is clear that all flavour jet algorithms solve that original IRC safety issue.

However, the tests reveal new issues for all algorithms other than our
IFN-based procedure. 
In two cases, flavour-$k_{t,\Omega}$ and CMP$_\Omega$, we will propose
modifications that seem to resolve the problem(s).
For the interested reader, the summary of the issues is as follows:
\begin{itemize}
\item \textbf{Initial-state (IHC$\times$IDS) subtlety at $\as^3$ for flavour-$k_t$ and GHS.}
   The ``$\sim$'' for $\alpha=2$ flavour-$k_t$ and GHS at $\as^3$
  (IHC$\times$IDS) indicates a borderline case.
  It arises, for example, for a hard event consisting of a single
  energetic parton (and resulting hard jet), supplemented with 
  a hard-collinear initial-state splitting and a large-angle double-soft pair, which may be IDS or FDS (see
  Fig.~\ref{fig:IRsafety-as3-IHC-DS-diagram}, together with the
  complete details in App.~\ref{sec:IRCapp-flavkt}).
  When one (anti)quark from the double-soft pair is somewhat softer
  than the other, its $d_{ij}$ distance with the hard-collinear
  particle can be smaller than that with the other (anti)quark from
  the soft pair, essentially because the $\Delta R_{ij}^2$ distance
  goes as $\Delta y_{ij}^2$, which is only logarithmically
  large.
  The large-angle soft (anti)quark and the initial-state collinear quark
  cluster, leaving a lone large-angle soft quark, which can contaminate
  the flavour of the hard jet.
  At $\order{\as^3}$ one ends up with an integral that goes as $\int
  d\ln p_t / (\ln p_t)^2$.
  This integral converges for $p_t \to 0$, however the way in which the integrand
  (multiplying $d\ln p_t$) vanishes as $p_t \to 0$ is not a power-law
  in $p_t$.
  One may thus consider the algorithm to be marginally safe at this
  order, however at the next order one would expect to see additional
  logarithmic enhancements.
  These might arise, e.g.\ from the running of the QCD coupling or
  evolution of the parton distribution functions (PDF),
  and would ultimately cause the integral to
  diverge.
  Indeed, our study identified a problem in the
  IHC$^2\times$IDS channel at order $\as^4$.
  However, a conclusive understanding of this configuration requires
  inclusion also of the virtual and PDF-counterterm contributions,
  which is beyond the scope of this study.
  A similar problem arises with $\alpha=1$, but with extra
  logarithms in the denominator of the corresponding
  integral. 
  This generic class of problem can be solved by replacing
  $\Delta R_{ij}^2 \to \Omega_{ij}^2$, and, as before, we will use
  Eq.~(\ref{eq:default-omega}) as our default choice for its $\omega$
  parameter.
  We refer to the modified algorithm as flavour-$k_{t,\Omega}$.
  This simple adaptation is possible because the issue is not with the
  original underlying strategy, but rather with the subtleties that
  arise in distance measures with QCD initial-state radiation (the
  same comment holds for related issues in other algorithms).
  As a consequence we do not expect to have to make any modifications to
  the $e^+e^-$ version of the flavour-$k_t$ algorithm.
  
\item  \textbf{Initial-state (IHC$^2$) issue at $\as^2$ for CMP.}
  This
  issue arises, for example, for a hard
  (Born) event consisting of a single hard parton, supplemented with
  two collinear initial-state quark and anti-quark emissions, one on
  each beam (see Fig.~\ref{fig:config-CMP-as2} and
  App.~\ref{sec:CMP-trouble-as2}).
  Those initial-state emissions cluster in the first step of the
  algorithm, producing a large-mass, low-$p_t$ flavourless pseudojet
  at central rapidities, which can then cluster with the hard parton,
  modifying its kinematics.
  The problem arises because in the CMP distance
  Eq.~(\ref{eq:flav-antikt-dij}), the small factor from the
  transverse-momenta dominates over the (only logarithmically large)
  factor from the rapidity separation between the pair.
  Ultimately this leads to an $\as^2 L^2$ divergence. 
  It can be resolved by replacing
  \begin{equation}
    \label{eq:CMP-fix}
    \cS_{ij} \to \overline{\cS}_{ij} =  \cS_{ij} \frac{\Omega_{ij}^2}{\Delta R^2_{ij}}
  \end{equation}
  for oppositely flavoured pairs and requiring the parameter $\omega >1$
  in the $\Omega_{ij}$ distance.
  In practice, we find that this modification has almost no impact on
  the phenomenological behaviour of the algorithm (e.g.\ $\lesssim1\%$
  in Fig.~\ref{fig:ttbar-tests} below).  
  \logbook{eb847975}{
    see figures/ttbar-cmp-checks.pdf.
  }
  We refer to this modified version of the CMP
  algorithm as CMP$_\Omega$ and unless otherwise specified we use
  $\omega=2$. 

\item \textbf{Initial-state (IHC$\times$IDS) issue at $\as^3$ for CMP.}
  This
  issue involves the same
  configuration and sequence that led to a marginal issue for
  flavour-$k_t$ (Fig.~\ref{fig:IRsafety-as3-IHC-DS-diagram}), but here
  it brings an $\as^3 L$ divergence (App.~\ref{sec:CMP-IHC-DS}),
  rather than $\as^3/L$.
  Recall that the problem here is that the quark from the large-angle
  double-soft pair can cluster with the IHC ($\bar q$) particle,
  leaving the large-angle soft $\bar q$ to contaminate the hard jet
  flavour.
  The different IRC behaviour of CMP versus flavour-$k_t$ can in part
  be attributed to the fact that the CMP algorithm retains the
  standard anti-$k_t$ form of the beam distance, causing the
  beam-clustering of the IHC particle to come at the end of the
  algorithm.
  In contrast, in the flavour-$k_t$ algorithm, the IHC beam clustering
  nearly always comes before the IHC particle can cluster with the
  soft large-angle quark, reducing the phase-space
  associated with IRC problems for this configuration.
  The fix of Eq.~(\ref{eq:CMP-fix}), i.e.\ CMP$_\Omega$, also solves
  this problem.
  
\item \textbf{Final-state (FHC$^2$) issue at $\as^2$ for GHS.}
  This problem
  appears with four hard particles such that each of two hard jets
  contains one flavoured ($q$ or $\bar q$) and one non-flavoured
  particle, as could arise in semi-leptonic $t\bar t$ decays when
  considering only $b$'s to be flavoured.
  The $\as^2$ modification of the event involves the hard-collinear
  splitting of one of the energetic quarks, followed by the
  hard-collinear splitting of that gluon into a $q'\bar q'$ pair
  (see Fig.~\ref{fig:config-GHS-as2} and App.~\ref{sec:GHS-trouble-as2}).
  The accumulation stage leaves the collinear $q'\bar q'$ as separate
  flavour clusters, and relative to the original Born event, the
  energy of the hard $q$'s cluster is modified.
  During the dressing stage, the collinear $q'\bar q'$ annihilate
  immediately, but the modification of the energy of the hard-$q$
  cluster means that that cluster can behave differently during the
  dressing stage. 
  We envisage that this problem could be solved by accounting for
  energies within each jet during the dressing stage, but have yet to
  formulate a concrete modification of the algorithm.
  This issue is present independently of the parameters of the
  algorithm and leads to an $\as^2 L^2$ divergence. 

\item \textbf{Mixed initial/final-state (IDS$\times$FDS) issue at $\as^4$ for GHS.}
  This problem
  involves a hard event with one Born gluon ($h$) leading to a single
  hard jet.\footnote{We are grateful to Simone Marzani for discussions
    that first led us to investigate this configuration analytically.}
  The issue arises with a final-state soft-collinear $q_1\bar q_2$
  pair emitted inside the jet and a large-angle $q_3\bar q_4$ pair
  (see Fig.~\ref{fig:GHStrouble} and App.~\ref{sec:GHS-trouble-as4}).
  If $\theta_{h1} < \theta_{12},\theta_{h2}$, then the $h1$ clustering
  will
  be the first step of the accumulation stage and may pass the
  SoftDrop condition, resulting in a pseudojet with the energy of $h$
  but the flavour of $1$, which goes on to form a flavour cluster
  separate from that of $q_2$.
  The two flavour clusters in the jet now have a large hierarchy of
  energy, and the softer one ($q_2$) may ultimately annihilate with a
  large-angle soft quark ($3$ or $4$) if the latter has a similar
  (or larger)
  $p_t$ with respect to the beam, resulting in the flavour of the hard
  jet being set by the $h1$ cluster.
  This issue gives an $\as^4 L^3$ divergence for $\alpha\beta>2$,
  and numerical results are consistent with an $\as^4 L$
  divergence for $\alpha\beta=2$.
  The analytical study of App.~\ref{sec:GHS-trouble-as4} indicates
  that the problem should be resolved when one takes $\alpha\beta < 2$
  (if one additionally replaces $\Delta R_{ij}^2 \to \Omega_{ij}^2$),
  though the above FHC$^2$ issue remains.
\end{itemize}

The discovery of the above issues highlights the importance of having a
systematic framework for testing IRC safety.
Indeed, some of these issues 
were first discovered with the testing framework, as were
the identification of $\alpha \beta < 2$ as a potential solution for
the GHS $\as^4$ (IDS$\times$FDS) issue, and the requirement
Eq.~(\ref{eq:omega-constraint}) for the IFN algorithm.
Such tests also led us to suspend study of more general $u_{ij}$
distances involving
$u_{ij} = [\max(p_{ti},p_{tj})]^{2p} [\min(p_{ti},p_{tj})]^{2q}\,
\Omega_{ij}^2$, specifically a dimensionless form with $p=-q=1$.

A final comment is that it is important to remember that the IRC tests
cover many cases, but are not totally exhaustive.
Specifically, as
discussed in Sec.~\ref{sec:IRC-methodology}, we have at most one
power of $L$ per power of $\as$, and only up to $\as^6$, the events
that we have generated have a band gap,\footnote{By band gap we mean
  the white region between the upper hard tip of the Lund diagrams in
  Fig.~\ref{fig:IRtests-sampling} and the upper edge of the shaded
  region.
  The concern that one might have is that of an IRC unsafety mechanism
  whereby an emission at some momentum scale $\epsilon$ clusters with an
  emission at a still soft, but much larger scale $\epsilon^{1/p}$
  (for some $p>1$) in the white band-gap region, and only after that
  clustering can it cause the IRC unsafety.
  Our test procedure is only sensitive to values of $p <
  \lnptmin/\lnptmax$.
  One would therefore like to take as large a
  value of that ratio as possible, keeping in mind however that larger
  values of $\lnptmin/\lnptmax$ are numerically more challenging, both
  because of the higher precision and the need for higher statistics,
  cf.\ footnote~\ref{footnote:lnptMaxMin}. 
} and a couple of nestings are still missing,
notably as regards double-soft emissions.
Thus our tests should not be considered an ultimate proof of IRC
safety, but merely a strong indication.

%======================================================================
\section{Phenomenological illustrations}
\label{sec:pheno}

In this section, we present three phenomenological test cases, intended
to convey some of the main features of our IFN algorithms.
We include comparisons to standard anti-$k_t$ clustering and also
to those prior flavour algorithms for which we have been able to identify
an IRC-safe adaptation, namely flavour-$k_{t,\Omega}$ and CMP$_\Omega$.

The first two tests will be specific to heavy flavour, which is the main
experimental application of flavoured jet algorithms.
The third test will be for generic flavour and can be seen as a stress
test of the algorithm's practical performance with light flavour at
parton level.

% ------------------------------------------
\subsection{Heavy flavour in $pp \to WH(\to \mu\nu b\bar b)$}
\label{sec:pheno-WH}

We begin with the case of Higgs production in association with a $W$ boson
at hadron colliders, $pp \to WH$, where the Higgs boson decays to a pair of
$b$-quarks and the $W$ decays leptonically.
This process is of interest for obvious phenomenological reasons,
e.g.\ because of the sensitivity to the $HWW$ and $Hb\bar b$ couplings,
and it has been measured by both ATLAS and
CMS~\cite{CMS:2017odg,ATLAS:2020fcp}. 
Additionally, it is one of the processes in which one can probe
high-$p_t$ Higgs production~\cite{ATLAS:2020jwz,CMS:2020zge},
especially in
conjunction with jet substructure
tools~\cite{Butterworth:2008iy,Marzani:2019hun}, bringing particular
sensitivity to new physics. 
For a long time, calculations at NNLO QCD were performed with massless
$b$ quarks, which prohibited the use of the standard anti-$k_t$ algorithm
to cluster the final state.
Only recently~\cite{Behring:2020uzq} was the calculation
performed with massive $b$-quarks.

Here, we examine a classic resolved-jet analysis of this process,
similar to that of Ref.~\cite{Behring:2020uzq}.
We use Pythia 8.306~\cite{Sjostrand:2014zea,Bierlich:2022pfr}
with the 4C tune~\cite{Corke:2010yf}
to 
generate $pp \to W(\to \mu \nu_{\mu})H(\to b\bar b)$.
Following Ref.~\cite{Behring:2020uzq}, we require the presence of a
muon satisfying
\begin{subequations}
  \begin{equation}
    \label{eq:WH-muon-cuts}
    |\eta_\mu | < 2.5\,,\quad p_{t\mu} > 15 \mathrm{\,GeV}\,. \\
  \end{equation}
  We cluster the event with a given jet algorithm, using a jet radius
  of $R=0.4$, and identify $b$-flavoured jets that satisfy
  \begin{equation}
    |y_{j_b}| < 2.5\,,\quad p_{tj_b} > 25 \mathrm{\,GeV}\,.
  \end{equation}
  We require the event to have at least two such
  jets. 
\end{subequations}
Finally, the reconstructed Higgs boson is defined as the 4-momentum sum of
the two $b$-jets whose invariant mass is closest to the Higgs mass.

\begin{figure}
  \centering
  \includegraphics[width=0.9\columnwidth,page=4]{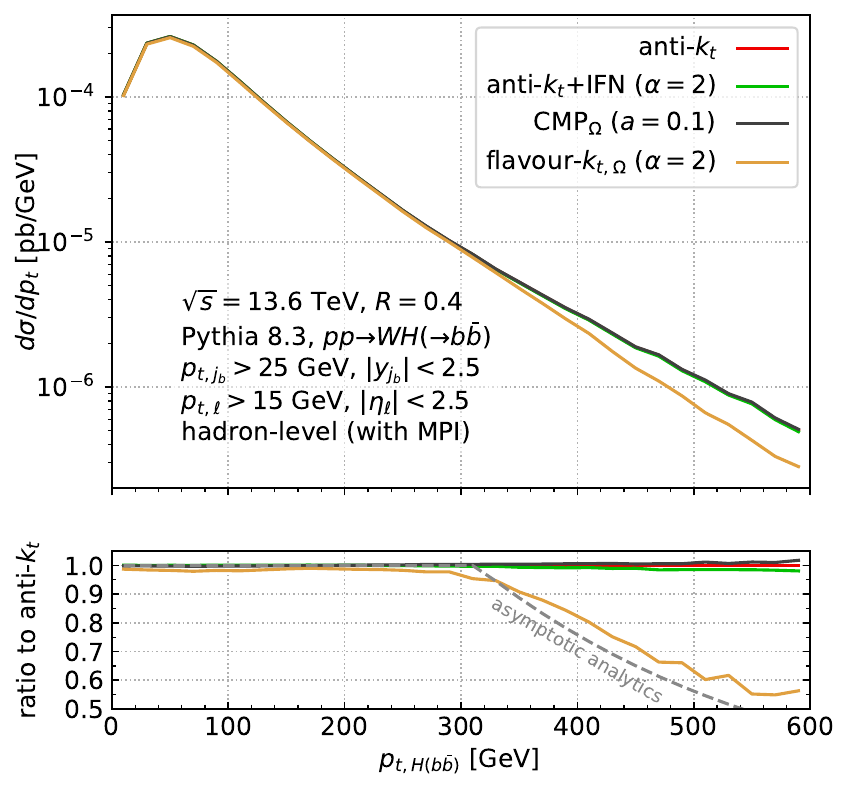}
  \caption{
    The transverse momentum spectrum of the reconstructed Higgs boson in $WH(\to
    \mu\nu b\bar b)$ at centre-of-mass energy $\sqrt{s}=13.6$ TeV, at hadron level
    (with stable $B$-hadrons).
    The upper panel shows the spectrum for four jet algorithms: anti-$k_t$
    with net flavour of the jet constituents (red), our
    IFN version of anti-$k_t$ (with $\alpha=2$, green), the CMP$_\Omega$
    algorithm (as adapted from Ref.~\cite{Czakon:2022wam} with a fix 
    of the angular measure, see Eq.~(\ref{eq:CMP-fix}), black) and the
    flavour-$k_{t,\Omega}$ algorithm (with $\alpha=2$, gold).
    The lower panel shows the ratio to standard anti-$k_t$. 
    CMP$_\Omega$ and our IFN algorithms all give very similar results to those
    from the plain anti-$k_t$ algorithm.
    In contrast, as already pointed out in
    Ref.~\cite{Behring:2020uzq}, flavour-$k_{t,\Omega}$ 
    jets can differ significantly from anti-$k_t$ kinematics at large
    transverse momentum, because they start clustering the $b$ and
    $\bar b$ together into a single jet well before the scale of $p_t \simeq 2m_H/R =
    625\GeV$ where this occurs with the normal anti-$k_t$ algorithm.
    This is reflected in Eq.~(\ref{eq:flavkt-alpha2-acceptance}), which
    is used to generate the ``asymptotic analytics'' curve in the
    lower panel.
  }
  \label{fig:WHbb-tests}
\end{figure}

The distribution of the transverse momentum of the reconstructed Higgs boson
is presented in Fig.~\ref{fig:WHbb-tests} at hadron level (with multi-parton
interactions turned on), for four algorithms:
\begin{itemize}
\item standard anti-$k_t$ with net flavour summation (red),
\item anti-$k_t$ with our IFN algorithm ($\alpha=2$, in green),
\item the CMP$_\Omega$ algorithm ($a=0.1$, where the angular part of the distance measure
is corrected as in Eq.~(\ref{eq:CMP-fix}), in black), and
\item the flavour-$k_{t,\Omega}$ algorithm ($\alpha=2$, in gold).
\end{itemize}
The flavour-$k_{t,\Omega}$ algorithm leads to a reconstructed Higgs
spectrum that is markedly different from that of the anti-$k_t$
algorithm.
In particular, for $p_{tH} \gtrsim 300\GeV$, the distribution starts
to drop relative to that with anti-$k_t$, reaching about 60\% of the
latter's value at $p_{tH} \sim 600$ GeV.
As noted in Ref.~\cite{Behring:2020uzq}, this occurs 
because the flavour-$k_t$ algorithm starts clustering the $b$ and
$\bar b$ together at lower values of $p_{tH}$ than for the anti-$k_t$
algorithm.
When the $b$ and $\bar b$ end up in a single jet, the event fails the
selection requirement of having at least two $b$-jets. 
Specifically for the decay of a scalar particle with invariant mass $m$ and
transverse momentum $p_t$, for small $R$ and in the limit of
$p_{t}R \ll m$, the efficiency for having two separate jets (without
any $p_{t}$ or rapidity cut on the jets) is $1$ at low $p_t$.
Above some threshold in $x = p_t R/m > x_{\min}$, it becomes
\begin{subequations}
  \label{eq:alg-acceptances}
\begin{align}
  \text{gen-$k_t$}&:\quad 1 - \frac{\sqrt{x^2-4}}{x}, &x_{\min} &= 2\,,
  \\
  \text{$\alpha=1$ flav-$k_t$}&:\quad \frac{2}{x^2}, &x_{\min} &= \sqrt{2}\,,
  \label{eq:flavkt-alpha1-acceptance}
  \\
  \text{$\alpha=2$ flav-$k_t$}&:\quad \frac{2}{1+x^2},& x_{\min} &= 1\,.
  \label{eq:flavkt-alpha2-acceptance}
\end{align}
\end{subequations}
The last of these, in particular, explains the qualitative behaviour
seen in Fig.~\ref{fig:WHbb-tests}, cf.\ the ``asymptotic analytics''
dashed line in the lower panel.
Note that Eqs.~(\ref{eq:flavkt-alpha1-acceptance}) and
(\ref{eq:flavkt-alpha2-acceptance}) are independent of the
$\Delta R_{ij}^2 \to \Omega_{ij}^2$ change, because they are evaluated
in a small-angle limit, where the two distance measures are identical.

In Fig.~\ref{fig:WHbb-tests},
the CMP$_\Omega$ algorithm, as well as anti-$k_t$+IFN, give results that are very similar to
those of plain anti-$k_t$ jets, to within about a percent.
This result is not entirely trivial: while it was expected that the
new generation of flavour algorithms should be kinematically more
similar or identical to anti-$k_t$, there was still a possibility that
flavour assignments could modify cross sections, e.g.\ because the
original anti-$k_t$ jets' flavours would have been subject to
modifications from soft $b\bar b$ pairs, while any such effect should
be substantially reduced for the new algorithms.
The absence of a numerically significant difference between jet
algorithms other than flavour-$k_{t,\Omega}$ suggests that for signal processes
like that shown here, with $R=0.4$ jets, the contribution of soft
$g\to b\bar b$ contamination is relatively small.
On one hand, this means that certain experimental signal measurements with
standard anti-$k_t$ jets may not require much unfolding in order to be
compared to the flavoured jet definitions.
On the other hand, if one is to perform a higher-order calculation
with the approximation of massless $b$ quarks, a flavoured jet
algorithm will still be required in order to obtain a finite
result.

% ------------------------------------------
\subsection{Heavy flavour in $pp \to t\bar t \to \ell\nu+\text{jets}$}
\label{sec:pheno-ttbar}

\begin{figure*}
  \centering
    \subfloat[]{
    \includegraphics[width=0.48\textwidth,page=1]{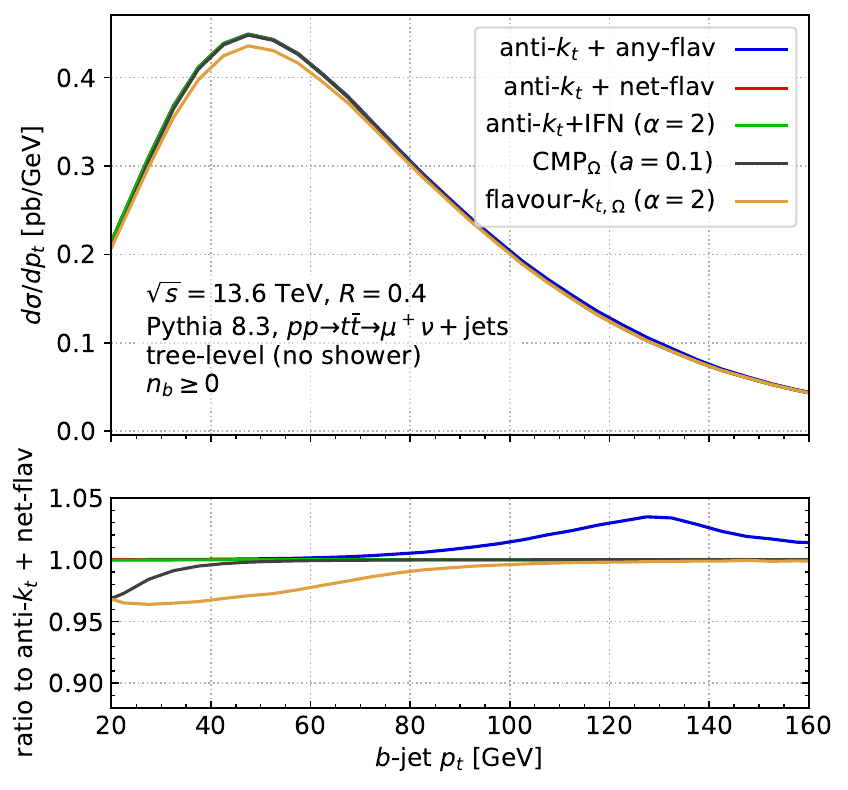}
      \label{fig:ttbar-tests_a}
  }
  \hfill
      \subfloat[]{
\includegraphics[width=0.48\textwidth,page=4]{figures/ttbar-tests.pdf}
  \label{fig:ttbar-tests_b}
}
  \caption{
  Inclusive $b$-jet spectrum from Pythia 8.3 in
    $pp \to t\bar t + X \to b\mu^+\nu \bar b q\bar q'  +
    X$ events at $\sqrt{s}=13.6$ TeV, at
    (a) partonic tree level (i.e.\ no showering or hadronisation)
    and (b) hadron level (with stable $B$-hadrons).
    The distribution is shown in the upper panels, for four jet
    algorithms (as in Fig.~\ref{fig:WHbb-tests}) and additionally for
    anti-$k_t$ with ``any-flavour'' recombination (i.e.\ a $b\bar b$
    jet counts as $b$-tagged).
    The lower panels show the ratio to anti-$k_t$ jets with net
    flavour summation.
    The anti-$k_t$+IFN algorithm  yields
    a $b$-jet spectrum that is almost identical to that from
    the net-flavour anti-$k_t$ algorithm,  across the whole
    $p_t$ range. 
    The closeness to anti-$k_t$ holds both at tree level and after
    showering and hadronisation (with the spectrum differing maximally
    by less than a percent at $p_t=20$ GeV, at hadron level).
    See text for further details.  }
  \label{fig:ttbar-tests}
\end{figure*}

For our second test case with heavy flavour, we consider top-quark
pair production in hadron collisions with semileptonic top decays
$pp \to t\bar t \to \ell + {\rm jets}$.
We select events by requiring at least one muon with
\begin{equation}
p_{t\mu} > 30 {\,\rm GeV}\,,\quad |\eta_{\mu}| < 2.4\,,
\end{equation}
and additionally $p_{t,{\rm miss}} > 30 {\,\rm GeV}$.
We then run a jet algorithm and examine the $p_t$ distribution of jets
that are considered $b$-flavoured according to a given jet algorithm,
with a requirement of $p_t > 20\GeV$ applied to the jets.
We again use Pythia~8.306, but now with the Monash13
tune~\cite{Skands:2014pea}.
It will be instructive to examine results
at both tree-level (where we use $b$ quarks as the flavoured inputs to
the algorithms) and hadron level including parton showering (where we
use stable $B$-hadrons as the flavoured inputs).

The results are shown in Fig.~\ref{fig:ttbar-tests}. 
The inclusive $b$-jet $p_t$ spectrum is shown in the upper panels,
on the left at partonic tree level, i.e.\ without shower or
hadronisation, and on the right at hadron level including showering.
We examine the same algorithms as in Fig.~\ref{fig:WHbb-tests}
and since we are again interested in the similarity of the distributions to
that of standard anti-$k_t$ with net flavour summation (in red), the
lower panels show the ratio to that result.
Additionally, we include a line for standard anti-$k_t$
clustering with ``any-flavour'' tagging, i.e.\ counting a
$b\bar b$-jet as a $b$-tagged jet, which is more in line with
experimental procedures than net flavour recombination.

Let us start by examining the tree-level results in Fig.~\ref{fig:ttbar-tests_a}.
When each jet contains at most one parton, the IFN algorithm is, by
design, intended to give identical results to the plain net-flavour
anti-$k_t$ algorithm.
Note that the tree-level Pythia sample does not guarantee this
property, since our analysis does not require each of the four
tree-level partons to be in four separate jets and, sometimes, two
tree-level partons may cluster together.
Nevertheless, we see that the IFN algorithm (green) gives results that
are essentially indistinguishable from the plain net-flavour
anti-$k_t$ net-flavour (red) results.
The flavour-$k_{t,\Omega}$ algorithm (gold) is expected to show differences,
but these are relatively modest, typically a few percent.
Finally the CMP$_\Omega$ results (with $a=0.1$, including the IRC fix as
in Eq.~(\ref{eq:CMP-fix}), in black) show a few-percent depletion of
low-$p_t$ $b$-jets.
We believe that this is a consequence of the small clustering distance
for pairs of two low-$p_t$ $b$-flavoured particles, which enhances the
likelihood that such pairs will cluster, even when well separated.

At hadron-level in Fig.~\ref{fig:ttbar-tests_b}, including parton showering and multi-parton
interactions, the qualitative pattern is broadly similar, but
with some effects enhanced relative to what is seen at tree level.
Now there are very small differences between anti-$k_t$ and 
IFN, below a percent --- this once again suggests that soft
$g \to b\bar b$ induced contamination is a small effect, as we noted
in the $WH$ case of Sec.~\ref{sec:pheno-WH}.
In contrast, the relative differences of flavour-$k_{t,\Omega}$ and CMP$_\Omega$ as
compared to anti-$k_t$ now reach $8{-}10\%$.
Further examining the results we have identified two effects that
contribute to this:
(1) a small reduction of the ratios due the shower and hadronisation,
in events where there are two $B$-hadrons, perhaps because
fragmentation of the $b$ quarks enhances the impact of modified
clustering distances for flavoured particles;
and (2) substantially smaller ratios, especially at low $p_t$, in the
$\sim 8\%$ of events with an additional $b\bar b$ pair from the
showering.
\logbook{c30d5df}{look at MC-tests/ttbar-results-2023-03-21/ttbar_2b_v_4b.pdf
  for further info.
  The $n_b$ fractions are a little hard to read exactly from there,
  but ``mergeidx.pl -f py8-rts13.6-allrseq.dat event.nb'' gives the
  bins and one gets $4-jet/all = 0.00145/0.0188 = 0.077$.
}

Finally, regarding anti-$k_t$ with ``any-flavour'' recombination
(in blue), we see that it differs only by a few percent from
the net-flavour tagging and the IFN algorithms.
The difference appears mostly to be associated with events where the
$b$ and $\bar b$ from the $t$ and $\bar t$ decays end up in a single
jet.
\logbook{167301ee9}{Look at figures/ttbar-tests-allalgs.pdf and note that with
  the $nbjet>=2$ requirement, at tree level, the difference goes away}
Insofar as any-flavour recombination is a good stand-in for standard
experimental tagging, the similarity of net-flavour and any-flavour
recombination indicates only a limited need for unfolding corrections in order for
experimental $t\bar t$ results to be presented unfolded to an
IFN-style flavour truth level.
Note, nevertheless, that there are other processes for which this
would not be true, e.g.\ inclusive $b$-jet
production~\cite{Banfi:2007gu}, and a case-by-case study is needed to
establish whether any-flavour and net-flavour recombination are
similar for a given process.
%

% ------------------------------------------
\subsection{Full flavour at parton level in $pp \to Z+j$}
\label{sec:pheno-Zj}

Our final hadron-collider test is carried out at parton level (after showering) and
applies jet flavour algorithms to all flavours of partons in the
context of events with a hard jet recoiling against a high-$p_t$ $Z$
boson. 
This study is not intended to be of direct experimental relevance, but
rather to test the flavour algorithm's performance and limitations for
addressing more theoretical questions such as the fractions of quark
v.\ gluon jets.
In particular, knowledge of the quark v.\ gluon fractions in a given
sample is important when assessing the performance of approaches that
attempt to distinguish quark v.\ gluon-induced jets from jet
substructure and energy flow observables~\cite{Gras:2017jty}.
To do so we study  $pp \to Z+j$ events.
We focus here on the $Z(\to \mu^+\mu^-)+q$ final state, where we
require exactly two muons to reconstruct a high-$p_t$ $Z$ candidate:
\begin{subequations}
  \begin{align}
&|\eta_{\mu}| < 2.4\,,\qquad\qquad p_{t\mu} > 20 {\,\rm GeV}\,,\\
&p_{t,\mu^+\mu^-} > 1 {\,\rm TeV}\,,\,\quad m_{\mu^+\mu^-} \in [80,102] {\,\rm GeV}\,.
\end{align}
\end{subequations}
We find qualitatively consistent results for the $Z+g$ case.

\begin{figure*}
  \centering
  \includegraphics[width=0.99\textwidth,page=1]{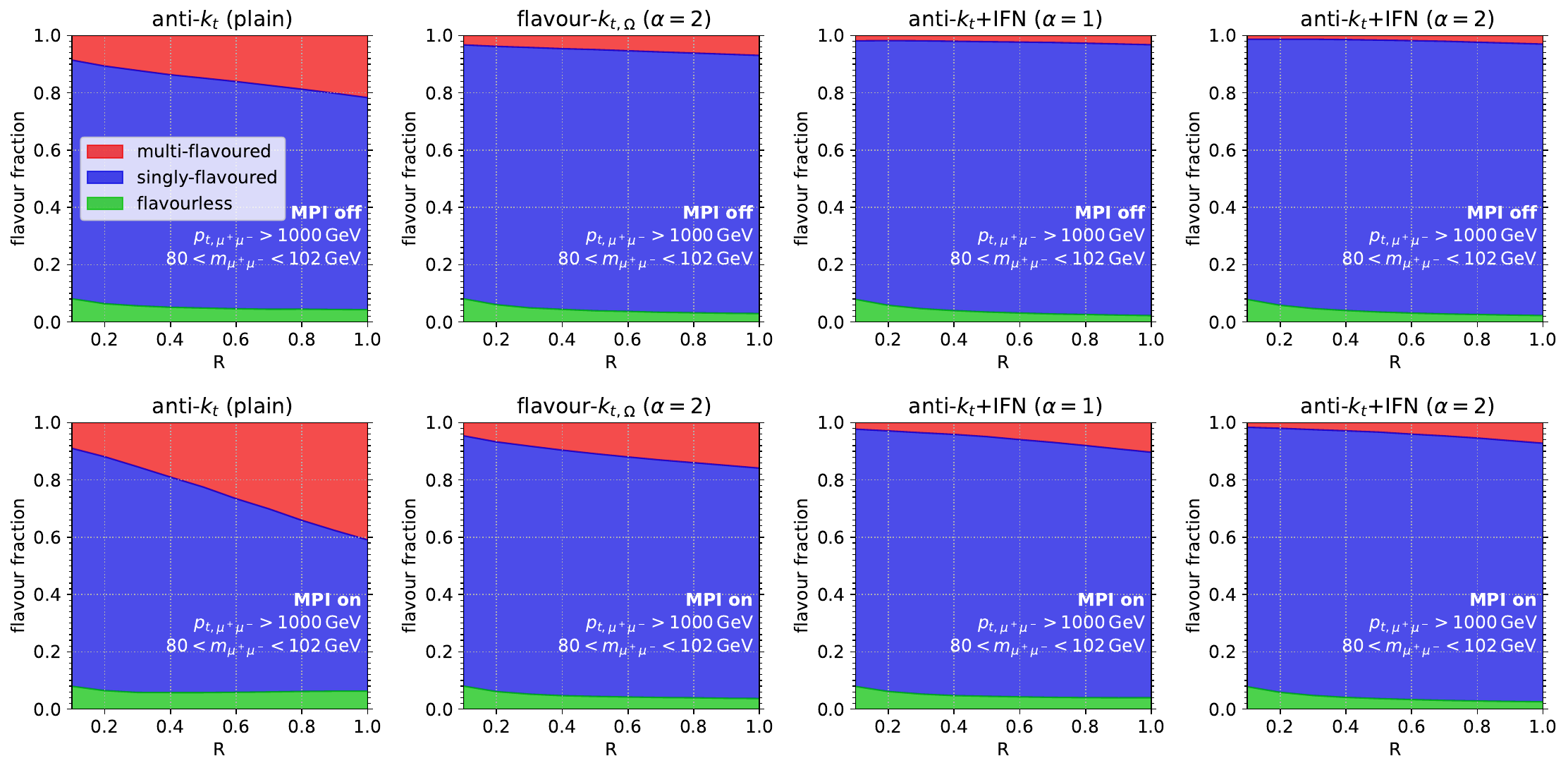}
  \caption{
    Stress-tests of the performance of the plain anti-$k_t$ algorithm (with
    net flavour summation, left column), the flavour-$k_{t,\Omega}$ algorithm
    (middle left column), and the anti-$k_t$ algorithm with flavour 
    neutralisation (with $\alpha=1$, middle right column, and $\alpha=2$, right column).
    The
    stress-tests are performed in $pp \to Z+q$ collisions with $p_{tZ}>1 \TeV$,
    as simulated with Pythia~8.3 at parton level with multi-parton interactions
    disabled (enabled) on the upper row (lower row).
    As a function of the jet radius parameter $R$, the plots show the
    fraction of leading jets that are 
    multi-flavoured, i.e.\ whose flavour is neither that of a gluon
    nor a single quark or anti-quark (red band), singly flavoured
    (blue band) and flavourless (green band).
    The key observation is the large fraction of multi-flavoured jets
    with the standard anti-$k_t$ algorithm, which occur due to
    contamination of the hard jet flavour from low-momentum
    particles.
    With the flavour-$k_{t,\Omega}$ algorithm, we see some reduction, while
    anti-$k_t$ with IFN shows a further reduced rate, especially for
    $\alpha=2$.
  }
  \label{fig:Zj-tests}
\end{figure*}

We use Pythia~8.306 with the Monash13 tune to generate the events, and specifically consider
its $pp \to Z+q$ process.
We cluster the events with a given jet algorithm, and examine the
flavour of the leading-$p_t$ jet.
At leading order, we expect the hard recoiling jet to always carry the
flavour of the underlying quark or antiquark, and the question that we
examine is that of how often the leading jet in the full showered
sample has a flavour other than that of a single quark or anti-quark.

Schematically, it is useful to think of two mechanisms that can cause
the flavour to differ.
One is that the quark can split to $q+g$ with a separation $\Delta R_{qg} >
R$.
If the gluon carries more energy than the quark, then the leading jet
will actually be a gluon jet.
The rate for this to happen is logarithmically enhanced in the
small-$R$ limit~\cite{Dasgupta:2014yra}.
The second mechanism to keep in mind is the contamination of the
flavour of a hard quark jet from a soft $g \to q\bar q$
splitting (i.e.\ the issue of Fig.~\ref{fig:classic-config}, which
flavoured jet algorithms are supposed to mitigate against). 
This can have two effects: if the soft $q\bar q$ pair's flavour
coincides with that of the jet, then it can cancel the jet's flavour;
much more often, a fraction $\sim 1 - 1/(2n_f)$ of the time, it will
lead to a multi-flavour jet. 
To a first approximation, this effect is expected to grow with
increasing jet radius.
We show results both with and without multi-parton interactions (MPI),
and we expect the flavour contamination to be worsened by MPI, insofar
as it adds significant numbers of additional low-$p_t$ $q\bar q$
pairs.

In Fig.~\ref{fig:Zj-tests}, we show the fraction of leading-$p_t$ jets
that are flavourless (green), singly-flavoured (quark or antiquark,
blue) or multi-flavoured (neither flavourless or singly-flavoured,
red), as a function of the jet radius parameter $R$ used in the
clustering.
We perform this comparison with Pythia at parton level, where the underlying
event is turned off (upper row), and with MPI turned on (lower row).
From left to right, the columns show results with the standard
anti-$k_t$ algorithm, flavour-$k_{t,\Omega}$ ($\alpha=2$), and
anti-$k_t$ with our IFN algorithm for two values of $\alpha=\{1,2\}$
(and $\omega=3-\alpha$).
A first point to observe is the large multi-flavoured contribution for
the plain anti-$k_t$ algorithm, about $14\%$ at $R=0.4$ without MPI,
increasing to $19\%$ with MPI.
Increasing $R$ substantially worsens the situation with over $40\%$
multi-flavoured jets for $R=1$ when MPI is on.

Flavour-$k_{t,\Omega}$ improves the situation somewhat, giving a multi-flavoured contribution of $5\%$ ($10\%$)
with MPI off (on) at $R=0.4$.
The anti-$k_t$ algorithm with IFN brings a more
substantial improvement, yielding $2\%$ ($4\%$)
 for $\alpha=1$ and
$1.5\%$ ($3\%$)
 for $\alpha=2$.%
 \footnote{For the CMP$_\Omega$
  algorithm there is freedom in how one extends it to multi-flavoured
  events, and accordingly we defer study of multi-flavoured events
  with that algorithm to future work.}

Examining instead the unflavoured (``gluon'') jet fractions, we find
that all flavour algorithms give a $\sim 4\%$ gluon-jet fraction at
$R=0.4$, relatively unaffected by the presence of MPI. 
This figure is important to keep in mind for quark/gluon
discrimination studies~\cite{Gras:2017jty}: the fact
that a jet was initiated by a quark in Pythia does not mean that the
corresponding jet observed after showering is always a quark jet.
In particular, Fig.~\ref{fig:Zj-tests} implies that if one is
attempting to tag gluon-jets and reject quark-jets, and one is using
Pythia's $Z+q$ and $Z+g$ samples as the sources of quark and gluon
jets, then even a perfect gluon tagger will still show an acceptance
of about $4\%$ on the $Z+q$ sample.

Ultimately, we would argue that the ``truth'' flavour labels should be
derived not from the generation process, but by running a jet flavour
algorithm such as anti-$k_t$+IFN.
Nevertheless the anti-$k_t$+IFN labelling remains subject to some
ambiguities, and the multi-flavoured jet fraction discussed above is
probably a good measure of those ambiguities.
As a future direction, one might wish to investigate whether one can develop jet
flavour algorithms that further reduce the multi-flavoured jet
fraction, while maintaining other good properties.

%======================================================================
\section{Exploration of IFN algorithm for $e^+e^-$ collisions}
\label{sec:ee-generalisations}

\begin{figure*}
  \centering
  \includegraphics[width=0.49\textwidth,page=2]{figures/rts-scaling-2023-05-ssbar.pdf}\hfill
  \includegraphics[width=0.49\textwidth,page=1]{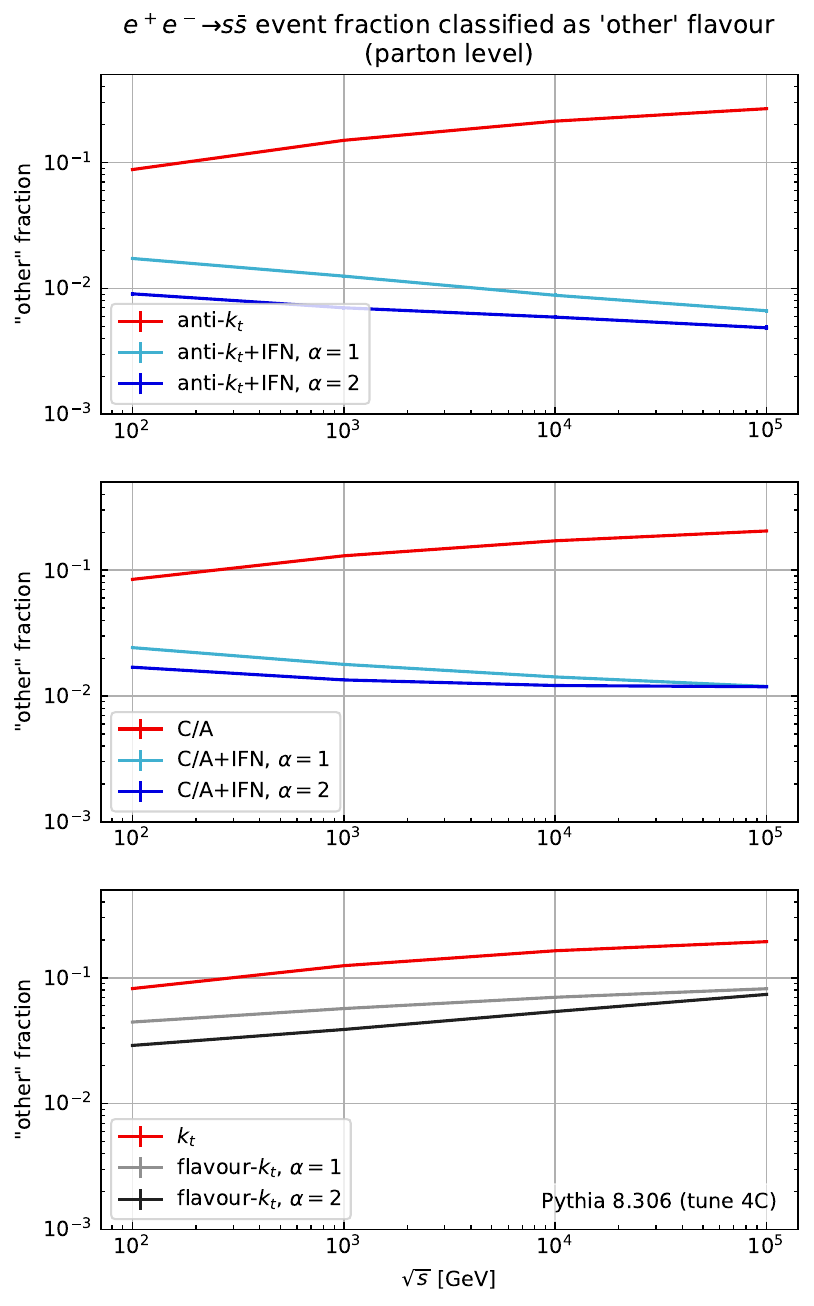}
  \caption{
    The fraction of $e^+e^- \to q\bar q$ events (after
    parton showering and 
    clustering to two jets), in which the flavour of the two jets is classified as being
    $gg$ (left column) or any other combination that is not the original $q\bar q$ (right column).
    The results are shown as a function of $\sqrt{s}$ for
    algorithms in the anti-$k_t$ family (top row), the C/A family (middle
    row) and $k_t$ family (lower row).
    The results have been obtained using Pythia 8.306 at parton level,
    with tune 4C.
  }
  \label{fig:ee-rts-scaling}
\end{figure*}

The IFN algorithm for $e^+e^-$ collisions follows the same set of
rules as the $pp$ version in Sec.~\ref{sec:genkt-FN}, adapting
the $u_{ij}$ distances so that they coincide with the $e^+e^-$
flavour-$k_t$ distances, specifically
\begin{equation}
  \label{eq:uik-ee}
  u_{ik} \equiv
  [\max\left(E_{i}, E_{k}\right)]^{\alpha}
  [\min\left(E_{i}, E_{k}\right)]^{2-\alpha}
  \times
  2(1-\cos\theta_{ik})\,.
\end{equation}
In contrast with the hadron-collider case, there is no need to use a
modified angular distance in the $e^+e^-$ form of the $u_{ij}$.
Additionally in step~1 of the part of the algorithm in
Sec.~\ref{sec:IFN-overview}, we define $i$ to be
the particle with lower energy.

We have not explicitly performed the same full set of IRC safety tests on the $e^+ e^-$ algorithm that
we carried out in the $pp$ case.
The issues identified across various algorithms in the $pp$ case fall
into two classes: those involving initial-state hard-collinear
radiation, which is irrelevant in the $e^+e^-$ case, and those
involving just the interplay between final-state soft large-angle and
hard-collinear branchings.
The analyses of these latter issues in App.~\ref{sec:list-ir-risky}
are expected to be insensitive to the differences between
Eqs.~(\ref{eq:uik-ee}) and (\ref{eq:uik-all}).
For this reason, we do not anticipate IRC safety issues in the
$e^+ e^-$, though a detailed study would ultimately be desirable.

The relatively clean environment of $e^+e^-$ collisions allows for a
further exploration of the performance of IFN-style algorithms.
Specifically, we take a parton shower simulation of
$e^+e^- \to q\bar q$ events, cluster each event into two jets, and examine the
flavours of those two jets.
Typically one expects each jet to have the flavour of the parton in
the Born event.
In analogy with our $pp$ study of Fig.~\ref{fig:Zj-tests}, we
examine the fraction of events where this does not happen, breaking it
into two components, one where each jet is flavourless (``gg'') and
the other where the two jets have neither the original 
$q\bar q$ flavour nor $gg$ flavour (``other'').
We plot these fractions as a function of $\sqrt{s}$.
For a well-behaved flavour algorithm, we expect the rate of $gg$
configurations to decrease with increasing $\sqrt{s}$, as
$\as(\sqrt{s})$, associated with the probability of producing a
$q\bar q g$ configuration where the $q\bar q$ pair ends up
back-to-back with respect to the gluon.
Similarly, the rate of ``other'' configurations should decrease as
$\as^2(\sqrt{s})$, since one expects to have to generate a hard
$q\bar q q'\bar q'$ configuration to obtain an ``other'' flavour.

We carry out the parton shower simulation with Pythia 8.306 (4C
tune) at parton level. 
The jet clustering is performed as follows: for the $k_t$ and C/A
algorithms, we use a large radius $R=2\pi$ and then decluster the
event back to two jets by undoing the last stage of the clustering in
order to obtain the two hard jets.
This is equivalent to asking for two hard jets in the normal exclusive
$k_t$ or Cambridge algorithms.
For the anti-$k_t$ algorithm we use a jet radius of $R=3\pi/4$ and
take the two highest energy jets.
The specific jet radius choice is designed to be large enough that
events with multiple well separated hard partons still only give two
hard jets, but small enough that opposite hemispheres of the event do not
cluster together.

Fig.~\ref{fig:ee-rts-scaling} shows the $gg$ (left column) and
``other'' (right column) rates as a function of $\sqrt{s}$ in the
range $100\GeV$ to $10^5\GeV$.
This broad (and today unrealistic) energy range is intended to help
visualise the scaling behaviour of the rates.
Each row corresponds to one underlying jet algorithm, with different
curves showing results for different flavour approaches.
Let us start by examining the $gg$ rate for the anti-$k_t$ algorithm
and its IFN variants (top-left panel).
Over the energy range being considered, $\as$ decreases by almost a
factor of two.
\logbook{}{
  # alphas values for variable flavour-number scheme
  # so MC would be a bit different
  alphas(91.1876) = 0.117900
  alphas(1e+04) = 0.060433 (3-loops), 0.061168 (2-loops)
  }%
The anti-$k_t$+IFN algorithms show a $gg$ rate that is more or less
consistent with $\as(\sqrt{s})$ scaling.
In contrast the plain anti-$k_t$ algorithm (with net flavour
summation) features a rate that slowly increases.
It is natural to ascribe the growth to the IRC unsafety of the
algorithm, however the differences between the safe (IFN) and unsafe
(plain anti-$k_t$) variants remain relatively modest.

The situation becomes clearer when looking at the ``other''
flavour combinations (top-right panel).
Here the plain anti-$k_t$ algorithm gives a rate that increases from
about $10\%$ at $\sqrt{s}=100\GeV$ to almost $30\%$ at $\sqrt{s} =
10^5\GeV$.
In contrast, the IRC-safe IFN variants give much smaller rates, well
below $2\%$ across the whole energy range, i.e.\ a huge improvement on
the plain anti-$k_t$ algorithm.
With the $\alpha=1$ IFN choice, the rate decreases more or less
consistently with the expectation of $\as^2$ scaling.
The $\alpha=2$ IFN algorithm shows a lower rate, but also a slower scaling.
The situation is broadly similar for the C/A algorithm, with the IFN
rate a little higher.
For the $k_t$ family, we show only the plain $k_t$ algorithm and the
flavour-$k_t$ algorithm, since we have not conclusively validated the
IRC safety of the $k_t$+IFN combination.
Interestingly the flavour-$k_t$ algorithm shows only modest
improvement in the ``other'' rate relative to the plain $k_t$
algorithm, and a scaling that is no better.

It is intriguing that different IRC safe algorithms lead to ``other''
rates that have varying degrees of consistency with the expected $\as^2$
scaling.
While we do not yet have a complete understanding of this phenomenon,
detailed investigations into the events have revealed all-order
mechanisms that operate across multiple scales and that, in some
situations, cause soft flavour to be successively associated with
harder and harder momenta, ultimately transferring soft flavour to the
hard jets.
Further study would require detailed analysis of the interplay between
the main jet algorithm's clustering sequence and the IFN flavour
neutralisation scales.
Still, despite these observations, the IFN algorithms clearly perform
much better than plain flavour unsafe ones, indicating the substantial
benefits for detailed flavour studies in using a suitably chosen
flavour-safe algorithm.

%======================================================================
\section{Conclusions and outlook}
\label{sec:conclusions}

In this article, we introduced an approach to jet clustering that
maintains the kinematics of the original anti-$k_t$ and C/A
algorithms, while also providing IRC-safe jet-flavour identification.
Our IFN algorithm has passed a battery of fixed-order IRC safety
tests, which revealed a number of unexpected and subtle issues in
prior jet-flavour proposals.
While not an absolute guarantee, these tests do provide a
  reasonable degree of confidence in the IRC safety of our
  approach.
On three benchmark jet flavour tasks, IFN exhibits the desired phenomenological behaviour.
These studies suggest that IFN can yield a theoretically sound
meaning to the concept of a flavoured jet in the majority of
heavy-flavour related applications that can be envisaged at the LHC.

There are various experimental considerations that should be noted before deploying IFN in a full analysis.
Our algorithm, like all other attempts at IRC-safe flavour jet
algorithms, requires the complete flavour information in the event for
those flavours under consideration, e.g.\ $b$-flavour.
This is highly challenging in an experimental environment, because of
the difficulties of tagging low-momentum single $B$-hadrons, as well
as quasi-collinear pairs of $B$-hadrons.
The question remains, however, whether recent advances in
machine-learning can help reveal the information that is needed, and
more generally whether experimentally one can unfold detector-level
results to particle-level jet definitions such as ours.
Furthermore, for certain signal processes, the practical impact of
this issue may only be moderate, cf.\ the $\lesssim 3\%$ difference
between the any-flavour and net-flavour anti-$k_t$ results for $t\bar
t$ (Fig.~\ref{fig:ttbar-tests} right).

Theoretically, we stress that the concept of jet flavour remains
subtle also beyond the scope of the discussion in this article.
We focused on the fixed-order behaviour, but there can be non-trivial interplay with the still perturbative but
complex structure induced by all-order showered events.
Beyond a perturbative analysis, there are even more difficult issues of jet flavour in the presence of the high
densities of flavoured particles that result from hadronisation. 
These questions warrant more investigation.
Nevertheless, the IFN algorithm developed here already shows clear and
substantial benefits both with respect to standard unflavoured
algorithms and to prior incarnations of flavoured algorithms.

Code implementing the IFN algorithm is available from
\url{https://github.com/jetflav/IFNPlugin}, in the form of a FastJet
Plugin.

\section*{Acknowledgements}

We are grateful to Andrea Banfi, Simone Marzani and Giulia Zanderighi
for helpful discussions and comments on the manuscript.
We are also grateful to the authors of Refs.~\cite{Czakon:2022wam} and
\cite{Gauld:2022lem} for discussions about their respective
algorithms and more generally on these topics.
GPS and LS would also like to thank their PanScales collaborators for
joint work
on the high-precision adaptations of the fjcore code used in the IRC
safety tests.
GPS would also like to thank Matteo Cacciari and Gregory Soyez for
joint work on updates to FastJet to facilitate the inclusion of
flavour in plugins.

This work has been funded 
by a Royal Society Research Professorship
(RP$\backslash$R1$\backslash$180112) (GPS+LS),
by the European Research Council (ERC) under the European Union’s
Horizon 2020 research and innovation programme (grant agreement No.\
788223, PanScales, GPS+LS+MH; and grant agreement 804394, HipQCD,
FC) and
by the Science and Technology Facilities Council (STFC) under grants
ST/T000864/1 and by Somerville College (LS).
JT is supported by the U.S. DOE Office of High Energy Physics under
Grant Contract No.~DE-SC0012567.
RG is supported by the STFC, a Wolfson Harrison UK Research Council
Physics Scholarship and by a Clarendon Scholarship, and in the early
stages of this work benefited from support from Merton College.
Part of this work benefited from the support and hospitality of the
Munich Institute for Astro-, Particle and BioPhysics (MIAPbP) which is
funded by the Deutsche Forschungsgemeinschaft (DFG, German Research
Foundation) under Germany's Excellence Strategy -- EXC-2094 --
390783311.

% Please send a short note to workshop@munich-iapp.de with title and
% author information if a MIAPbP paper is published or submitted, or
% simply tell us at http://www.munich-iapbp.de/publication

\appendix

%======================================================================
\section{Asymmetric double-soft branching}
\label{sec:double-soft-structure}

As discussed in Fig.~\ref{fig:classic-config}, the classic IRC safety issue involves configurations with a double-soft $g \to q\bar q$ pair.
One surprising feature that we will encounter in
Apps.~\ref{sec:IFN-omega}, \ref{sec:IRCapp-flavkt}
and \ref{sec:CMP-IHC-DS}
is that the structure of higher-order IRC divergences in
some algorithms is sensitive to configurations where one of the quarks is significantly softer than the other.
Consequently, it is important to understand the asymptotic behaviour of
double soft $q\bar q$ production in such limits.

We know that collinear $g \to q\bar q$ splitting comes
with a $P_{qg}(z)dz = T_R (z^2 + (1-z)^2)dz$ structure, which is
finite and non-zero when $z \to 0$.
One question one might ask, though, is whether the splitting
probability remains finite and non-zero for $z \to 0$ when the pair is
not collinear but instead separated by an angle that is commensurate
with the emission angle of the parent gluon.
In this appendix, we find that the splitting probability is indeed finite in that limit, and we derive a simple approximate expression for its behaviour.

Consider a process with $n$ hard
massless QCD partons with momenta $\{p_i\}$ and study the emission of
two additional soft quarks with momenta $q_{1,2}$. The double-soft
emission probability can be written as a sum over dipole contributions
as
\begin{equation}
  d\mathcal P_{\rm d.s.} = -\sum_{i\ne j=1}^{n} {\bf T}_i \cdot {\bf T}_j\,
  d\mathcal P^{(i,j)}_{\rm d.s.},
  \label{eq:ds_dipoles}
\end{equation}
where ${\bf T}_i$ are the standard colour operators (see
e.g. Ref.~\cite{Catani:1996vz} for details) and $d\mathcal P^{(i,j)}_{\rm}$ only
depends on the momenta of the soft quarks and of the hard partons $i$
and $j$. To write an explicit representation for $d\mathcal
P^{(i,j)}_{\rm}$, it is convenient to work in the dipole
centre-of-mass frame. Specifically, we write
\begin{align}
  & p_i = E(1;0,0,1),
  \nonumber\\
  & p_j = E(1;0,0,-1),
  \nonumber\\
  & q_1 = p_{t,1} (\cosh y_1;1,0,\sinh y_1),
  \nonumber\\
  & q_2 = p_{t,2} (\cosh y_2;\cos\Delta\phi,\sin\Delta\phi,\sinh y_2).
\end{align}
We stress that $p_{t,{1,2}}$, $y_{1,2}$ and $\Delta\phi$ are dipole-specific
and not global variables.
In terms of these variables, the double-soft emission probability reads
\begin{align}
  \label{eq:dPSds}
  &
  d\mathcal P^{(i,j)}_{\rm d.s.} = 
  \left(\frac{\as}{2\pi}\right)^2
  4  T_R\,
  d p_{t,1} d p_{t,2} d y_1 d\Delta y \frac{d\Delta\phi}{2\pi}\times
  \\
  &
  \frac{2 p_{t,1}p_{t,2} - (p^2_{t,1}+p^2_{t,2})\cos\Delta\phi +
    |\vec p_{t,1}-\vec p_{t,2}|^2 \cosh\Delta y}
       {(p_{t,1}^2 + p_{t,2}^2 + 2 p_{t,1}  p_{t,2} \cosh\Delta y)^2
         (\cosh\Delta y-\cos\Delta\phi)^2},
       \nonumber
\end{align}
where $\Delta y = y_2-y_1$ and $T_R= 1/2$. In the limit when the soft quark
pair is also collinear to parton $i$, Eq.~(\ref{eq:ds_dipoles}) simplifies
to
\begin{equation}
  d\mathcal P_{\rm d.s.} = -\sum_{i\ne j=1}^{n} {\bf T}_i \cdot {\bf T}_j\,
  d\mathcal P^{(i,j)}_{\rm d.s.}
  \to
  C_i \,d\mathcal P^{(i,j)}_{\rm d.s.}\,,
\end{equation}
with $d\mathcal P^{(i,j)}_{\rm d.s.}$ still given by Eq.~(\ref{eq:dPSds}) and
where $C_i = C_A$ if parton $i$ is a gluon and $C_i=C_F$ if it is a quark.

We can now study the asymmetric $p_{t,2}\ll p_{t,1}$ configuration. We write
$p_{t,2} = z \, p_{t,1}$. In the small-$z$ region, Eq.~(\ref{eq:dPSds})
becomes
\begin{equation}
  d\mathcal P^{(i,j)}_{\rm d.s.} \sim
  \left(\frac{\as}{2\pi}\right)^2
  \frac{4 T_R dp_{t,1} dz\, d y_1 d\Delta y\, d\Delta\phi}
  {2\pi\, p_{t,1}(\cosh\Delta y-\cos\Delta\phi)}.
\end{equation}
We see that, as for the $P_{qg}$ splitting function, this probability
is finite and non-zero for $z\to 0$. For our analysis, we also find it
useful to consider Eq.~(\ref{eq:dPSds}) in the limit of large rapidity
separation between the two soft quarks, $\Delta y \gg 1$. We obtain
\begin{align}
  \label{eq:dPSds_largey}
  d\mathcal P^{(i,j)}_{\rm d.s.} \sim
  &
  \left(\frac{\as}{2\pi}\right)^2\, 8 T_R\,
  \frac{dp_{t,1}}{p_{t,1}} \frac{dp_{t,2}}{p_{t,2}}
  \, d y_1 d\Delta y\, \frac{d\Delta\phi}{2\pi}
  \times
  \\
  \nonumber
  &
  \frac{p_{t,1} p_{t,2}\, e^{-\Delta y}}{(p_{t,1}+e^{\Delta y} p_{t,2})^2}.
\end{align}
In the asymptotic regime, the second line of Eq.~(\ref{eq:dPSds_largey}) is
well approximated by the following expression
\begin{equation}
  \frac{p_{t,1} p_{t,2}\, e^{-\Delta y}}{(p_{t,1}+e^{\Delta y} p_{t,2})^2}
  \approx \min\left[\frac{p_{t,1}}{p_{t,2}}e^{-3\Delta y},
    \frac{p_{t,2}}{p_{t,1}}e^{-\Delta y}\right],
\end{equation}
which interpolates between the $1\ll \ln(p_{t,1}/p_{t,2})\ll \Delta y $  and
$1\ll \Delta y \ll \ln(p_{t,1}/p_{t,2})$ limits.

\begin{figure}
  \centering
  \includegraphics[width=\columnwidth]{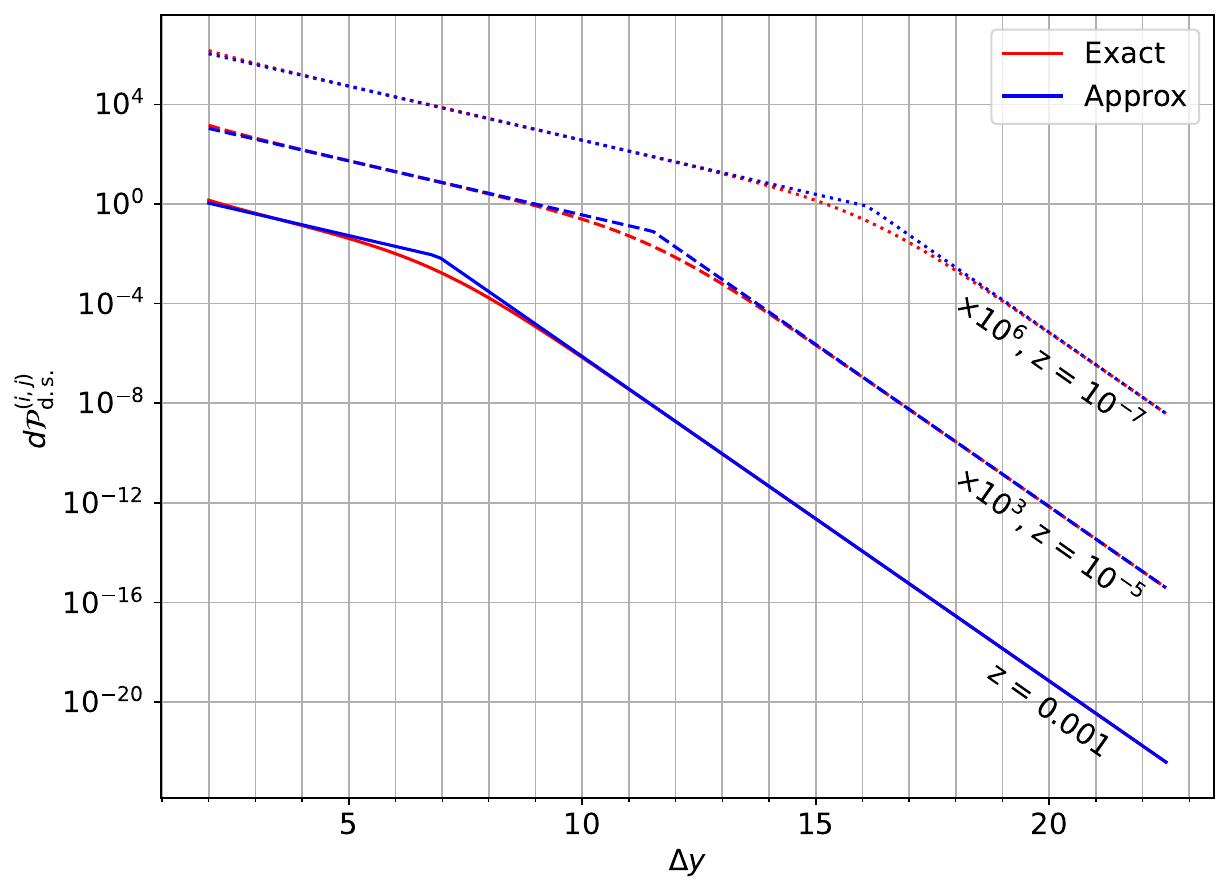}
  \caption{
  The double-soft matrix element $d\mathcal P^{(i,j)}_{\rm d.s.}$ in (red) its
  exact form from Eq.~(\ref{eq:dPSds}), and in (blue) the approximate
  form from Eq.~(\ref{eq:dPSds_int}) that we use in our numerical
  IRC-safety tests, as a
  function of the rapidity separation $\Delta y$ of the two soft partons in
  the dipole centre-of-mass frame.
  The matrix element is plotted for $p_{t,1}=1\,\GeV$, and $\Delta \phi = 0$, for several values of $z$
  (shown as full, dashed and dotted lines).
  The two regimes with $\sim e^{-\Delta y}$ and $\sim e^{-3\Delta y}$
  scaling are clearly visible.  }
  \label{fig:DS-approx}
\end{figure}

In practice, we find that
this interpolation works well across the whole phase space.
As shown in Fig.~\ref{fig:DS-approx},
we find that 
\begin{align}
  d\mathcal P^{(i,j)}_{\rm d.s., approx} \equiv
  &\left(\frac{\as}{2\pi}\right)^2\, 8 T_R\,
  \frac{dp_{t,1}}{p_{t,1}} \frac{dp_{t,2}}{p_{t,2}}
  \, d y_1 d\Delta y\, \frac{d\Delta\phi}{2\pi}
  \times
  \nonumber\\
  &
  \min\left[\frac{p_{t,1}}{p_{t,2}}e^{-3\Delta y},
    \frac{p_{t,2}}{p_{t,1}}e^{-\Delta y}\right]
  \label{eq:dPSds_int}
\end{align}
provides a good approximation of the exact $d \mathcal P^{(i,j)}_{\rm
  d.s.}$ result down to values of $\ln (p_{t,1}/p_{t,2})$ and $\Delta
y$ of order $1$.
Furthermore, it is free of the collinear divergence when the $q$ and
$\bar q$ go close in angle, a collinear divergence that we
deliberately wish to leave out, because of our approach of allowing at
most one divergence (or logarithm) per power of $\as$. 
In our studies, we therefore
use the convenient interpolation Eq.~(\ref{eq:dPSds_int}) rather than the
exact result Eq.~(\ref{eq:dPSds}).

\section{Numerical tests of IFN}
\label{sec:IRCapp-IFN}

In the process of developing the IFN algorithm, we tested a number of possible variants.
In this appendix, we provide numerical support for the analytic arguments made in Sec.~\ref{sec:genkt-FN} to justify our design choices.

%......................................................................
\subsection{Relation between $\alpha$ and $\omega$}
\label{sec:IFN-omega}

\begin{figure}
  \centering
  \includegraphics[width=\columnwidth]{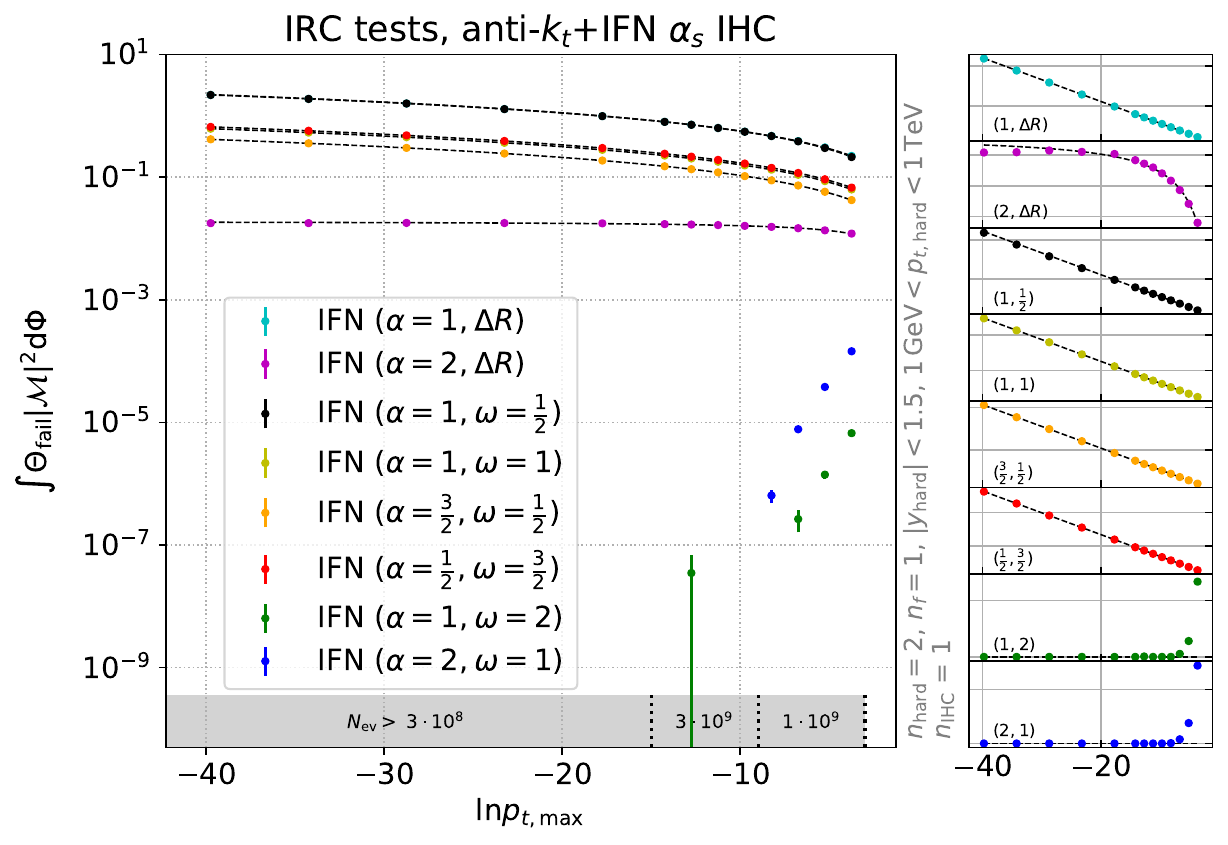}
  \caption{
  IRC safety tests of anti-$k_t$+IFN for variants with different angular scaling factors. 
  The tested configuration from Fig.~\ref{fig:config-as2-IHC-FmaybeC} features two hard partons clustered
  together by the anti-$k_t$ algorithm and one initial-state hard-collinear
  splitting.
  With a $\Delta R_{ij}^2$ angular factor, the IFN algorithms diverge for all choices of $\alpha$.
  Switching to the $\Omega_{ij}^2$ angular distance, the cases where $\alpha+\omega \leq 2$ also diverge,
  whereas for $\alpha+\omega > 2$, they converge to zero as
  a power law, as expected from Eq.~(\ref{eq:omega-constraint}).
  The right-hand side-panels show the results on a linear scale,
  to help visualise the scaling for the IRC-unsafe variants.
  Note that here and in some of the following figures, certain curves differ
  in the number of generated events.
  Because the multiple curves feature
  different scaling behaviours, they require a varying number of events
  to make a conclusive statement about the form of the divergence (or
  the absence thereof).
  The smallest number of events generated among all runs performed for a
  given figure is shown in the three shaded regions at the bottom of the plot.
  }
  \label{fig:IFN-omegas}
\end{figure}

According to the arguments in Sec.~\ref{sec:neutralisation-distance-choice}, we need to take particular
care in choosing the values of the parameters $\alpha$ and $\omega$ in the
IFN algorithms.
A potentially dangerous configuration is that presented in
Fig.~\ref{fig:config-as2-IHC-FmaybeC}.
In that diagram, two partons (one flavoured, one flavourless) at central
rapidity are clustered together by the anti-$k_t$ algorithm. In our IFN
algorithms, the IRC safety issue arises from an initial-state hard-collinear
splitting, which can act as a possible neutralisation partner for the
flavoured ``hard'' particle.
As argued in Eq.~(\ref{eq:omega-constraint}), the condition
$\alpha+\omega > 2$ ensures that such a neutralisation does not happen.

To test this argument numerically, we integrate uniformly
over the momentum of each of a central hard quark and hard gluon (each
in the range $1\GeV$ to $1\TeV$) and sample an IHC emission as
described in Sec.~\ref{sec:IRC-methodology}.
The results are presented in Fig.~\ref{fig:IFN-omegas}, for various
values of the parameters $\alpha$ and $\omega$.
As expected, in cases where $\alpha+\omega < 2$, as well as for IFN
variants that use a $\Delta R_{ij}^2$ type angular distance instead
of our $\Omega_{ij}^2$, the failure rate typically diverges
for $\ptmax \to 0$,
and conversely falls off as a power law when
$\alpha + \omega > 2$ (green and blue curves).
We observe numerically that the border cases, $\alpha + \omega = 2$,
are all unsafe.

Let us see analytically why
$\alpha+\omega=2$ is problematic for the specific case of $\alpha=2$
(and $\omega = 0$).
We note that in the limit where $\omega \to 0$, the angular
factor $\Omega_{ik}^2$ in Eq.~(\ref{eq:omegaik-gen}) differs from
$\Delta R_{ik}^2$ at most by a factor of $\order{1}$, which we can
typically neglect in the discussions below.
We take the configuration shown in Fig.~\ref{fig:config-as2-IHC-FmaybeC}
with $p_{t2} = z_2p_{t3}$ with $z_2 \ll 1$, $p_{t3} = 1$.
There are two competing distances in the neutralisation step,
\begin{subequations}
  \begin{align}
    u_{12} &= z_2^2 \Delta y_{12}^2 \simeq z_2^2 \ln^2 1/p_{t1}\,,
    \\
    u_{23} &= \Delta R_{23}^2 \sim 1\,.
  \end{align}
\end{subequations}
The IFN algorithm will neutralise the flavours of $1$ and $2$ when
$z_2 \ln 1/p_{t1} < \Delta R_{23}$.
If we integrate over the momentum of $2$ and assume a $dz_2$
distribution (see e.g.\ \cite{vanBeekveld:2019prq}) for finite
$\Delta R_{23}$ and take $z_2 \to 0$, then the resulting integral is
given by $\int d\ln p_{t1} \int_0^{1/\ln p_{t1}} dz_2$, which
diverges.
The analytic argument shown here does not apply to generic values of
$\alpha$ and $\omega$, but as mentioned above, we find numerically that all
cases that we have tested with $\alpha+\omega=2$ diverge.
%

% ......................................................................
\subsection{Recursive v.\ non-recursive}
\label{sec:IFN-non-recursive}

\begin{figure}
  \centering
  \includegraphics[width=\columnwidth]{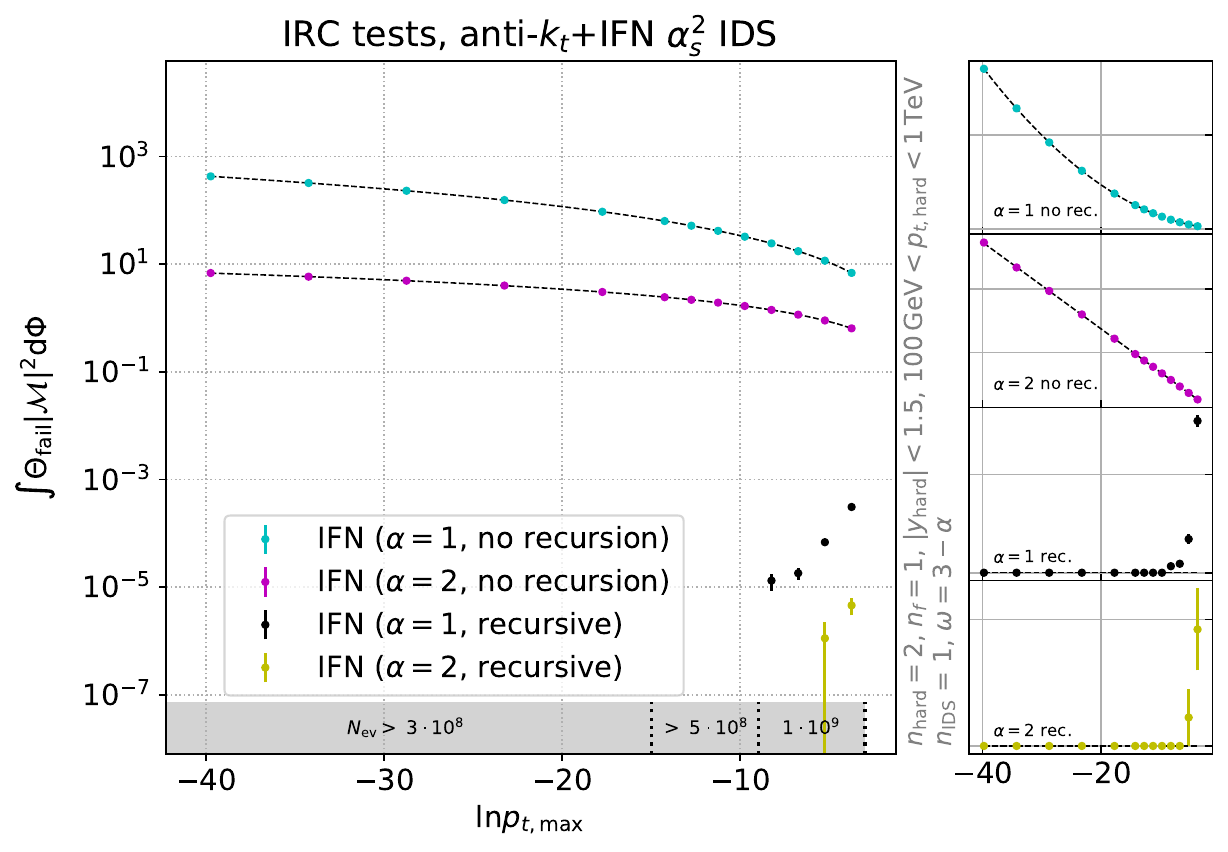}
  \caption{IRC safety test of anti-$k_t$+IFN for variants with and
    without the recursion step.
    The tested events consist of two hard partons supplemented with one
    initial-state double soft pair, as in Fig.~\ref{fig:need-for-recursion}.
  }
  \label{fig:IFN-no-recursion}
\end{figure}

In Sec.~\ref{sec:need-recursion}, we presented an analytic
argument to explain why the IFN algorithms need a recursion step.
Fig.~\ref{fig:IFN-no-recursion} shows the failure rate events with two
hard partons and one IDS pair, which includes configurations such as
that of Fig.~\ref{fig:need-for-recursion}.
It clearly shows that without recursion, the algorithm shows a growing
failure rate for $\ptmax \to 0$, while the failure vanishes for
$\ptmax \to 0$ with the recursive step turned on.
The side-figures help illustrate that the failure rate goes as
$\ln^2\ptmax$ for $\alpha=1$ and as $\lnptmax$ for $\alpha=2$.
The stronger power for $\alpha=1$ arises because failures can happen
even when the IDS pair is collinear to the beams.

% ======================================================================
\section{IRC-unsafe configurations}
\label{sec:list-ir-risky}

In this appendix, we analyse the specific IRC-unsafe configurations identified in Sec.~\ref{sec:IRC-results} for the flavour-$k_t$, CMP and GHS algorithms.
For each of the configurations that we have identified, we present both
analytic and numerical results to demonstrate why they are problematic.
Throughout this section we define $p_{ti} \lesssim p_{tj}$ to mean that
$p_{ti} < p_{tj}$ but that they are of similar orders of magnitude.

% ----------------------------------------------------------------------
\subsection{IHC$\times$IDS subtlety at $\as^3$ for flavour-$k_t$}
\label{sec:IRCapp-flavkt}

\begin{figure}
  \centering
  \includegraphics[width=0.9\columnwidth]{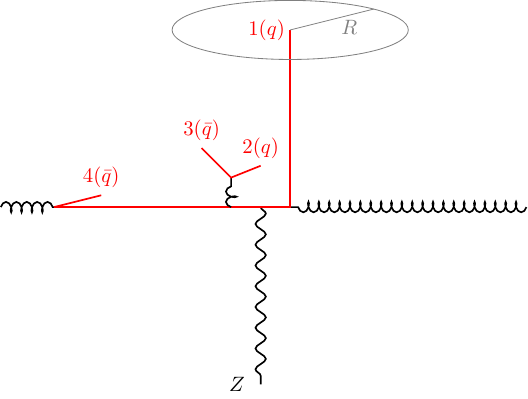}
  \caption{Example configuration to illustrate issues that arise
    across multiple algorithms when using a standard
    $\Delta R$ type angular measure in inter-particle distances.
  }
  \label{fig:IRsafety-as3-IHC-DS-diagram}
\end{figure}

The flavour-$k_t$ (and GHS) algorithms encounter a problematic configuration at order $\as^3$, shown in Fig.~\ref{fig:IRsafety-as3-IHC-DS-diagram}, associated with the choice of angular measure.
There is a hard parton ($1$, with flavour $q$) that produces the
only hard jet in the event, together with a soft gluon $g$ that
splits to a soft large-angle $q \bar q$ pair ($2$ and $3$), and
additionally an initial-state collinear gluon splitting that produces
an energetic small-angle quark of flavour $\bar q$ ($4$).
For the sake of the discussion, we assume that the transverse momentum of $2$ is
smaller than that of $3$, by a factor $z_{23}$,
\begin{equation}
  \label{eq:flavkt-pt2-pt3}
  p_{t2} = z_{23} p_{t3}.
\end{equation}
We take the rapidity and azimuth of $1$ to be zero,
$y_1 = \phi_1 = 0$.
Additionally, we take $E_4 = z_{41} E_1$, which implies
\begin{equation}
  y_4 = \ln \frac{2z_{41} p_{t1}}{p_{t4}}.
\end{equation}

For concreteness, we work with the $\alpha=2$, $R=1$ variant of the
flavour-$k_t$ algorithm.
The inter-particle distances for the flavour-$k_t$ algorithm were
given in Eq.~(\ref{eq:flavkt-dij}), and additionally we will need to
take into account a beam distance with the right-moving beam.
When $i$ is flavoured (as it will always be in the example here),
\begin{equation}
  \label{eq:diB-flavkt-partial}
  d_{iB}^\text{flav-$k_t$} =
  [\max(p_{ti},p_{tB}(y_i))]^\alpha
  [\min(p_{ti},p_{tB}(y_i))]^{2-\alpha}\,,
\end{equation}
with
\begin{equation}
  \label{eq:ktB-flavkt-partial}
  p_{t B}(y) = \sum_i p_{ti}\left[
    \Theta(y - y_i)e^{y_i - y}  + \Theta(y_i - y)
  \right]\,,
\end{equation}
so that
\begin{subequations}
  \begin{align}
    \label{eq:ptB-flavkt}
    p_{t B}(0) &\simeq p_{t1}\,,
    \\
    p_{t B}(y_4) &\simeq p_{t4} + p_{t1} e^{-y_4} \simeq p_{t4}\left(1 +\frac{1}{2z_{41}}\right)\,,
  \end{align}
\end{subequations}
where we have used the fact that $p_{t4} \ll p_{t1}$.
In the absence of the soft quark pair, there are three distances,
\begin{subequations}
  \label{eq:dij-flavkt-hard-coll}
  \begin{align}
    d_{14} &= p_{t1}^2 \Delta R_{14}^2 \simeq p_{t1}^2 y_4^2 \,,
    \\
    d_{1B} &\simeq p_{t1}^2 \,,
    \\
    d_{4B} &\simeq p_{t4}^2\left(1 +\frac{1}{2z_{41}}\right)^2\,.
  \end{align}
\end{subequations}
The smallest is $d_{4B}$, since $p_{t4}/z_{41}\ll p_{t1}$ and so
initial-state collinear particle $4$ clusters first, leaving a
flavoured jet consisting of particle $1$.

Now we examine the additional distances that arise when the soft
$q\bar q$ (23) pair is present.
The beam distances $d_{2B}$ and $d_{3B}$ are both similar to
$d_{1B} = p_{t1}^2$, since they are at central rapidities where
$p_{tB} \sim p_{t1}$.
The distances that will matter for the clustering are 
\begin{subequations}
  \label{eq:dij-flavkt-soft-pair}
  \begin{align}
    d_{23} &= p_{t3}^2 \Delta R_{23}^2 \sim \frac{p_{t2}^2}{z_{23}^2} ,
    \\
    d_{24} &= \max(p_{t4}^2,p_{t2}^2) \Delta R_{24}^2 \sim
             \max(p_{t4}^2,p_{t2}^2) \ln^2 \frac{z_{41}^2p_{t1}^2}{p_{t4}^2}\,,
             \label{eq:d24-flavkt}
  \end{align}
\end{subequations}
where $\sim$ implies that we leave out factors of $\order{1}$, e.g.\
from $\Delta R_{23}^2 \sim 1$.
We neglect $d_{34}$, since in the moderately small $z_{23}$ limit where we
will be working (cf.\ Eq.~(\ref{eq:flavkt-pt2-pt3})),
$d_{34} > d_{24}$.

If $d_{23}$ is the smallest of the distances across
Eqs.~(\ref{eq:dij-flavkt-hard-coll}) and
(\ref{eq:dij-flavkt-soft-pair}), particles $2$ and $3$ annihilate,
then $4$ clusters with the beam, and the hard jet has flavour $q$.
If $d_{4B}$ is the smallest, $4$ clusters with the beam, then $2$ and
$3$ annihilate and the hard jet has flavour $q$.
The problematic situation is when $d_{24}$ is the smallest of the
distances, causing $2$ and $4$ to annihilate.
This leaves $3$, which can cluster with $1$, resulting in a
flavourless hard jet.

\begin{figure}
  \centering
  \includegraphics[width=\columnwidth]{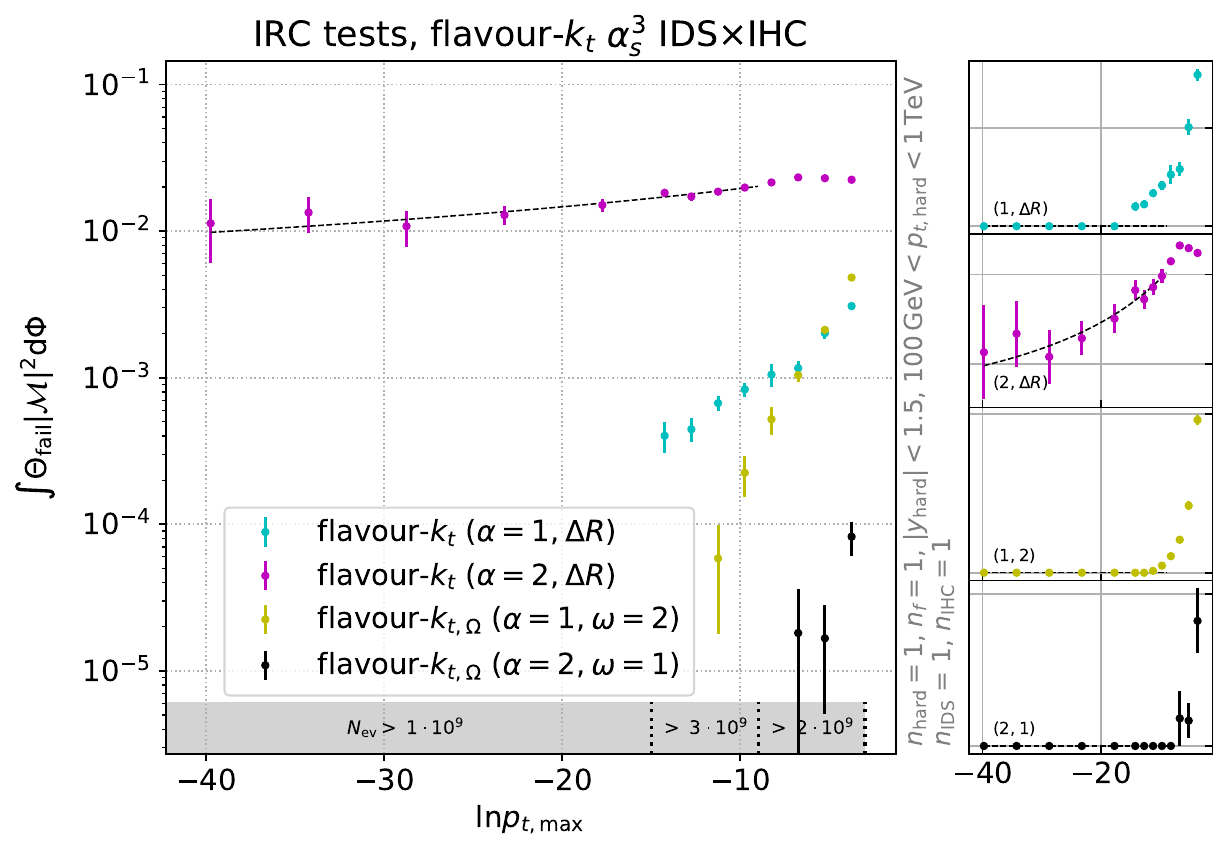}
  \caption{Failure rate of the flavour-$k_t$ algorithm 
    for the configuration of 
    Fig.~\ref{fig:IRsafety-as3-IHC-DS-diagram}, in particular illustrating 
    results for $\alpha=2$ (magenta points) that are qualitatively
    consistent with the expected $1/\lnptmax$ behaviour 
    (dashed line).
  Also shown are results for $\alpha=1$, as well as the results for
  flavour-$k_{t,\Omega}$, i.e.\ the adaptation with an $\Omega_{ij}$ angular
    distance, illustrating the much faster drop of the failure rate.
   }
  \label{fig:flavkt-IHC-DS-rate}
\end{figure}

To understand the likelihood of this occurring,
we first introduce the shorthand
\begin{equation}
  \label{eq:ell-shorthands}
  \ell_{ij} = \ln \frac{p_{ti}}{p_{tj}}\,.
\end{equation}
Let us first consider $p_{t4} < p_{t2}$ ($\ell_{24}>0$).
The $d_{24}$ will be the smallest one when
\begin{subequations}
  \label{eq:flavkt-l24gt0}
  \begin{align}
    d_{24} < d_{23} &\to \ell_{14} + \order{\ln z_{41}} < \frac1{z_{23}},
    \\
    d_{24} < d_{4B} &\to \ell_{14} + \order{\ln z_{41}} < \frac{e^{-\ell_{24}}}{z_{41}}.
  \end{align}
\end{subequations}
Ignoring azimuthal integrals (and the rapidity of $2$ and $3$), we now
have to integrate over four phase-space variables, which we take to be
$\ell_{14}$, $\ell_{24}$, $z_{23}$ and $z_{41}$.
We have found that we can ignore the $\order{\ln z_{41}}$ terms in
Eq.~(\ref{eq:flavkt-l24gt0}) and rewrite the limits as 
\begin{equation}
  z_{23} < \frac{1}{\ell_{14}}\,,
  \quad
  z_{41} < \frac{e^{-\ell_{24}}}{\ell_{14}}\,,
  \quad
  [\ell_{24} > 0] \,.
\end{equation}
Since both $z$ fractions will be small, we will perform the
integrations over $z_{23}$ and $z_{41}$ using the constant small-$z$ limit of
$P_{g\to q\bar q}(z)dz$.
The overall rate of $24$ clustering (with $l_{24}>0$) is then given by
\begin{multline}
  \label{eq:N24-flavkt-a}
  N_{24,\text{flav-}k_t}^{(\ell_{24}>0)}
  \sim \as^3
  \int_0^\infty \!\! d\ell_{14}
  \int_0^{\ell_{14}} \!\! d\ell_{24}
  \int_0^{\frac{1}{\ell_{14}}} \!\! dz_{23}
  \int_0^{\frac{e^{-\ell_{24}}}{\ell_{14}}} \!\! dz_{41}
  \\
  \sim \as^3 \int^\infty \frac{d\ell_{14}}{\ell_{14}^2}\,,
\end{multline}
where in setting the lower limits of the $\ell_{14}$ and $\ell_{24}$
integrals to zero, we are ignoring any constraints from the interplay, e.g.,
with the $z_{23}$ integral.
A similar analysis can be carried out for $\ell_{24} < 0$ (or
equivalently $\ell_{42}>0$), giving 
\begin{equation}
  z_{41} < \frac{1}{\ell_{14}}\,,
  \quad
  z_{23} < \frac{e^{-\ell_{42}}}{\ell_{14}}\,,
  \quad
   [\ell_{42} > 0]\,,
\end{equation}
and yielding
\begin{equation}
  \label{eq:N24-flavkt-b}
  N_{24,\text{flav-}k_t}^{(\ell_{42}>0)}
  \sim \as^3 \int^\infty \frac{d\ell_{14}}{\ell_{14}^2}\,.
\end{equation}
This and Eq.~(\ref{eq:N24-flavkt-a}) both converge in the infrared,
i.e.\ for $\ell_{14} \to \infty$, however this convergence is
extremely slow.
In particular, if one places an upper limit $p_{t4} > \epsilon$, the
result converges as $(\ln 1/\epsilon)^{-1}$, which is consistent also
with what we find in our numerical tests, cf.\ the magenta
($\alpha=2$) points in Fig.~\ref{fig:flavkt-IHC-DS-rate}.
While this is strictly IRC safe at this order, one should worry that
at the next order there may be logarithmic enhancements proportional
to $\ell_{14}$ (for example from running-coupling effects), which
would be sufficient to make the integral diverge.
Accordingly, it would seem wise for future uses of the flavour-$k_t$
algorithm to
adopt the same kind of $\Delta R_{ij}^2 \to \Omega_{ij}^2$ replacement
as used in our IFN algorithm, and similarly for any other algorithms
that make use of similarly defined distances, e.g.\ the GHS
flavour-dressing algorithm.

A final comment concerns the $\alpha=1$ case.
The analysis is somewhat more involved than for $\alpha=2$, and it is
also clear from Fig.~\ref{fig:flavkt-IHC-DS-rate} that the issue is
reduced with $\alpha=1$.
Our investigations are consistent with a $1/\ln^p \epsilon$ scaling,
with a larger value of $p$ than for the $\alpha=2$  case.
One might wish to investigate this point further, however it would
anyway seem wise to use the $\Delta R_{ij}^2 \to \Omega_{ij}^2$
replacement also for $\alpha=1$.

%......................................................................
\subsection{IHC$^2$ issue at $\as^2$ for CMP}
\label{sec:CMP-trouble-as2}

\begin{figure}[t]
  \centering
  \includegraphics[width=.6\columnwidth]{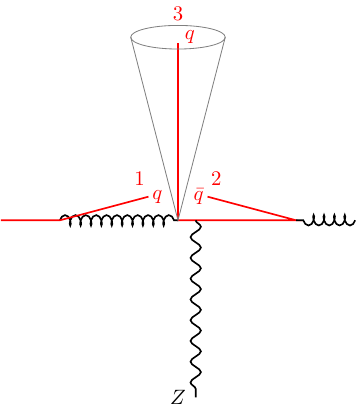}
  \caption{
  \label{fig:config-CMP-as2}
  Example $\mathcal{O}(\as^2)$ configuration that yields an issue for the CMP algorithm.
  There are two oppositely-flavoured initial-state hard-collinear splittings ($b$ and $\bar b$, labelled 1 and 2), and a hard
  particle $3$ at central rapidity.
  }
\end{figure}

An issue arises in the CMP algorithm at order $\as^2$ for a configuration
like the one shown in Fig.~\ref{fig:config-CMP-as2}.
We consider two initial-state hard-collinear emissions, a $q$ and a
$\bar q$ (labelled 1 and 2), from the forward and backward beams
respectively, and additionally one hard large-angle particle (numbered 3 in the
figure).
The initial-state hard-collinear emissions have a very small transverse momentum
($p_{t1}, p_{t2} \ll p_{t3}$) but large energies ($E_1, E_2 \sim E_3$).
Let us assume $p_{t3} = E_3 = 1$, and particle $3$ is simply aligned
along the $x$-axis. Then we have that $y_1 \sim -\ln p_{t1}$
and $y_2 \sim \ln p_{t2}$ in the hard-collinear limit.
For simplicity, we will work with $R=1$.

The CMP algorithm will strongly favour clustering $1$ and $2$ together first.
The global scale ($p_{t,\text{global-max}}$ in
Eq.~(\ref{eq:flav-antikt-Sij})) is set by the $p_t$ of the hardest pseudojet currently
available, $p_{t3}$, so the value of $\kappa_{12}$ is small,
\begin{equation}
\kappa_{12} = \frac{1}{2a} \frac{p_{t1}^2 + p_{t2}^2}{p_{t3}^2} \ll 1\,,
\end{equation}
and the distance between the oppositely-flavoured
particles $1$ and $2$ is thus given by
\begin{subequations}
  \begin{align}
    d_{12} &= \frac{1}{\max (p_{t1}^2, p_{t2}^2 )} \Delta R_{12}^2
             \left(1 - \cos \left( \frac{\pi}{2} \kappa_{12} \right) \right)
             \label{eq:CMP-as2issue-d12-a}\\
           &\simeq \frac{1}{\max (p_{t1}^2, p_{t2}^2 )} \Delta R_{12}^2
             \frac{1}{2} \left( \frac{\pi}{2}\kappa_{12} \right)^2\\
           &\simeq \frac{\pi^2}{32 a^2} \Delta R_{12}^2 \frac{\max (p_{t1}^2, p_{t2}^2)}{p_{t3}^4},\\
           &\simeq \frac{\pi^2}{32 a^2} \left(\ln \frac{p_{t3}}{p_{t1}} + \ln  \frac{p_{t3}}{p_{t2}} \right)^2
             \frac{\max (p_{t1}^2, p_{t2}^2)}{p_{t3}^4},
             \label{eq:CMP-as2issue-d12-d}
  \end{align}
\end{subequations}
where $\Delta R_{12}$ is dominated by the large rapidity difference.
The other distances are
\begin{subequations}
  \begin{align}
    d_{iB} &= \frac{1}{p_{ti}^2}\,,     \qquad i = \{1,2,3\}\,,
    \\
    d_{i3} &\sim \frac{y_i^2}{p_{t3}^2}\,, \qquad i = \{1,2\}\,.
  \end{align}
\end{subequations}
When $p_{t1}, p_{t2} \ll p_{t3}$, it is straightforward to see that
$d_{12} < d_{1B},d_{2B}$ and $d_{12} < d_{13},d_{23}$ (the logarithms
in Eq.~(\ref{eq:CMP-as2issue-d12-d}) have no impact on this).
Therefore the first step of the algorithm will be to cluster particles
$1$ and $2$, giving a flavourless pseudojet with transverse momentum,
rapidity and squared invariant mass of
\begin{subequations}
  \begin{align}
    p_{t,(1+2)} &\sim \max(p_{t1},p_{t2})\,,
    \\
    \label{eq:CMP-y12}
    y_{(1+2)} &\simeq
                \frac{1}{2} \ln \frac{E_1}{E_2}\,,
    \\
    m_{(1+2)}^2 &\simeq 4E_1E_2\,.
  \end{align}
\end{subequations}
From the point of view of standard jet clustering, the $(1+2)$
pseudojet is unusual, because its transverse momentum is much smaller
than its invariant mass.

The $(1+2)$ pseudojet will cluster with particle $3$ if it is within a
distance $\Delta R_{(1+2),3} < R=1$.
For any $p_{t1},p_{t2} \ll p_{t3}$, there is always a finite azimuthal
and $E_1,E_2$ phase space region such that that condition is
satisfied, and the resulting $1+2+3$ cluster will have significantly different
kinematics than particle $3$, because of the extra energy brought by
particles $1$ and $2$.
Thus the jets can differ between the $1,2,3$ event and the event with
just particle $3$.
The rate for this to happen is given by constants from the azimuthal
and energy integrations multiplying divergent integrals over
$p_{t1}$ and $p_{t2}$,
\begin{equation}
  N \sim \as^2
      \int^{p_{t3}}_\epsilon \frac{dp_{t1}}{p_{t1}}
      \int^{p_{t3}}_\epsilon \frac{dp_{t2}}{p_{t2}}
  = \as^2 \ln^2 \epsilon,
  \label{eq:CMP-as2issue-failure}
\end{equation}
where we have explicitly included a cutoff scale $\epsilon$ in order
to make the nature of the divergence manifest.

In Sec.~\ref{sec:IRC-results}, we proposed a modification of the
CMP algorithm, Eq.~(\ref{eq:CMP-fix}).
For a generic $\omega$, Eq.~(\ref{eq:CMP-as2issue-d12-d}) in
particular is replaced by
\begin{subequations}
  \label{eq:CMP-d12-Omega}
  \begin{align}
    d_{12}^{(\Omega)}
    &\sim \left(\frac{p_{t3}^2}{p_{t1}p_{t2}}\right)^{\omega}
      \frac{\max(p_{t1}^2,p_{t2}^2)}{p_{t3}^4},
    \\
  \label{eq:CMP-d12-Omega-b}
    &\sim
      \frac{\max(p_{t1}^2,p_{t2}^2)}{p_{t1} p_{t2} p_{t3}^2}
      \cdot \left(\frac{p_{t3}^2}{p_{t1}p_{t2}}\right)^{\omega-1}
      \,.
  \end{align}
\end{subequations}
When $\omega=1$, $d_{12}^{(\Omega)}$ will be of same order as $d_{3B}$
if $p_{t1} \sim p_{t2}$.
Schematically that suggests that there can still be a divergence of
the form
\begin{equation}
  N \sim \as^2
      \int^{p_{t3}}_\epsilon \frac{dp_{t1}}{p_{t1}}
      \int^{p_{t3}}_\epsilon \frac{dp_{t2}}{p_{t2}}
      \delta\left(\ln p_{t2}
      - \ln p_{t1}\right)
  = \as^2 \ln \epsilon,
  \label{eq:CMP-as2issue-failure-omega1}
\end{equation}
i.e.\ with one power of the logarithm.
With a more complete calculation one can verify that that divergence
is indeed present for $\omega=1$.
\logbook{}{see logbook/2023-04-CMP-IHCIHC.tex}%
For $\omega > 1$, the second factor in Eq.~(\ref{eq:CMP-d12-Omega-b})
instead ensures that $d_{12}^{(\Omega)} \gg d_{3B}$, thus ensuring
particle $3$ becomes a jet before any $1+2$ clustering, and resolving
the IRC safety issue with the kinematics of jet $3$.\footnote{%
  Once particle $3$ has been declared a jet and removed from the
  clustering, the $1+2$ cluster, if formed, would become a jet in its
  own right.
  Standard jet analyses place a cut on the jet $p_t$, which ensures
  that a residual lone massive, but low-$p_t$ $1+2$ cluster does not
  count as a hard jet.
  However, one could imagine a scenario where one cuts not on $p_t$, but on
  $p_t^2+m^2$, and in this case the $1+2$ cluster would count as an
  additional hard jet.
  In such a case, one could envisage that this would cause IRC
  unsafety even for $\omega>1$.
  We have not explored this question further, insofar as analyses do
  not normally cut on $p_t^2+m^2$.  }

\begin{figure}
  \centering
  \includegraphics[width=.99\columnwidth]{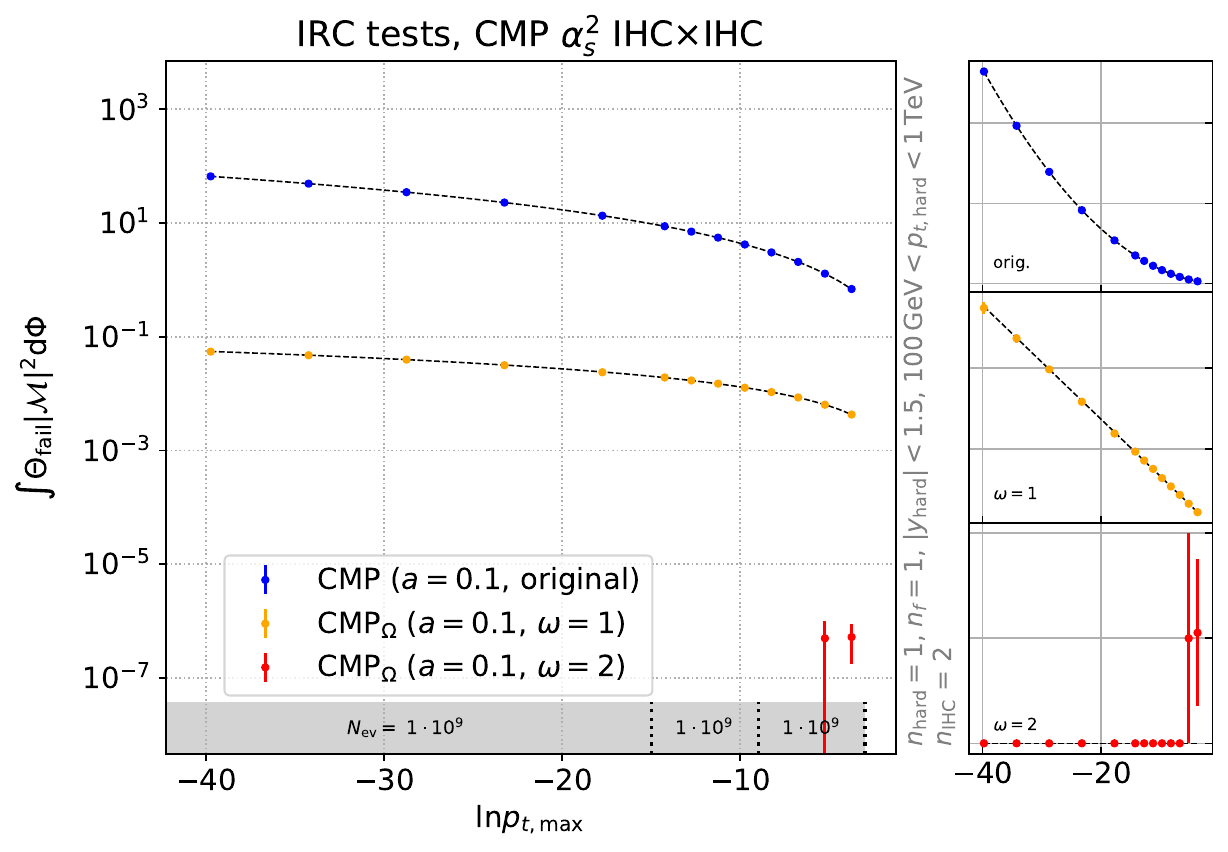}
  \caption{
    Results of the numerical tests for CMP for a configuration as
    in Fig.~\ref{fig:config-CMP-as2}, with the $\Delta R^2$ angular
    factor in the distance measure, as in the original algorithm (in blue), and
    with a corrected angular factor as in Eq.~(\ref{eq:CMP-fix}) where
    $\omega=1$ (in orange), and $\omega=2$ (in red).
    Miniature plots on the right depict the integrated failure
    rate on a linear scale to help read the functional form
    of the divergence.
    The original algorithm suffers from a quadratic divergence, the
    corrected measure with $\omega=1$ from a linear divergence, and
    the fix with $\omega=2$ is IRC safe.}
  \label{fig:config-CMP-as2-num}
\end{figure}

Results of the numerical IRC safety tests are presented in
Fig.~\ref{fig:config-CMP-as2-num} for the original algorithm (with an
angular factor $\Delta R^2$ in the distance measure, in blue) and for
the algorithm with an IRC-safe angular distance (for values of the
parameter $\omega=1$, in orange, and $\omega=2$, in red).  The results
confirm the above analyses, showing a quadratic divergence in the
original algorithm, while the divergence is linear for $\omega=1$ and
fully lifted for $\omega = 2$.%
\footnote{Note that Fig.~1 (left) of Ref.~\cite{Czakon:2022wam} has
  studied $pp \to \ell^+\ell^- bb\bar b$ (with massless $b$'s), which
  should include the configuration of our Fig.~\ref{fig:config-CMP-as2}.
  That figure does not appear to show a divergence as the technical
  cutoff is reduced.
  At first sight that may seem surprising, however two considerations
  should be kept in mind.
  Firstly, the configuration of Fig.~\ref{fig:config-CMP-as2}
  requires an initial-state gluon to originate from a $b \to gb$ splitting,
  which is responsible for only a small fraction of incoming
  gluons. 
  Secondly, the impact of the IRC unsafety is to smear the rapidity
  distribution of the $b$-jet and a smearing of broad distribution
  tends to have a limited impact on the integral of the broad
  distribution within some window (Ref.~\cite{Czakon:2022wam} used $|y_b|<2.4$).
  As a result, it is conceivable that the expected squared logarithmic
  divergence in Fig.~1 (left) of Ref.~\cite{Czakon:2022wam} might be
  too small to clearly see in that figure. }

Finally, note that the flavour-$k_t$ algorithm does not suffer from
the issue presented here thanks to the form of its beam distance: the
initial-state emissions $1$ and $2$ would be declared as beam jets and so be
removed from further consideration early in the clustering sequence.

\subsection{IHC$\times$IDS issue at $\as^3$ for CMP}
\label{sec:CMP-IHC-DS}

\begin{figure}
  \centering
  \includegraphics[width=\columnwidth]{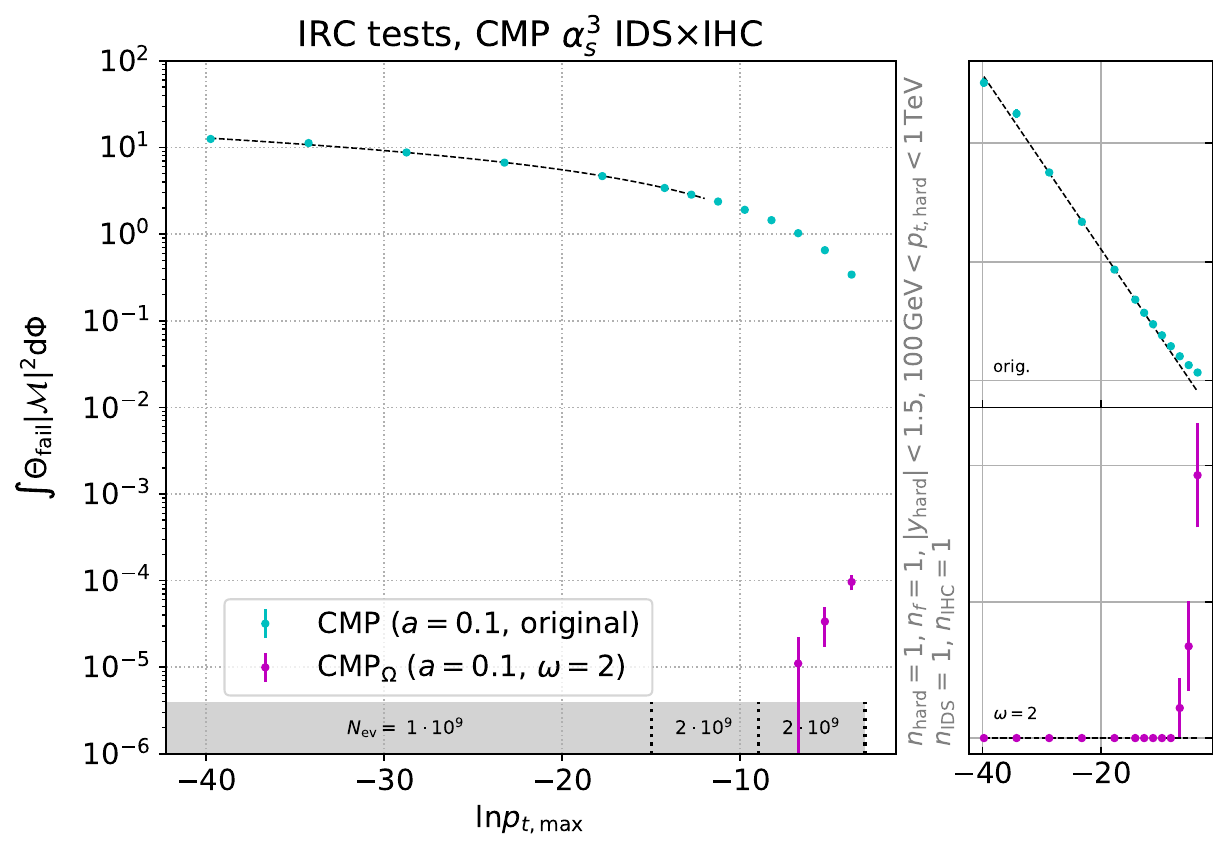}
  \caption{Failure rate of the CMP algorithm for the configuration of
    Fig.~\ref{fig:IRsafety-as3-IHC-DS-diagram}, both for the original
    formulation and our modification, Eq.~(\ref{eq:CMP-fix}), showing
    a divergence for the former and none for the latter.}
  \label{fig:CMP-IHC-DS-rate}
\end{figure}

The subtlety from Sec.~\ref{sec:IRCapp-flavkt} for the flavour-$k_t$
and GHS algorithms has an interesting manifestation in the CMP
algorithm. 
We consider again the scenario of
Fig.~\ref{fig:IRsafety-as3-IHC-DS-diagram}, with the same set of
variables and in a configuration where $p_{t4}\ll p_{t2}$.
We will concentrate on the two distances that are smallest, which,
neglecting $\mathcal O(1)$ factors, read
\begin{equation}
  d_{23} \sim \frac{p_{t2}^2}{z_{23}^2 p_{t1}^4},
  \quad
  d_{24} \sim \frac{p_{t2}^2}{p_{t1}^4} y_4^2\,.
\end{equation}
As before, the probability for a 2+4 recombination must be finite
for IRC safety. The 2+4 recombination will occur if
\begin{equation}
  z_{23}\lesssim \frac{1}{y_4} = \frac{1}{\ell_{14} + \ln 2z_{41}}.
\end{equation}
We neglect the $\ln 2z_{41}$ term in the denominator (and take the
$z_{41}$ integral to give a constant), integrate over all $p_{t4}$
values, over $p_{t2} > p_{t4}$ and over the allowed $z_{23}$ range
(with the same constant splitting function approximation as in
App.~\ref{sec:IRCapp-flavkt}).
We then obtain the probability for a $2+4$ clustering 
\begin{equation}
  N_{24} \sim \as^3\int_0^\infty d\ell_{14}\int_0^{\ell_{14}}
  d\ell_{24}
  \int_0^{1/\ell_{14}} d z_{23},
\end{equation}
which is divergent. If we regulate the upper integration region of the
$\ell_{14}$ integral with $\infty \to \ln \frac{1}{\epsilon}$, the
probability scales as
\begin{equation}
  N_{24} \sim \as^3 \ln\frac{1}{\epsilon}.
  \label{eq:CMP-as3-divergence}
\end{equation}
As with the other occurrences of this kind of issue, the replacement
Eq.~(\ref{eq:CMP-fix}) solves the problem, as can be seen from the numerical IRC
safety tests for this configuration shown in
Fig.~\ref{fig:CMP-IHC-DS-rate}.

%----------------------------------------------------------------------
\subsection{FHC$^2$ issue at $\as^2$ for GHS}
\label{sec:GHS-trouble-as2}

For the discussion here and in App.~\ref{sec:GHS-trouble-as4}, we assume a version of the GHS algorithm
with an $\Omega_{ij}^2$ style angular distance in the dressing
(flavour-$k_t$-like) phase, since we know from
Sec.~\ref{sec:IRCapp-flavkt} that this is required for the
flavour-$k_t$ algorithm, which is the basis of the algorithm's
dressing step.
To make this clear, in plots where this modification is used, we
refer to the algorithm as GHS$_\Omega$.

\begin{figure}
  \centering
  \includegraphics[width=.86\columnwidth]{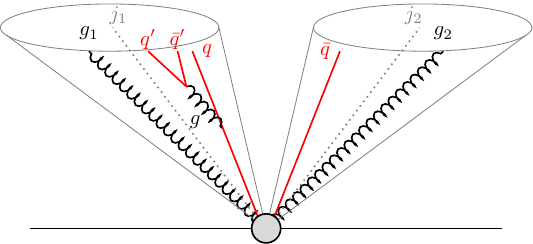}
  \caption{
    Example $\mathcal{O}(\as^2)$ configuration that yields an issue for the GHS algorithm.
    There are four
    hard particles (that one can imagine recoiling against a hard
    gluon or electroweak system on the other side of the event), a
    collinear emission of a hard gluon $g$ from one of the flavoured
    particles (the $q$), which then splits collinearly to a flavoured
    pair $q'\bar q'$.}
  \label{fig:config-GHS-as2}
\end{figure}

Let us consider a hard event as that in Fig.~\ref{fig:config-GHS-as2}.
The event has four particles, which we will call $g_1$, $q$,
$\bar{q}$ and $g_2$, starting from the left.
We will work through the algorithm
to see what happens if a hard, but collinear gluon $g$, is emitted from the $q$,
and splits collinearly to $q' \bar q'$ (with $\Delta R_{q' \bar q'} \ll 1$
much smaller than any other scale in the problem).

Focusing on the hard event first (i.e.\ without the emission of the
hard collinear gluon $g$), we assume that the anti-$k_t$ algorithm clusters the four particles
into two jets ($j_1$ and $j_2$), as indicated in the figure.
We can further assume that $\Delta R_{g_1 q}, \Delta R_{g_2 \bar{q}} > 
R_{\text{cut}} \sim 0.1$, so that the hard gluons $g_1$ and $g_2$
are not accumulated into $q$ and $\bar q$ in that phase of the algorithm.

First we will consider the case $\alpha < 2$.
For any angular structure of the event satisfying the above limits, we
take the momenta of $g_1, g_2, q$ and $\bar{q}$ such that the event
without the $g\to q'\bar q'$ emission has the following properties
\begin{subequations}
  \begin{align}
    \label{eq:GHS-4hard-initial-distances-alphaLT2}
    &d_{{q}j_1} > d_{{q}{\bar{q}}} > d_{{\bar{q}}j_2}\,, \\
    &p_{tq} < p_{t\bar{q}}\,. \label{eq:GHS-hard-pts}
  \end{align}
\end{subequations}
As a result, the first dressing step is for the $\bar q$ flavour to be
assigned to jet $j_2$, followed by the $q$ flavour being assigned to
jet $j_1$.
Thus both $j_1$ and $j_2$ are flavoured.
Note that for a full analysis, one should also take into account
$d_{iB}$ beam distances for all flavoured particles $i$.
To help understand why we can ignore it, suppose that all the hard
particles have rapidities close to zero, which results in $p_{tB}(0)$
in Eq.~(\ref{eq:ktB-flavkt-partial}) being approximately the scalar
sum of all the particles' transverse momenta.
That scale will tend to be a few times larger than the transverse
momenta of any of the individual particles, which ensures that the
distance of any cluster to its jet will be smaller than the $d_{iB}$,
as will the $d_{q\bar q}$ if the two jets are not too far away in
angle.

Next we consider the impact of the emission of the collinear hard gluon
from $q$ with $p_{tg} = z p_{tq}$, followed by its splitting into a
collinear $q'\bar q'$ pair.
Recall that we work with $\Delta R_{q'\bar q'} \ll R_{\text{cut}}$ so
that it is the smallest angular distance in the event.
The algorithm goes through the accumulation step, and will identify four flavour
clusters: $\hat q'$, $\hat{\bar q}'$, and the original $\hat q$ and $\hat{\bar q}$.
The angular structure is otherwise unchanged, so we
get no further flavour accumulation.
To lighten the notation, below we will leave out the explicit ``hats''
for the flavour clusters, especially as the flavour clusters coincide
with the original particles.

The final step is the flavour dressing:
the $\nohat{q}', \nohat{\bar{q}}'$ pair will annihilate first, as it should
because the pair came from a common parent gluon.
These flavour clusters (including their kinematics) are discarded from
further consideration, and any distance involving them is removed from
the list.
The remaining distances ($d'$) for the event with the
$g\to q' \bar q'$ splitting are then given in terms of the
hard-event's distances ($d$) as
\begin{subequations}
  \begin{align}
    &d'_{\nohat{q}j_1} = (1 - z)^{2-\alpha} d_{\nohat{q}j_1}\,, \\
    &d'_{\nohat{q}\nohat{\bar{q}}} = (1 - z)^{2-\alpha} d_{\nohat{q}\nohat{\bar{q}}}\,, \\
    &d'_{\nohat{\bar{q}} j_2} = d_{\nohat{\bar{q}}j_2}\,,
  \end{align}
\end{subequations}
where the $(1-z)^{2-\alpha}$ factor arises because of the reduction in
transverse momentum of the $q$ after emission of the $g\to q'\bar q'$
(which carries a fraction $z$ of its original $q$ momentum).
The potentially dangerous scenario is that where the ordering of
distances, Eq.~(\ref{eq:GHS-4hard-initial-distances-alphaLT2}), is
modified,
\begin{equation}
  \label{eq:GHS-4hard-danger}
  d'_{\nohat{q}\nohat{\bar{q}}} < \min\left(d'_{\nohat{q}j_1},d'_{\nohat{\bar{q}} j_2}\right)\,,
\end{equation}
because then $q$ and $\bar q$'s flavours will annihilate, leaving
flavourless hard jets, associated with a squared logarithmic
divergence from the two nested hard collinear divergences.
There is a finite range of $z$ in which this occurs,
\begin{equation}
  \label{eq:GHS-alpha-less2-z}
  1-z < 
      \left(\frac{d_{\nohat{\bar{q}}j_2}}{d_{\nohat{q}\nohat{\bar{q}}}}\right)^{\frac{1}{2
      - \alpha}}\,,
\end{equation}
thus confirming the presence of IRC unsafety from the configuration of
Fig.~\ref{fig:config-GHS-as2} for $\alpha<2$.

When $\alpha = 2$, we instead consider a hard event satisfying
$p_{tq} > p_{t\bar{q}}$ rather than the inequality in
Eq.~(\ref{eq:GHS-hard-pts}), in which case we have
\begin{subequations}
  \begin{align}
    &d'_{\nohat{q}j_1} = d_{\nohat{q}j_1}\,, \\
    &d'_{\nohat{q}\nohat{\bar{q}}} = \max\left((1 -
      z)^{2},\frac{p_{t\bar q}^2}{p_{tq}^2}\right)
      d_{\nohat{q}\nohat{\bar{q}}}\,,
    \\
    &d'_{\nohat{\bar{q}} j_2} = d_{\nohat{\bar{q}}j_2}\,.
  \end{align}
\end{subequations}
Again, there is the possibility of $d'_{\nohat{q}\nohat{\bar{q}}}$
becoming the smallest of the three distances, with the outcome that
the $q$ and $\bar q$ flavours would annihilate, leaving flavourless
hard jets, with a squared logarithmic divergence associated with the
collinear splittings.

\begin{figure}
  \centering
  \includegraphics[width=\columnwidth]{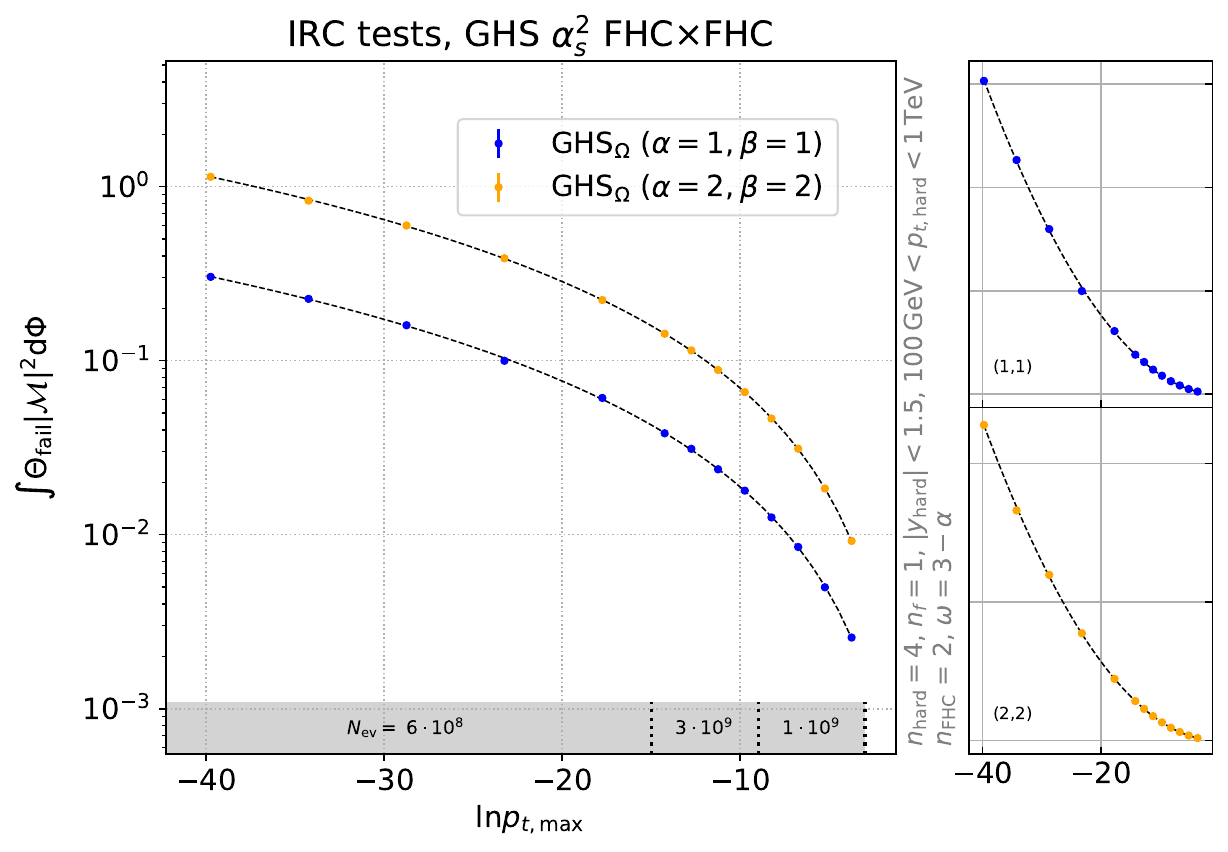}
  \caption{Failure rate of the GHS algorithm for the FHC$\times$FHC
    configuration of Fig.~\ref{fig:config-GHS-as2}, illustrating the
    quadratic divergence, specifically for $\alpha=1$, $\beta=1$ and
    $\alpha=2$, $\beta=2$.
    Other parameters are $R_\text{cut}=0.1$,
    $z_\text{cut}=0.1$ and $p_{t,\text{cut}}=100\GeV$.
    The jet radius $R$ has been sampled in the range $0.3{-}1.57$.
    The version of the GHS algorithm used is one where $\Delta
    R_{ij}^2$ in the dressing stage has been replaced with
    $\Omega_{ij}^2$ using $\omega = 3-\alpha$ (the original $\Delta
    R_{ij}^2$ similarly gives a squared logarithmic divergence). 
  }
  \label{fig:GHS-4hard-as2-rate}
\end{figure}

The set of distances in the argument above is perhaps somewhat
complicated, with angular factors to consider, the beam distances and
the extra subtleties of the $\alpha=2$ case.
Therefore, in Fig.~\ref{fig:GHS-4hard-as2-rate} we show the outcome of
our IRC safety tests, illustrating that the divergence is indeed
present for the two combinations $\alpha=1$, $\beta=1$ and
$\alpha=2$, $\beta=2$.
We leave to future work the possibility of identifying a concrete
modification of the algorithm that solves this problem, nevertheless
we anticipate that one line of investigation could be to allow
accumulation of kinematics within a jet during the dressing stage.

A final comment is that this configuration can appear at NNLO for a
process such as fully hadronic $t\bar t$ production, however only if
one asks for two massive $b$-tagged jets.
It also appears at N$^4$LO for a process such as $Zb\bar b$ production.

%----------------------------------------------------------------------
\subsection{IDS$\times$FDS issue at $\as^4$ for GHS}
\label{sec:GHS-trouble-as4}

\begin{figure}
  \centering
  \includegraphics[width=0.7\columnwidth]{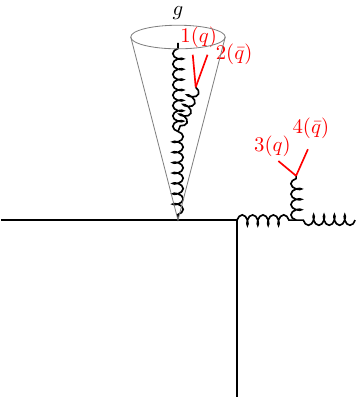}
  \caption{ An FDS$\times$IDS kinematic configuration that causes problems for GHS
    algorithms for $\alpha\beta \ge 2$. }
  \label{fig:GHStrouble}
\end{figure}

The GHS algorithm exhibits an interesting interplay between initial-state and final-state double-soft emissions at order $\as^4$ if $\alpha\beta \ge 2$.
The configuration that we consider here is that represented in
Fig.~\ref{fig:GHStrouble}, involving a hard Born event with one or more
unflavoured jets.
That event is then supplemented with a double-soft pair ($1$,$2$) that
is collinear to an (originally) unflavoured jet and an additional
large-angle double-soft pair ($3$,$4$) outside the jet.

We are specifically interested in the situation where
\begin{subequations}
  \begin{align}
    \theta_{1g} &\lesssim \{ \theta_{2g} , \theta_{12}\} \ll 1\,,
    \\
    p_{t1} &\sim p_{t2} \ll p_{tg}\,,
    \\
    p_{t3} &\sim p_{t4} \ll p_{tg}\,,
    \\
    \theta_{23} &\sim \theta_{34} \sim 1\,.
  \end{align}
\end{subequations}
During the accumulation step, there is a possibility that $1$ clusters
with $g$, giving a hard $\widehat{g1}$ flavour cluster, leaving an
unclustered, much softer $\hat 2$ flavour cluster.
If during the subsequent dressing phase, $\hat 2$ goes on to
annihilate with $\hat 3$ rather than with $\widehat{g1}$, then the
resulting hard jet will be flavoured.

The SoftDrop condition for $1$ to cluster with $g$ is given by
\begin{equation}
  \label{eq:GHS-DSDS-SDcondition}
  \frac{p_{t1}}{p_{tg}} > \theta_{1g}^{\beta}\,,
\end{equation}
where throughout our discussion here we neglect factors of order $1$
(e.g.\ $R_\text{cut}$ and $z_\text{cut}$).
There is no further accumulation since all particles are now flavoured
and the flavoured clusters will be $\widehat{g1}$, $\hat{2}$,
$\hat{3}$ and $\hat{4}$.
The angle between the $\widehat{g1}$ and the jet direction will be
given by
\begin{equation}
  \label{eq:eq:GHS-DSDS-theta-j-g1}
  \theta_{j\widehat{g1}} \sim \frac{p_{t2}}{p_{tg}} \theta_{2g}\,.
\end{equation}
Without loss of generality we can consider the case where
$d_{\hat{2}\hat{3}} < d_{\hat{2}\hat{4}}$, and then the distances to
take into account during the dressing phase are 
\begin{subequations}
  \label{eq:eq:GHS-DSDS-annihilation-conditions}
  \begin{align}
    d_{j\widehat{g1}} &\sim p_{t2}^2 \theta_{2g}^2\,,
    \\
    d_{\widehat{g1}\hat 2}
                      &
                        \sim d_{j\hat 2}
                        \sim p_{tg}^2 \left(\frac{p_{t2}}{p_{tg}}\right)^{2-\alpha} \theta_{2g}^2\,,
    \\
    d_{\hat2\hat3}
                      &
                        \sim \max(p_{t2},p_{t3})^\alpha \min(p_{t2},p_{t3})^{2-\alpha},
    \\
    d_{\hat3\hat4}
                      &
                        \sim p_{t3}^2,
  \end{align}
\end{subequations}
where, again we have ignored factors of order $1$, e.g.\ from angular
distances. 
The hard jet will acquire a flavour if the SoftDrop condition
of Eq.~(\ref{eq:GHS-DSDS-SDcondition}) is satisfied and if additionally
$\hat 2$ fails to annihilate the flavour of the $\widehat{g1}$
cluster.
This will occur if $d_{23} < d_{34}$ and
$d_{23} < d_{j\hat 2}\simeq d_{\widehat{g1} 2}$.%
\footnote{ Note that the $d_{23} < d_{34}$ condition implies that
  $p_{t2}$ cannot be substantially larger than $p_{t3}$,
  which leads to $ d_{j\widehat{g1}}$ being the
  smallest of all the distances.
  Consequently, the first step of the dressing is that the flavour of
  $\widehat{g1}$ is assigned to the jet and the $\widehat{g1}$ cluster
  is removed from consideration.  }
In determining whether these conditions are satisfied, it is helpful
to introduce shorthands
\begin{subequations}
  \begin{align}
    \ell_i &= \ln \frac{p_{tg}}{p_{ti}}, \\
    \ell_\theta &= \ln \frac{1}{\theta_{2g}},
  \end{align}
\end{subequations}
and to observe that in the SoftDrop condition
Eq.~(\ref{eq:GHS-DSDS-SDcondition}), we can replace $1 \to 2$, since
this only affects $\order{1}$ terms.

\begin{figure}
  \centering
  \includegraphics[width=\columnwidth]{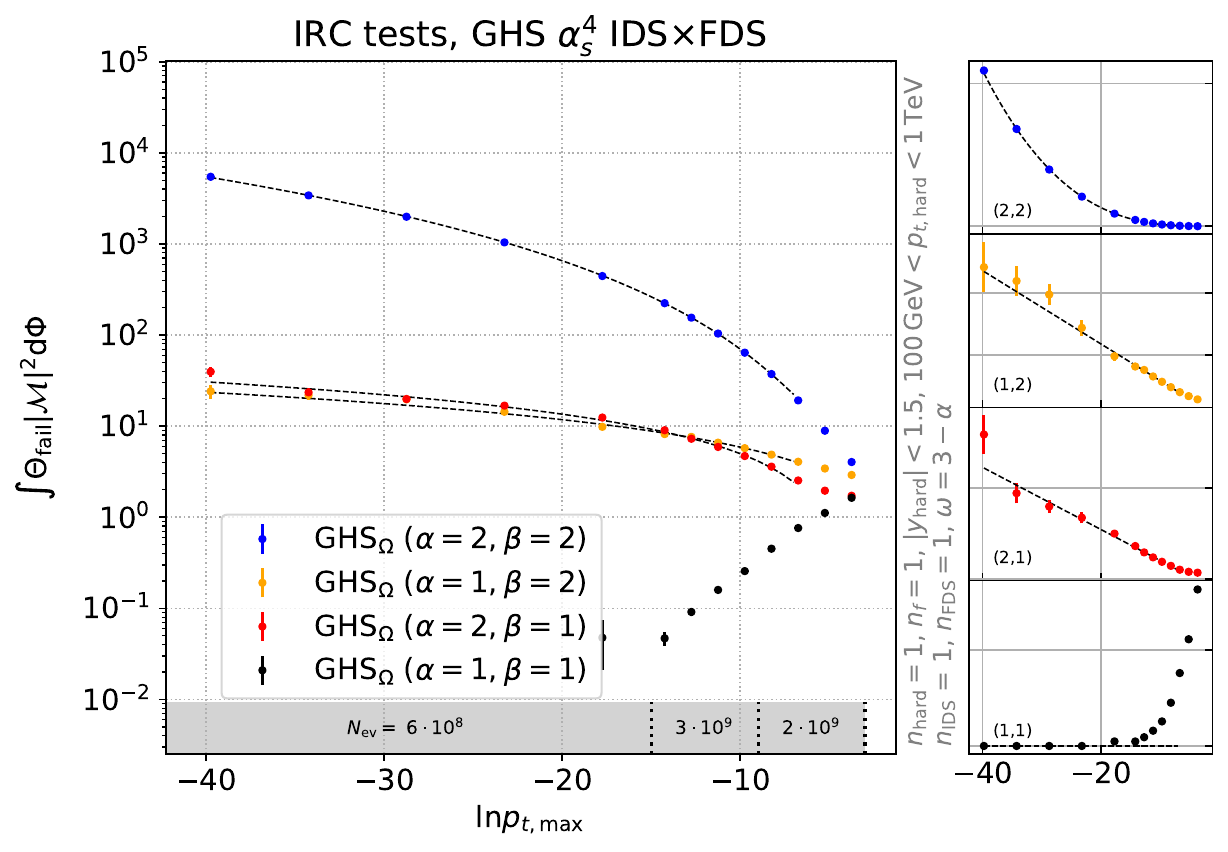}
  \caption{Failure rate of the GHS algorithm   for the $\as^4$
    FDS$\times$IDS configuration of Fig.~\ref{fig:GHStrouble},
    illustrating the cubic divergence for parameter choices involving
    $\alpha\beta>2$, the linear divergence for $\alpha\beta=2$ and
    convergence for $\alpha\beta<2$.
    In its dressing stage, the GHS implementation for these runs uses an
    $\Omega_{ij}^2$ angular distance instead of
    $\Delta R_{ij}^2$, where $\omega=3-\alpha$.
    Other parameters are $R_\text{cut}=0.1$,
    $z_\text{cut}=0.1$ and $p_{t,\text{cut}}=100\GeV$.
  }
  \label{fig:GHS-DSDS-tests}
\end{figure}

Our conditions then become (to within $\order{1}$ offsets)
\begin{subequations}
  \begin{align}
    \text{SD}&:\quad \ell_2 < \beta \ell_\theta\,,
    \\
    d_{34} > d_{23}&:\quad\ell_2 > \ell_3\,,
    \\
    d_{g2} > d_{23}&:\quad (2-\alpha)\ell_2 +2\ell_\theta < \alpha
                     \ell_3 + (2-\alpha)\ell_2\,,
                     \label{eq:GHS-DSDS-dg2-d23-full}
  \end{align}
\end{subequations}
where the last line already underwent some simplification (using the
second line), and can then be further simplified to read
\begin{equation}
  2\ell_\theta < \alpha \ell_3\,.
\end{equation}
Assembling all inequalities, we obtain
\begin{equation}
  \label{eq:eq:GHS-DSDS-conditions}
  2 \ell_\theta < \alpha \ell_3 < \alpha \ell_2 < \alpha \beta \ell_\theta\,.
\end{equation}
We immediately see that if $\alpha \beta < 2$, there is no available
logarithmic integration region, and so no IRC divergence from this
configuration.
Conversely, if $\alpha \beta > 2$, we expect to see a cubic
logarithmic divergence from integrals over $\ell_\theta$, $\ell_3$
and $\ell_2$.
For $\alpha \beta = 2$, the $\order{1}$ factors become critical and it
is easiest to carry out a numerical study, but it is reasonable to
expect a divergence with a single logarithm.

The results of the numerical study are shown in
Fig.~\ref{fig:GHS-DSDS-tests} for four combinations of $\alpha$ and
$\beta$.
They confirm our expectations and suggest that if one wishes to employ
a GHS-style algorithm, one should use it with $\alpha\beta < 2$.
Nevertheless, one would still need to find a solution to the separate
issue identified in App.~\ref{sec:GHS-trouble-as2} (which cannot be
resolved just through parameter choices), and then verify that the
resulting algorithm passes a full set of IRC safety tests.

\section{Summary plots for IRC-safe algorithms}
\label{sec:final-summary-plots}

\begin{figure*}
  \centering
  \includegraphics[width=0.32\textwidth,page=1]{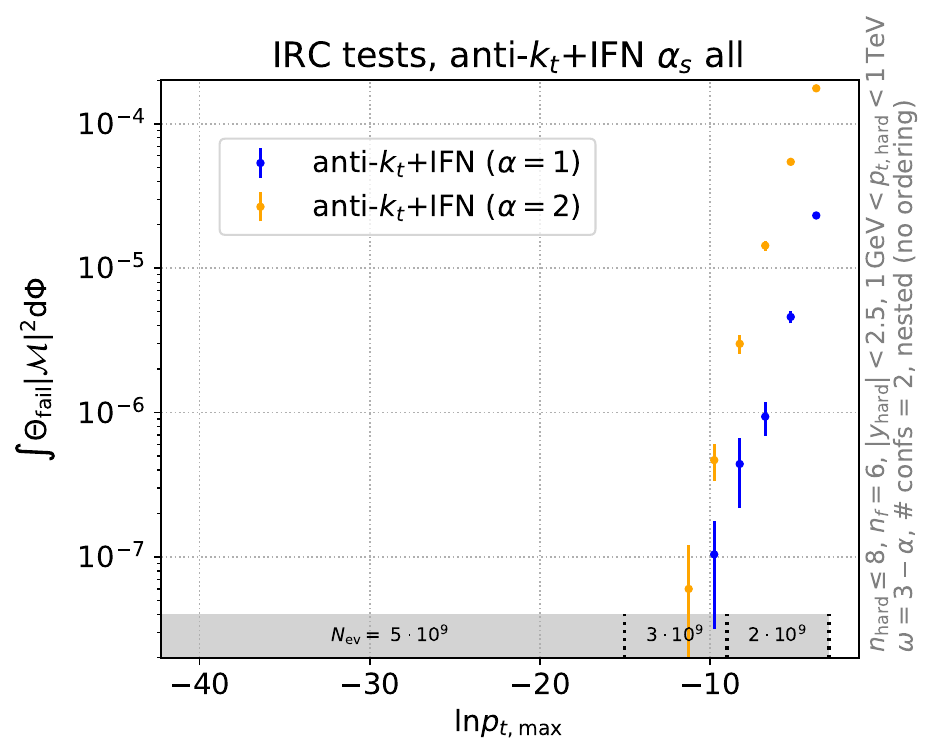}
  \includegraphics[width=0.32\textwidth,page=2]{figures/plot-akt-IFN-full.pdf}
  \includegraphics[width=0.32\textwidth,page=3]{figures/plot-akt-IFN-full.pdf}
  \includegraphics[width=0.32\textwidth,page=4]{figures/plot-akt-IFN-full.pdf}
  \includegraphics[width=0.32\textwidth,page=5]{figures/plot-akt-IFN-full.pdf}
  \includegraphics[width=0.32\textwidth,page=6]{figures/plot-akt-IFN-full.pdf}
  \\
  \caption{Summary of IRC safety test results at
    orders $\as$ to $\as^6$ for the anti-$k_t$
    algorithm with IFN.
  }
  \label{fig:IRC-safety-summary-IFN-akt}
\end{figure*}

\begin{figure*}
  \centering
  \includegraphics[width=0.32\textwidth,page=1]{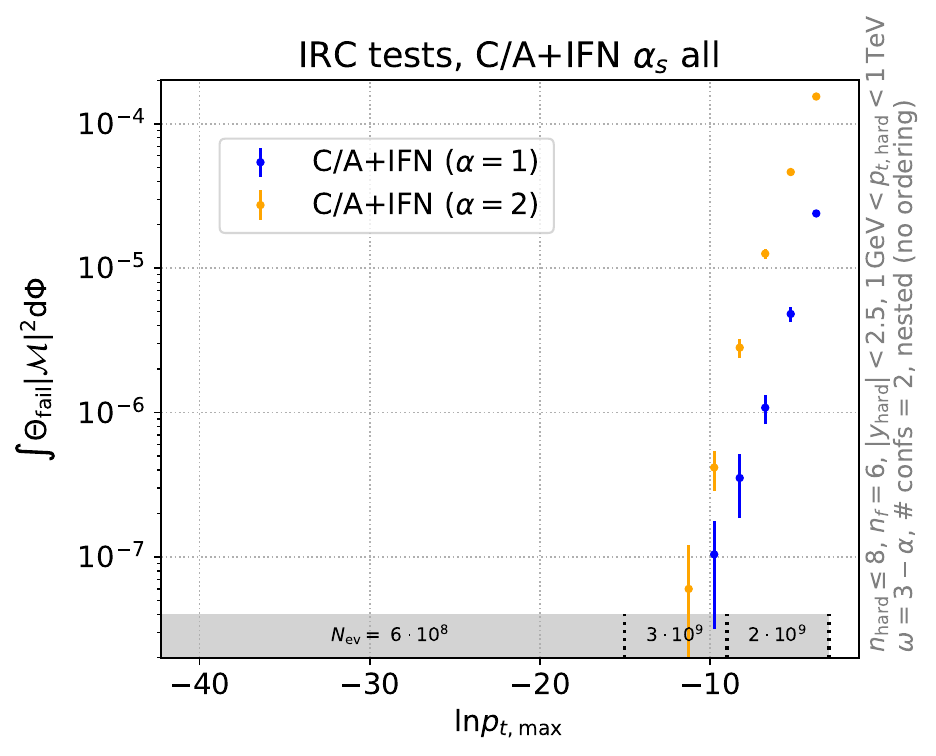}
  \includegraphics[width=0.32\textwidth,page=2]{figures/plot-cam-IFN-full.pdf}
  \includegraphics[width=0.32\textwidth,page=3]{figures/plot-cam-IFN-full.pdf}
  \includegraphics[width=0.32\textwidth,page=4]{figures/plot-cam-IFN-full.pdf}
  \includegraphics[width=0.32\textwidth,page=5]{figures/plot-cam-IFN-full.pdf}
  \includegraphics[width=0.32\textwidth,page=6]{figures/plot-cam-IFN-full.pdf}
  \\
  \caption{Same as Fig.~\ref{fig:IRC-safety-summary-IFN-akt}, for the
    C/A algorithm with IFN.
  }
  \label{fig:IRC-safety-summary-IFN-cam}
\end{figure*}

\begin{figure*}
  \centering
  \includegraphics[width=0.32\textwidth,page=1]{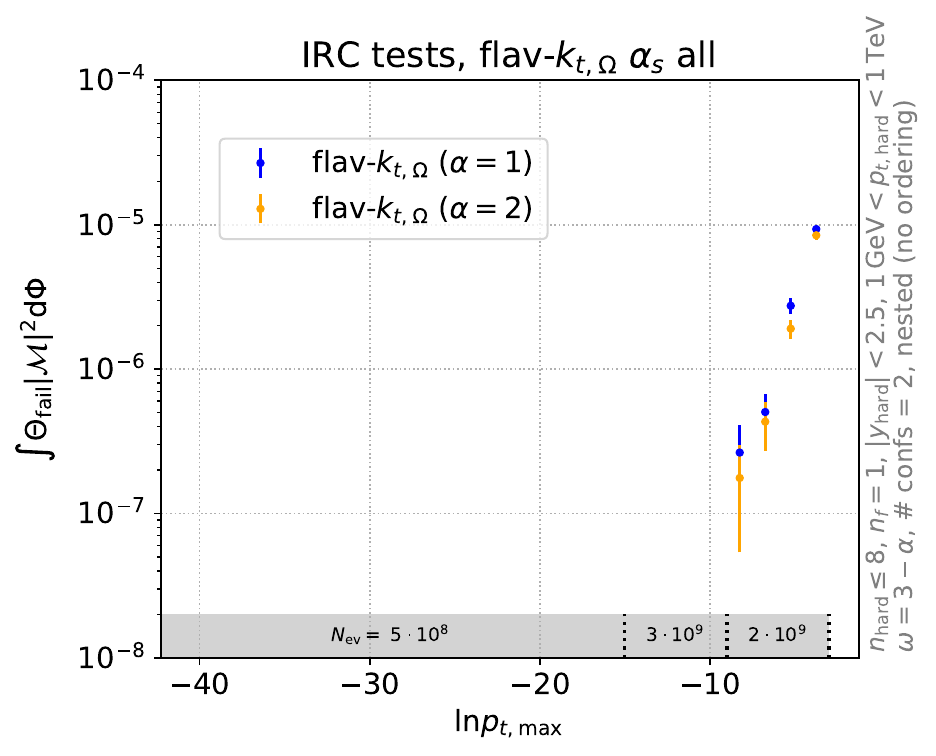}
  \includegraphics[width=0.32\textwidth,page=2]{figures/plot-flavktOmega-full.pdf}
  \includegraphics[width=0.32\textwidth,page=3]{figures/plot-flavktOmega-full.pdf}
  \includegraphics[width=0.32\textwidth,page=4]{figures/plot-flavktOmega-full.pdf}
  \includegraphics[width=0.32\textwidth,page=5]{figures/plot-flavktOmega-full.pdf}
  \includegraphics[width=0.32\textwidth,page=6]{figures/plot-flavktOmega-full.pdf}
  \\
  \caption{Same as Fig.~\ref{fig:IRC-safety-summary-IFN-akt}, for the
    flavour-$k_{t,\Omega}$ algorithm.
  }
  \label{fig:IRC-safety-summary-FlavKtOmega}
\end{figure*}

\begin{figure*}
  \centering
  \includegraphics[width=0.32\textwidth,page=1]{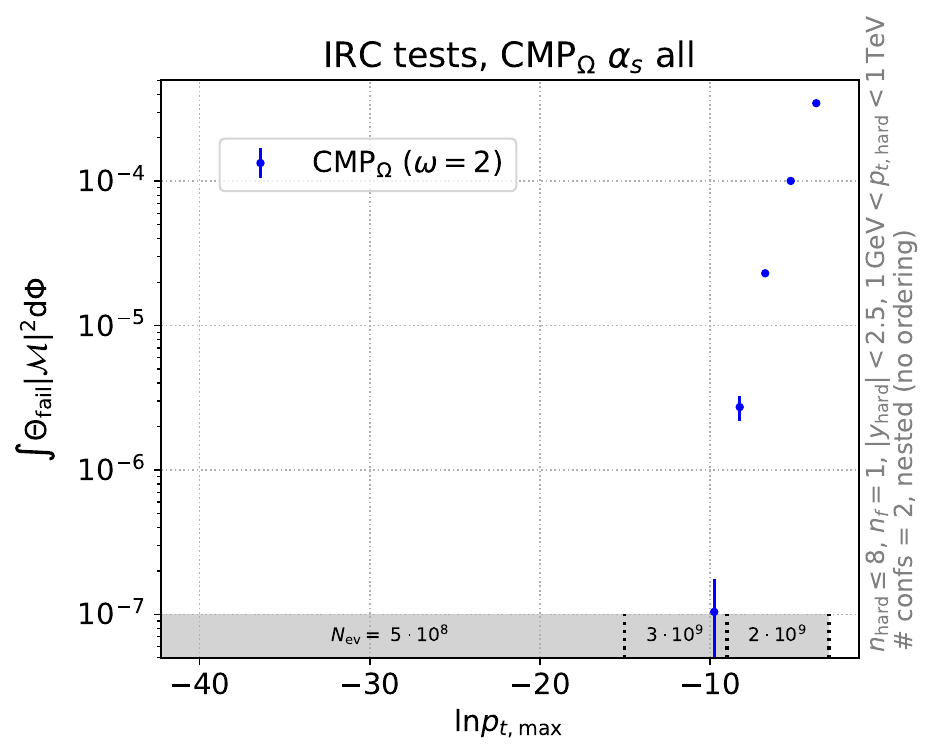}
  \includegraphics[width=0.32\textwidth,page=2]{figures/plot-CMPOmega-full.pdf}
  \includegraphics[width=0.32\textwidth,page=3]{figures/plot-CMPOmega-full.pdf}
  \includegraphics[width=0.32\textwidth,page=4]{figures/plot-CMPOmega-full.pdf}
  \includegraphics[width=0.32\textwidth,page=5]{figures/plot-CMPOmega-full.pdf}
  \includegraphics[width=0.32\textwidth,page=6]{figures/plot-CMPOmega-full.pdf}
  \\
  \caption{Same as Fig.~\ref{fig:IRC-safety-summary-IFN-akt}, for the
    CMP$_{\Omega}$ algorithm.
  }
  \label{fig:IRC-safety-summary-CMPOmega}
\end{figure*}

In this appendix, we present summary plots from our IRC safety tests
for the three approaches that have passed all those tests:
IFN, CMP$_\Omega$ and flavour-$k_{t,\Omega}$.
In Figs.~\ref{fig:IRC-safety-summary-IFN-akt} and
~\ref{fig:IRC-safety-summary-IFN-cam}, we show the results from the 
IRC-safety tests for the anti-$k_t$ and C/A algorithms with IFN, i.e.\ each
of the algorithms labelled as safe in Table~\ref{tab:IRC-test-results}.
Figs.~\ref{fig:IRC-safety-summary-FlavKtOmega} and \ref{fig:IRC-safety-summary-CMPOmega} show corresponding results
for our adaptations of the flavour-$k_t$ and CMP algorithms,
supporting the conclusion that they too are IRC safe.

\clearpage

%======================================================================
\bibliographystyle{bibliostyle}
\bibliography{flavour}

\providecommand{\href}[2]{#2}\begingroup\raggedright\begin{thebibliography}{10}

\bibitem{Banfi:2006hf}
A.~Banfi, G.~P. Salam, and G.~Zanderighi, {\it {Infrared safe definition of jet
  flavor}},  {\em Eur. Phys. J. C} {\bf 47} (2006) 113--124,
  [\href{http://arxiv.org/abs/hep-ph/0601139}{{\tt hep-ph/0601139}}].

\bibitem{Banfi:2007gu}
A.~Banfi, G.~P. Salam, and G.~Zanderighi, {\it {Accurate QCD predictions for
  heavy-quark jets at the Tevatron and LHC}},  {\em JHEP} {\bf 07} (2007) 026,
  [\href{http://arxiv.org/abs/0704.2999}{{\tt arXiv:0704.2999}}].

\bibitem{Catani:1991hj}
S.~Catani, Y.~L. Dokshitzer, M.~Olsson, G.~Turnock, and B.~R. Webber, {\it {New
  clustering algorithm for multi-jet cross-sections in $e^+ e^-$
  annihilation}},  {\em Phys. Lett. B} {\bf 269} (1991) 432--438.

\bibitem{Catani:1993hr}
S.~Catani, Y.~L. Dokshitzer, M.~H. Seymour, and B.~R. Webber, {\it
  {Longitudinally invariant $K_t$ clustering algorithms for hadron hadron
  collisions}},  {\em Nucl. Phys. B} {\bf 406} (1993) 187--224.

\bibitem{Ellis:1993tq}
S.~D. Ellis and D.~E. Soper, {\it {Successive combination jet algorithm for
  hadron collisions}},  {\em Phys. Rev. D} {\bf 48} (1993) 3160--3166,
  [\href{http://arxiv.org/abs/hep-ph/9305266}{{\tt hep-ph/9305266}}].

\bibitem{Cacciari:2008gp}
M.~Cacciari, G.~P. Salam, and G.~Soyez, {\it {The anti-$k_t$ jet clustering
  algorithm}},  {\em JHEP} {\bf 04} (2008) 063,
  [\href{http://arxiv.org/abs/0802.1189}{{\tt arXiv:0802.1189}}].

\bibitem{Gauld:2020deh}
R.~Gauld, A.~Gehrmann-De~Ridder, E.~W.~N. Glover, A.~Huss, and I.~Majer, {\it
  {Predictions for $Z$-Boson Production in Association with a $b$-Jet at
  $\mathcal {O}(\alpha_s^3)$}},  {\em Phys. Rev. Lett.} {\bf 125} (2020),
  no.~22 222002, [\href{http://arxiv.org/abs/2005.03016}{{\tt
  arXiv:2005.03016}}].

\bibitem{Caletti:2022hnc}
S.~Caletti, A.~J. Larkoski, S.~Marzani, and D.~Reichelt, {\it {Practical jet
  flavour through NNLO}},  {\em Eur. Phys. J. C} {\bf 82} (2022), no.~7 632,
  [\href{http://arxiv.org/abs/2205.01109}{{\tt arXiv:2205.01109}}].

\bibitem{Caletti:2022glq}
S.~Caletti, A.~J. Larkoski, S.~Marzani, and D.~Reichelt, {\it {A fragmentation
  approach to jet flavor}},  {\em JHEP} {\bf 10} (2022) 158,
  [\href{http://arxiv.org/abs/2205.01117}{{\tt arXiv:2205.01117}}].

\bibitem{Czakon:2022wam}
M.~Czakon, A.~Mitov, and R.~Poncelet, {\it {Infrared-safe flavoured
  anti-k$_{T}$ jets}},  {\em JHEP} {\bf 04} (2023) 138,
  [\href{http://arxiv.org/abs/2205.11879}{{\tt arXiv:2205.11879}}].

\bibitem{Gauld:2022lem}
R.~Gauld, A.~Huss, and G.~Stagnitto, {\it {Flavor Identification of
  Reconstructed Hadronic Jets}},  {\em Phys. Rev. Lett.} {\bf 130} (2023),
  no.~16 161901, [\href{http://arxiv.org/abs/2208.11138}{{\tt
  arXiv:2208.11138}}].

\bibitem{Weinzierl:2006yt}
S.~Weinzierl, {\it {The Forward-backward asymmetry at NNLO revisited}},  {\em
  Phys. Lett. B} {\bf 644} (2007) 331--335,
  [\href{http://arxiv.org/abs/hep-ph/0609021}{{\tt hep-ph/0609021}}].

\bibitem{Trocsanyi:2015zma}
Z.~Tr\'ocs\'anyi, G.~Somogyi, and F.~Tramontano, {\it {Fully Differential Decay
  Rate of a Standard Model Higgs Boson into Two $b$-jets at NNLO}},  {\em Acta
  Phys. Polon. B} {\bf 46} (2015), no.~11 2097.

\bibitem{Ferrera:2017zex}
G.~Ferrera, G.~Somogyi, and F.~Tramontano, {\it {Associated production of a
  Higgs boson decaying into bottom quarks at the LHC in full NNLO QCD}},  {\em
  Phys. Lett. B} {\bf 780} (2018) 346--351,
  [\href{http://arxiv.org/abs/1705.10304}{{\tt arXiv:1705.10304}}].

\bibitem{Caola:2017xuq}
F.~Caola, G.~Luisoni, K.~Melnikov, and R.~R\"ontsch, {\it {NNLO QCD corrections
  to associated $WH$ production and $H \to b \bar b$ decay}},  {\em Phys. Rev.
  D} {\bf 97} (2018), no.~7 074022,
  [\href{http://arxiv.org/abs/1712.06954}{{\tt arXiv:1712.06954}}].

\bibitem{Gauld:2019yng}
R.~Gauld, A.~Gehrmann-De~Ridder, E.~W.~N. Glover, A.~Huss, and I.~Majer, {\it
  {Associated production of a Higgs boson decaying into bottom quarks and a
  weak vector boson decaying leptonically at NNLO in QCD}},  {\em JHEP} {\bf
  10} (2019) 002, [\href{http://arxiv.org/abs/1907.05836}{{\tt
  arXiv:1907.05836}}].

\bibitem{Czakon:2020coa}
M.~Czakon, A.~Mitov, M.~Pellen, and R.~Poncelet, {\it {NNLO QCD predictions for
  W+c-jet production at the LHC}},  {\em JHEP} {\bf 06} (2021) 100,
  [\href{http://arxiv.org/abs/2011.01011}{{\tt arXiv:2011.01011}}].

\bibitem{Hartanto:2022qhh}
H.~B. Hartanto, R.~Poncelet, A.~Popescu, and S.~Zoia, {\it
  {Next-to-next-to-leading order QCD corrections to $Wb\bar{b}$ production at
  the LHC}},  {\em Phys. Rev. D} {\bf 106} (2022), no.~7 074016,
  [\href{http://arxiv.org/abs/2205.01687}{{\tt arXiv:2205.01687}}].

\bibitem{Hartanto:2022ypo}
H.~B. Hartanto, R.~Poncelet, A.~Popescu, and S.~Zoia, {\it {Flavour
  anti-$k_\text{T}$ algorithm applied to $Wb\bar{b}$ production at the LHC}},
  \href{http://arxiv.org/abs/2209.03280}{{\tt arXiv:2209.03280}}.

\bibitem{Czakon:2022khx}
M.~Czakon, A.~Mitov, M.~Pellen, and R.~Poncelet, {\it {A detailed investigation
  of W+c-jet at the LHC}},  {\em JHEP} {\bf 02} (2023) 241,
  [\href{http://arxiv.org/abs/2212.00467}{{\tt arXiv:2212.00467}}].

\bibitem{Gauld:2023zlv}
R.~Gauld, A.~Gehrmann-De~Ridder, E.~W.~N. Glover, A.~Huss, A.~R. Garcia, and
  G.~Stagnitto, {\it {NNLO QCD predictions for Z-boson production in
  association with a charm jet within the LHCb fiducial region}},  {\em Eur.
  Phys. J. C} {\bf 83} (2023), no.~4 336,
  [\href{http://arxiv.org/abs/2302.12844}{{\tt arXiv:2302.12844}}].

\bibitem{Dokshitzer:1997in}
Y.~L. Dokshitzer, G.~D. Leder, S.~Moretti, and B.~R. Webber, {\it {Better jet
  clustering algorithms}},  {\em JHEP} {\bf 08} (1997) 001,
  [\href{http://arxiv.org/abs/hep-ph/9707323}{{\tt hep-ph/9707323}}].

\bibitem{Wobisch:1998wt}
M.~Wobisch and T.~Wengler, {\it {Hadronization corrections to jet
  cross-sections in deep inelastic scattering}},  in {\em {Workshop on Monte
  Carlo Generators for HERA Physics (Plenary Starting Meeting)}}, pp.~270--279,
  4, 1998.
\newblock \href{http://arxiv.org/abs/hep-ph/9907280}{{\tt hep-ph/9907280}}.

\bibitem{Hoeche:2009rj}
S.~Hoeche, F.~Krauss, S.~Schumann, and F.~Siegert, {\it {QCD matrix elements
  and truncated showers}},  {\em JHEP} {\bf 05} (2009) 053,
  [\href{http://arxiv.org/abs/0903.1219}{{\tt arXiv:0903.1219}}].

\bibitem{Karlberg:2021kwr}
A.~Karlberg, G.~P. Salam, L.~Scyboz, and R.~Verheyen, {\it {Spin correlations
  in final-state parton showers and jet observables}},  {\em Eur. Phys. J. C}
  {\bf 81} (2021), no.~8 681, [\href{http://arxiv.org/abs/2103.16526}{{\tt
  arXiv:2103.16526}}].

\bibitem{vanBeekveld:2022ukn}
M.~van Beekveld, S.~Ferrario~Ravasio, K.~Hamilton, G.~P. Salam, A.~Soto-Ontoso,
  G.~Soyez, and R.~Verheyen, {\it {PanScales showers for hadron collisions:
  all-order validation}},  {\em JHEP} {\bf 11} (2022) 020,
  [\href{http://arxiv.org/abs/2207.09467}{{\tt arXiv:2207.09467}}].

\bibitem{Salam:2007xv}
G.~P. Salam and G.~Soyez, {\it {A Practical Seedless Infrared-Safe Cone jet
  algorithm}},  {\em JHEP} {\bf 05} (2007) 086,
  [\href{http://arxiv.org/abs/0704.0292}{{\tt arXiv:0704.0292}}].

\bibitem{Gallicchio:2018elx}
J.~Gallicchio and Y.-T. Chien, {\it {Quit Using Pseudorapidity, Transverse
  Energy, and Massless Constituents}},
  \href{http://arxiv.org/abs/1802.05356}{{\tt arXiv:1802.05356}}.

\bibitem{Buckley:2015gua}
A.~Buckley and C.~Pollard, {\it {QCD-aware partonic jet clustering for
  truth-jet flavour labelling}},  {\em Eur. Phys. J. C} {\bf 76} (2016), no.~2
  71, [\href{http://arxiv.org/abs/1507.00508}{{\tt arXiv:1507.00508}}].

\bibitem{JADE:1986kta}
{\bf JADE} Collaboration, W.~Bartel et~al., {\it {Experimental Studies on
  Multi-Jet Production in e+ e- Annihilation at PETRA Energies}},  {\em Z.
  Phys. C} {\bf 33} (1986) 23.

\bibitem{JADE:1988xlj}
{\bf JADE} Collaboration, S.~Bethke et~al., {\it {Experimental Investigation of
  the Energy Dependence of the Strong Coupling Strength}},  {\em Phys. Lett. B}
  {\bf 213} (1988) 235--241.

\bibitem{Behring:2020uzq}
A.~Behring, W.~Bizo\'n, F.~Caola, K.~Melnikov, and R.~R\"ontsch, {\it {Bottom
  quark mass effects in associated $WH$ production with the $H \to b\bar{b}$
  decay through NNLO QCD}},  {\em Phys. Rev. D} {\bf 101} (2020), no.~11
  114012, [\href{http://arxiv.org/abs/2003.08321}{{\tt arXiv:2003.08321}}].

\bibitem{Larkoski:2014wba}
A.~J. Larkoski, S.~Marzani, G.~Soyez, and J.~Thaler, {\it {Soft Drop}},  {\em
  JHEP} {\bf 05} (2014) 146, [\href{http://arxiv.org/abs/1402.2657}{{\tt
  arXiv:1402.2657}}].

\bibitem{Andersson:1988gp}
B.~Andersson, G.~Gustafson, L.~Lonnblad, and U.~Pettersson, {\it {Coherence
  Effects in Deep Inelastic Scattering}},  {\em Z. Phys. C} {\bf 43} (1989)
  625.

\bibitem{Cacciari:2011ma}
M.~Cacciari, G.~P. Salam, and G.~Soyez, {\it {FastJet User Manual}},  {\em Eur.
  Phys. J. C} {\bf 72} (2012) 1896, [\href{http://arxiv.org/abs/1111.6097}{{\tt
  arXiv:1111.6097}}].

\bibitem{hida2000quad}
Y.~Hida, X.~S. Li, and D.~H. Bailey, {\it Quad-double arithmetic: Algorithms,
  implementation, and application},  in {\em 15th IEEE Symposium on Computer
  Arithmetic}, pp.~155--162, 2000.

\bibitem{Dasgupta:2020fwr}
M.~Dasgupta, F.~A. Dreyer, K.~Hamilton, P.~F. Monni, G.~P. Salam, and G.~Soyez,
  {\it {Parton showers beyond leading logarithmic accuracy}},  {\em Phys. Rev.
  Lett.} {\bf 125} (2020), no.~5 052002,
  [\href{http://arxiv.org/abs/2002.11114}{{\tt arXiv:2002.11114}}].

\bibitem{Hamilton:2020rcu}
K.~Hamilton, R.~Medves, G.~P. Salam, L.~Scyboz, and G.~Soyez, {\it {Colour and
  logarithmic accuracy in final-state parton showers}},  {\em JHEP} {\bf 03}
  (2021), no.~041 041, [\href{http://arxiv.org/abs/2011.10054}{{\tt
  arXiv:2011.10054}}].

\bibitem{CMS:2017odg}
{\bf CMS} Collaboration, A.~M. Sirunyan et~al., {\it {Evidence for the Higgs
  boson decay to a bottom quark\textendash{}antiquark pair}},  {\em Phys. Lett.
  B} {\bf 780} (2018) 501--532, [\href{http://arxiv.org/abs/1709.07497}{{\tt
  arXiv:1709.07497}}].

\bibitem{ATLAS:2020fcp}
{\bf ATLAS} Collaboration, G.~Aad et~al., {\it {Measurements of $WH$ and $ZH$
  production in the $H \rightarrow b\bar{b}$ decay channel in $pp$ collisions
  at 13 TeV with the ATLAS detector}},  {\em Eur. Phys. J. C} {\bf 81} (2021),
  no.~2 178, [\href{http://arxiv.org/abs/2007.02873}{{\tt arXiv:2007.02873}}].

\bibitem{ATLAS:2020jwz}
{\bf ATLAS} Collaboration, G.~Aad et~al., {\it {Measurement of the associated
  production of a Higgs boson decaying into $b$-quarks with a vector boson at
  high transverse momentum in $pp$ collisions at $\sqrt{s} = 13$ TeV with the
  ATLAS detector}},  {\em Phys. Lett. B} {\bf 816} (2021) 136204,
  [\href{http://arxiv.org/abs/2008.02508}{{\tt arXiv:2008.02508}}].

\bibitem{CMS:2020zge}
{\bf CMS} Collaboration, A.~M. Sirunyan et~al., {\it {Inclusive search for
  highly boosted Higgs bosons decaying to bottom quark-antiquark pairs in
  proton-proton collisions at $\sqrt{s} =$ 13 TeV}},  {\em JHEP} {\bf 12}
  (2020) 085, [\href{http://arxiv.org/abs/2006.13251}{{\tt arXiv:2006.13251}}].

\bibitem{Butterworth:2008iy}
J.~M. Butterworth, A.~R. Davison, M.~Rubin, and G.~P. Salam, {\it {Jet
  substructure as a new Higgs search channel at the LHC}},  {\em Phys. Rev.
  Lett.} {\bf 100} (2008) 242001, [\href{http://arxiv.org/abs/0802.2470}{{\tt
  arXiv:0802.2470}}].

\bibitem{Marzani:2019hun}
S.~Marzani, G.~Soyez, and M.~Spannowsky, {\em {Looking inside jets: an
  introduction to jet substructure and boosted-object phenomenology}},
  vol.~958.
\newblock Springer, 2019.

\bibitem{Sjostrand:2014zea}
T.~Sj\"ostrand, S.~Ask, J.~R. Christiansen, R.~Corke, N.~Desai, P.~Ilten,
  S.~Mrenna, S.~Prestel, C.~O. Rasmussen, and P.~Z. Skands, {\it {An
  introduction to PYTHIA 8.2}},  {\em Comput. Phys. Commun.} {\bf 191} (2015)
  159--177, [\href{http://arxiv.org/abs/1410.3012}{{\tt arXiv:1410.3012}}].

\bibitem{Bierlich:2022pfr}
C.~Bierlich et~al., {\it {A comprehensive guide to the physics and usage of
  PYTHIA 8.3}},  \href{http://arxiv.org/abs/2203.11601}{{\tt
  arXiv:2203.11601}}.

\bibitem{Corke:2010yf}
R.~Corke and T.~Sjostrand, {\it {Interleaved Parton Showers and Tuning
  Prospects}},  {\em JHEP} {\bf 03} (2011) 032,
  [\href{http://arxiv.org/abs/1011.1759}{{\tt arXiv:1011.1759}}].

\bibitem{Skands:2014pea}
P.~Skands, S.~Carrazza, and J.~Rojo, {\it {Tuning PYTHIA 8.1: the Monash 2013
  Tune}},  {\em Eur. Phys. J. C} {\bf 74} (2014), no.~8 3024,
  [\href{http://arxiv.org/abs/1404.5630}{{\tt arXiv:1404.5630}}].

\bibitem{Gras:2017jty}
P.~Gras, S.~H\"oche, D.~Kar, A.~Larkoski, L.~L\"onnblad, S.~Pl\"atzer,
  A.~Si\'odmok, P.~Skands, G.~Soyez, and J.~Thaler, {\it {Systematics of
  quark/gluon tagging}},  {\em JHEP} {\bf 07} (2017) 091,
  [\href{http://arxiv.org/abs/1704.03878}{{\tt arXiv:1704.03878}}].

\bibitem{Dasgupta:2014yra}
M.~Dasgupta, F.~Dreyer, G.~P. Salam, and G.~Soyez, {\it {Small-radius jets to
  all orders in QCD}},  {\em JHEP} {\bf 04} (2015) 039,
  [\href{http://arxiv.org/abs/1411.5182}{{\tt arXiv:1411.5182}}].

\bibitem{Catani:1996vz}
S.~Catani and M.~H. Seymour, {\it {A General algorithm for calculating jet
  cross-sections in NLO QCD}},  {\em Nucl. Phys. B} {\bf 485} (1997) 291--419,
  [\href{http://arxiv.org/abs/hep-ph/9605323}{{\tt hep-ph/9605323}}]. [Erratum:
  Nucl.Phys.B 510, 503--504 (1998)].

\bibitem{vanBeekveld:2019prq}
M.~van Beekveld, W.~Beenakker, E.~Laenen, and C.~D. White, {\it
  {Next-to-leading power threshold effects for inclusive and exclusive
  processes with final state jets}},  {\em JHEP} {\bf 03} (2020) 106,
  [\href{http://arxiv.org/abs/1905.08741}{{\tt arXiv:1905.08741}}].

\end{thebibliography}\endgroup

%======================================================================
%======================================================================
%======================================================================
%======================================================================
%======================================================================
%======================================================================
%======================================================================
%

\end{document}